\pdfoutput=1

\documentclass[aps,prd,%
onecolumn,%
10pt,%
tightenlines,%
eqsecnum,%
showkeys,%
showpacs,%
floats,%
superscriptaddress, 
notitlepage,%
nofootinbib,%
nobibnotes,%
amsfonts,amssymb,amsmath%
]{revtex4-2}

\usepackage{ifthen}
\makeatletter%
\usepackage{xstring}
\IfSubStr{\@classoptionslist}{preprint}%
{}%
{}%
\makeatother%

\newcommand{\arxiv}{0}

\usepackage{graphicx} 
\usepackage[caption=false]{subfig}             

\graphicspath{{./},{Figures/}}

\usepackage{hyperref}

\usepackage{enumitem}

\usepackage{mathtools}   

\usepackage{comment}
\includecomment{comment}  

\usepackage{array} 

\newcommand{\bra}[1]{\mbox{$\langle {#1} |$}}
\newcommand{\ket}[1]{\mbox{$| {#1} \rangle$}}
\newcommand{\bracket}[2]{\mbox{$\langle {#1} \!\mid\! {#2} \rangle$}}
\newcommand{\braket}[2]{\mbox{$\langle {#1} \!\mid\! {#2} \rangle$}}
\newcommand{\ketbra}[2]{\mbox{$| {#1} \rangle\langle {#2} |$}}
\newcommand{\melt}[3]{\mbox{$\langle {#1} | {#2} | {#3} \rangle$}}

\newcommand{\expct}[1]{\mbox{$\langle {#1} \rangle$}}

\newcommand{\brasub}[2]{\ensuremath{{}_{#2}\kern-1pt\langle {#1} |}}
\newcommand{\ketsub}[2]{\ensuremath{| {#1} \rangle_{\kern-1pt #2}}}
\newcommand{\bracketsub}[4]{\ensuremath{{}_{#3}\kern-1pt\langle {#1} \!\mid\! {#2} \rangle_{\kern-1pt #4}}}
\newcommand{\ketbrasub}[4]{\ensuremath{| {#1} \rangle_{\kern-1pt #3}{}_{#4}\kern-1pt\langle {#2} |}}
\newcommand{\meltsub}[5]{\ensuremath{{}_{#4}\kern-1pt\langle {#1} \!\mid\! {#2} \!\mid\! {#3} \rangle_{\kern-1pt #5}}}

\newcommand{\rbrasub}[2]{\ensuremath{{}_{#2}\kern-1pt( {#1} |}}
\newcommand{\rketsub}[2]{\ensuremath{| {#1} )_{\kern-1pt #2}}}
\newcommand{\rbracketsub}[4]{\ensuremath{{}_{#3}\kern-1pt( {#1} \!\mid\! {#2} )_{\kern-1pt #4}}}
\newcommand{\rketbrasub}[4]{\ensuremath{| {#1} )_{\kern-1pt #3}{}_{#4}\kern-1pt( {#2} |}}
\newcommand{\rmeltsub}[5]{\ensuremath{{}_{#4}\kern-1pt( {#1} \!\mid\! {#2} \!\mid\! {#3} )_{\kern-1pt #5}}}

\def\Re{\mathbb{R}}

\def\Co{\mathbb{C}}

\def\Int{\mathbb{Z}}

\usepackage{dsfont} 
\def\Id{\mathbb{I}}

\newcommand{\isrep}{\mathrel{:=}}

\def\Aop{\hat{A}}
\def\Bop{\hat{B}}
\def\Cop{\hat{C}}
\def\Dop{\hat{D}}
\def\xop{\hat{x}}
\def\pop{\hat{p}}
\def\qop{\hat{q}}

\def\Uop{\hat{U}}
\def\Vop{\hat{V}}
\def\Zop{\hat{Z}}
\def\Uopr{\hat{U}_{\mathrm{r}}}
\def\Vopr{\hat{V}_{\mathrm{r}}}

\def\Hilbert{\mathcal{H}}
\def\Hsch{\mathcal{H}_{\mathrm{Sch}}}
\def\Hpoly{\mathcal{H}_{\mathrm{poly}}}

\def\Hgraphmuo{\mathcal{H}_{\gamma_{\mu_0}}}

\def\Red{\Re_d}
\def\Redhat{\hat{\Re}_d}
\def\Reb{\Re_{\mathrm{B}}}
\def\Circ{U(1)}

\def\CircS{\mathbb{S}^1}

\def\CircT{\mathbb{T}}
\def\CircThat{\hat{\mathbb{T}}}
\def\IntB{\Int_{\mathrm{B}}}

\def\Gdhat{\hat{G}_d}
\def\Gb{G_{\mathrm{B}}}
\usepackage{accents}
\def\dhatG{\widehat{\vphantom{\rule{1pt}{7.5pt}}\smash{\hat{G}}}}
\def\dhatGd{\widehat{\vphantom{\rule{1pt}{7.5pt}}\smash{\hat{G}_d}}}

\begin{document}
\count\footins = 1000


\title{Polymer quantum mechanics on compact configuration spaces}


\author{Maxwell R.~Siebersma}
\email[]{E-mail: msiebe5@lsu.edu}
\affiliation{%
Department of Physics and Astronomy, Louisiana State University,
Baton Rouge, Louisiana, 70803, USA}

\author{Basie Seibert}
\email[]{E-mail: bseibert1@unm.edu}
\affiliation{%
Department of Physics and Astronomy, Quantum New Mexico Institute, University of New Mexico,
Albuquerque, New Mexico 87106, USA}

\author{Samuel Shuman}
\email[]{E-mail: samuels@bigbend.edu}
\affiliation{%
Division of Math \& Science, Big Bend Community College,
Moses Lake, Washington, 98837, USA}


\author{David A.~Craig}
\email[]{E-mail: craigda@oregonstate.edu}
\affiliation{%
Department of Physics, Oregon State University,
Corvallis, Oregon, 97331, USA}


\begin{abstract}
``Polymer quantum mechanics'' is the name given to a quantization scheme inspired by loop quantum gravity in which the configuration space of the theory is chosen to have a discrete topology. Polymer quantization yields a representation of the canonical commutation relations that is genuinely distinct from the conventional ``Schr\"{o}dinger'' representation. In this paper, we summarize the main features of polymer quantum mechanics and investigate in detail the polymer quantization of systems with configuration spaces that are classically compact. We show explicitly how using the standard construction of polymer states leads to a Hilbert space of states defined on a finite graph of points. By way of example, we find the exact energy eigenvalues and eigenfunctions for a particle on a ring 
and a particle in a box defined on such lattices,
and discuss similarities and differences from standard Schr\"{o}dinger quantum mechanics. 
We also explore the continuum limit of states in 
these systems, 
and demonstrate in detail how the exact eigenfunctions in the position representation approach their continuum counterparts.
\end{abstract}

%

\maketitle

\section{Introduction}
\label{sec:intro}

``Polymer quantization'' is a scheme for quantization of finite-dimensional systems originally inspired by loop quantum cosmology. 
Famously, loop-quantized gravitational models with a finite number of degrees of freedom, such as cosmological models and black holes, are generically non-singular \cite{liv-rev,ashsingh11}, whereas many of those same models quantized in the conventional fashion remain singular in the quantum theory \cite{aps,Ashtekar:2006wn,CS10c,dac13a}, contrary to the hope that quantum mechanics might resolve the spacetime singularities inevitable in classical general relativity. Polymer quantum mechanics (PQM) \cite{afwPQM,physhilbspace,continuumlimit,velhinho04,bohrpaper} was conceived as a simplified setting in which to explore some of the mathematical and physical features of the methods used to arrive at the loop quantization \cite{abl03,bohrpaper,bojo13,Thiemann}.%
\footnote{It is reasonable to wonder about the origins of the evocative name ``polymer representation''. The authors of \cite{afwPQM}, in which the term was first introduced, say it is because ``states are mathematically analogous to the polymer-like excitations of quantum geometry''.  The name ``polymer quantum mechanics'' is a reference to its inspiration, not anything intrinsic to the theory itself.
} %

The earliest quantum cosmological models analyzed in detail in the polymer quantization framework were highly symmetric homogeneous and isotropic Friedmann-Lemaitre-Robertson-Walker (FLRW) spacetimes \cite{abl03,Ashtekar:2006wn}. In these models, the gravitational degree of freedom was quantized under the polymer quantization paradigm, while the matter sector was quantized using standard quantization techniques. More recently, there has been interest in quantum cosmological models which involve polymer-quantized matter alongside polymer quantized gravitational dynamics \cite{propagator,Kreienbuehl:2013toa,Zulfiqar:2025aef}, as well as analyses of quantization ambiguities for polymer quantized cosmology \cite{LandscapePQC}. As a distinct method of quantization, techniques from PQM have also been applied to other problems such as the analysis of statistical mechanical systems \cite{polymerStatThermo}, gravitational wave detection \cite{gravitationalwave,Garcia-Chung:2022pdy}, and systems relevant to quantum computing \cite{crowe2025}.

The existence of fundamentally distinct quantizations of even simple classical systems may at first seem surprising due to the Stone-von Neumann theorem, which guarantees that representations of the canonical commutation relations (CCR) are unitarily equivalent. Like all theorems, however, the Stone-von Neumann theorem has \emph{assumptions}, and like all assumptions, in the face of significant questions about the applicability of the results implied by the theorem, it is prudent to critically evaluate their necessity. In the present case, recognition that conventional quantization methods applied to cosmological models yielded quantum theories which still contained spacetime singularities motivated that reassessment.

In polymer quantization, one of the assumptions crucial to the Stone-von Neumann theorem is relaxed, opening the door to a genuinely distinct quantum theory. This assumption 
turns out to be related to the topology of the underlying configuration space of the system. The effectively discrete configuration space that emerges mimics the discrete spacetime widely expected to emerge from quantum gravity \cite{Kiefer25}. Polymer quantization thus evades the uniqueness result provided by the Stone-von Neumann theorem for canonical quantization of the canonical commutation relations by violating one of that theorem's key assumptions, resulting in a genuinely distinct quantum theory that nonetheless has the expected classical limit.

In this paper we describe the polymer quantization of a simple one-dimensional quantum system, 
a spinless particle confined to motion on a ring, a system that has not previously been investigated in the literature on polymer quantum mechanics. (Previously solved systems in PQM include the simple harmonic oscillator and the free particle \cite{afwPQM,continuumlimit}. A system with a compact phase space in loop quantum gravity has been studied in \cite{rovvid15}.) Akin to the widely-studied ``particle in a box'', the particle on a ring admits representations of the one-dimensional rotation group because of the underlying two-dimensional nature of the system. Both systems have a compact configuration space, a feature we explore explicitly as a focal point of the paper.  We show that polymer quantization of such a system leads to a configuration space which is a spatial lattice effectively comprising a finite number of points. As such, dynamical equations expressed in terms of the configuration space variable are most naturally finite difference equations, rather than differential equations as is the case in standard quantum mechanics.

This work thus emphasizes the position space representation in PQM. Apart from the fact that the position representation is more familiar from the study of these examples in textbook quantum mechanics \cite{McIntyre}, this follows naturally from the need to impose periodic spatial boundary conditions when considering the particle on a ring. The quantum mechanics of polymer quantized systems bears a close relationship to quantum mechanics on discrete lattices, a subject of interest in its own right. Outside of the context of loop quantum gravity (LQG) and PQM, quantization on lattice spacetimes enjoys a long history, including early efforts in high energy physics to formulate lattice quantum chromodynamics \cite{Wilson:1974sk}. Quantum mechanics on discrete configuration spaces has continued to be a subject of interest in fundamental physics \cite{STOVICEK1984157,lorente1997quantum}, including investigations into quantum dynamics on spaces which are not only discrete, but also finite \cite{de2002quantum,vourdas04,carroll2023completely}. As will be seen in the sequel, polymer quantum mechanics is distinct from these models in that finite lattice structures can emerge directly from polymer quantization of the canonical commutation relations, without the need for imposing a lattice ad hoc. Therefore, this investigation is not only of interest because of its relevance to loop quantum cosmology, but also as a mechanism to motivate quantum dynamics on discrete spacetimes more generally.

Because both the Stone-von Neumann theorem and polymer quantization itself are unfamiliar to many physicists, we give a fairly extensive account of the foundations of polymer quantization in the hopes of making this material more broadly accessible by presenting the subject in a manner that, while still involving a significant degree of mathematics, is less formal than many existing treatments. Much of this part of the paper is pedagogical, but we believe also offers some fresh insight into the physics of polymer quantization.  While we briefly summarize much of the necessary mathematics and provide extensive references for the reader interested in learning more, we do nonetheless assume at least passing familiarity with some core mathematical concepts such as elementary point-set topology, representations of groups, the basic idea of a measure, Hilbert space, and the theory of operators in quantum mechanics. (For relevant background, see, for example, \cite{szekeres,Pal19,hallQM}.) 

The plan of the paper is as follows. In section \ref{sec:SvN}, we review canonical quantization, the canonical commutation relations and the Weyl algebra, and the Stone-von Neumann theorem, to provide background and motivation for polymer quantization. In section \ref{sec:pqm}, we outline the construction of PQM, beginning with a discussion  of how the choice of a discrete topology for space leads naturally to the definition of a Hilbert space $\Hpoly$ that reflects that underlying discreteness. After presenting details of the position and momentum representation of states on $\Hpoly$, we describe how to construct the  Hamiltonian dynamics of ``polymerized'' quantum theories that naturally live on a discrete lattice of spatial points.  Appendix \ref{app:bohr} discusses some of the technical details of what is known as the Bohr compactification of the real line that are needed for the mathematical formulation of the theory. The emphasis in these sections is pedagogical, and the presentation and calculations are thus more detailed and explicit than much of the existing literature on the subject. Section \ref{sec:pqmcompact} continues with an analysis of polymer quantum mechanics for systems with compact configuration spaces, the focal point of the paper. In section \ref{sec:pqmring}, we apply this analysis to the particle on a ring. We solve the time-independent Schrödinger equation for this system using two different methods, one relying on properties of the operator constituents of the Hamiltonian, and one using standard recurrence relation methods.  (A parallel analysis of the particle in a box is offered in appendix \ref{app:pqmbox}.) We end with discussion of the results.

\section{Quantization and the Stone-von Neumann theorem}
\label{sec:SvN}

Identifying a quantum theory that, in the limit $\hbar \rightarrow 0$, corresponds to a given classical theory, is referred to as ``quantization''.  Loosely put, this involves choosing an algebra of operators to represent the physical observables of the theory, and a suitable Hilbert space of states on which those observables act.  The algebra of observables should, in the limit $\hbar \rightarrow 0$, correspond to the ``Poisson algebra'' of the classical system, specifically, the algebra of functions $f(q,p)$ on the phase space $(q_i,p_i)$ of the classical system defined by the Poisson bracket
\begin{equation}
\{f,g\}_{q,p} = \sum_{i=1}^{n}
\left\{\frac{\partial f}{\partial q_i}\frac{\partial g}{\partial p_i}
-\frac{\partial f}{\partial p_i}\frac{\partial g}{\partial q_i}\right\} ~.
\label{eq:PB}
\end{equation}
Here $(q_i,p_i)$ are a set of canonical coordinates and their conjugate momenta.  The Poisson brackets of the canonical coordinates themselves are, of course, simply
\begin{equation}
\{q_i,p_j\}_{q,p} = \delta_{ij} ~.
\label{eq:CPB}
\end{equation}
It is well-known that the Poisson bracket is a canonical invariant i.e.~is invariant under canonical transformations (\emph{viz.}~transformations $(q,p) \rightarrow (Q(q,p),P(q,p))$ that leave the action invariant), and so is independent of the particular choice of coordinates on phase space \cite{HellSah21}. The Poisson bracket captures the so-called ``symplectic structure'' of the classical theory \cite{schutzgeo}.

\subsection{Canonical quantization}
\label{sec:canquant}

In order to explain what makes the polymer representation distinct, let us first briefly summarize the conventional procedure for quantizing a given classical theory.

Focusing on the one dimensional case, (very) roughly speaking, to ``quantize'' a classical physical system described by phase space variables $(q,p)$ we identify corresponding operators $\hat{q}, \hat{p}$ on a Hilbert space $\Hilbert$ that satisfy the so-called ``canonical commutation relations'' (CCR)
\begin{equation}
[\hat{q},\hat{p}]\equiv\hat{q}\hat{p}-\hat{p}\hat{q}=i\hbar\,\Id ~.
\label{eq:CCR}
\end{equation}
Clearly we also have
\begin{equation}
    [\hat{q},\hat{q}]=[\hat{p},\hat{p}]=0 ~.
\end{equation}
The Hilbert space $\Hilbert$ of vectors on which those operators act thus carries a representation of the CCR, and may be identified with the space of quantum states of the system. The Stone-von Neumann theorem, to be discussed further below, guarantees that (under certain assumptions) this representation is essentially unique, that is, is unique up to unitary transformations.

Classical physical observables represented by functions $A(q,p)$ on phase space become quantum operators $\hat{A}$ on $\Hilbert$ through the substitution $A(q,p) \rightarrow \hat{A} = \hat{A}(\hat{q},\hat{p})$. Because $\hat{q}$ and $\hat{p}$ do not commute whereas the classical coordinates $(q,p)$ do, this prescription is in general not unique, and other considerations must be used to resolve the resulting operator ordering ambiguities.

Because of the obvious correspondence between Eqs.(\ref{eq:CPB}) and (\ref{eq:CCR}), going back to Dirac it is often said that quantization amounts to the identification of the classical Poisson bracket with the quantum commutator \cite{DiracQM4e,StoneGoldbart}
\begin{equation}
 \{q,p\} \sim     \frac{[\hat{q},\hat{p}]}{i\hbar} ~.
\end{equation}
More generally, commutators of operators $\hat{A}(\hat{q},\hat{p})$ should mirror the Poisson brackets of the corresponding classical observables $A(q,p)$. The key point being made here is that ``quantization'' amounts to identifying a space of operators whose commutation relations mirror the classical Poisson bracket (symplectic) structure of the underlying classical theory.%
\footnote{For a general discussion of the freedom available to define canonically conjugate pairs of operators in separable Hilbert spaces, see \cite{vanenk25}.
} %

Clearly there is much more to ``quantization'' than this simple account. The usual approach focuses on finding representations of the Poisson algebra generated by the symplectic form $dq\wedge dp$ -- effectively the Poisson bracket, Eq.(\ref{eq:CPB}) -- on the tangent bundle of a continuous configuration space \cite{schutzgeo,hallQM}. 
As will be seen in the sequel, however, because in polymer quantum mechanics the underlying configuration space will be chosen to have a discrete topology, we take an approach inspired by harmonic analysis to identifying the quantum momentum space that is rooted in the idea that momentum is the generator of translations. 

All of this is, of course, backwards from the way Nature does it, but since the starting point for developing a theoretical model of a quantum system is most often its classical limit, quantization is a time-honored way of exploring the possible quantum theories that may underlie a given classical theory.   The quantum theory of gravitation is no exception.


\subsection{Translation operators}
\label{sec:translate}

The canonical commutation relations amount to the statement that -- as in the classical theory -- $\hat{q}$ and $\hat{p}$ act as the generators of translations for one another, that is to say, $\hat{p}$ generates translations in $q$ and vice-versa, as befits the classical relationship between a generalized coordinate and its conjugate momentum \cite{HellSah21}. 
In the quantum theory this is usually argued starting with infinitesimal translations (see, e.g., \cite{Sakurai3e}), but for our purposes it will be convenient to see this from the point of view of the operator for finite translations in $q$,
\begin{equation}
\mathbb{T}(a) \equiv e^{ia\hat{p}} ~,
\label{eq:T}
\end{equation}
where we temporarily set $\hbar=1$. Starting with a relation sometimes called the Hadamard formula and proceeding formally –– that is to say, without concerning oneself with operator domains, convergence, and related mathematical details –– for operators $\Aop$ and $\Bop$ we have 
\begin{equation}
e^{\lambda\Bop}\Aop e^{-\lambda\Bop} = \Aop + \lambda[\Bop,\Aop] 
   + \frac{\lambda^2}{2!}[\Bop,[\Bop,\Aop]] 
   + \frac{\lambda^3}{3!}[\Bop,[\Bop,[\Bop,\Aop]]] 
   + \cdots ~, 
\label{eq:Hadamard}
\end{equation}
where the exponential of an operator is defined as usual by its series expansion.%
\footnote{Formally, at least, the right-hand side of the expression Eq.~(\ref{eq:Hadamard}) is simply the Taylor expansion of the left.
} %
As an immediate consequence, 
\begin{subequations}
\begin{align}
\Aop e^{\lambda\Bop} &= 
e^{\lambda\Bop}\left(\Aop + \lambda [\Aop,\Bop] + \frac{\lambda^2}{2!}[[\Aop,\Bop],\Bop] 
  + \frac{\lambda^3}{3!}[[[\Aop,\Bop],\Bop],\Bop] + \cdots   \right) \\
  & \equiv e^{\lambda\Bop}\left(\Aop + \lambda\Cop\right) ~.
\end{align}
\label{eq:HadamardCommutator}%
\end{subequations}
For operators $\qop$ and $\pop$ that satisfy the CCR, $[\qop,\pop]= i\,\Id $; this tells us that 
\begin{subequations}
\begin{align}
\hat{q}\,e^{ia\pop} 
  &= e^{ia\pop}\left(\qop+ia\,[\qop,\pop] \right) \\
  & = e^{ia\pop}\left(\qop-a\,\Id \right) ~.
\end{align}
\label{eq:qexpp}%
\end{subequations}
Thus we find that on a basis of eigenstates $\ket{q}$ of $\qop$, $\qop\ket{q}=q\ket{q}$, we have
\begin{subequations}
\begin{align}
\hat{q}\,e^{ia\pop} \ket{q} 
  &= e^{ia\pop}\left(\qop-a\Id \right) \ket{q} \\
  &= (q-a) e^{ia\pop} \ket{q} ~, 
\end{align}
\label{eq:qexppq}%
\end{subequations}
so that
\begin{equation}
\mathbb{T}(a) \ket{q} = \ket{q-a} ~,
\label{eq:Tq}
\end{equation}
and the CCR tell us that $\mathbb{T}(a)$ is a translation operator in the coordinate $q$.  Conversely, if we start with the \emph{requirement} that $\mathbb{T}(a)$ is a translation operator for every $\ket{q}$, 
\begin{subequations}
\begin{align}
\hat{q}\,e^{ia\pop} \ket{q} 
  &= (q-a) e^{ia\pop} \ket{q} \\ 
  &= e^{ia\pop}\left(\qop+ia\,\Cop \right) \ket{q} ~,
\end{align}
\label{eq:qexppqC}%
\end{subequations}
we quickly find that
\begin{equation}
\Cop = i\,\Id ~.
\label{eq:CId}
\end{equation}
As
\begin{equation}
\Cop = [\qop,\pop] + \frac{a}{2}[[\qop,\pop],\pop] + \mathcal{O}(a^2) ~,
\end{equation}
if Eq.(\ref{eq:CId}) holds for every $a$, it must be that $\pop$ commutes with $[\qop,\pop]$ so that all the terms in $a$ vanish, and therefore that $[\qop,\pop]=i\,\Id$, or in other words that $\qop$ and $\pop$ satisfy the CCR.  And of course, the same arguments show that $\qop$ generates translations in $p$: with $\pop\ket{p}=p\ket{p}$,
\begin{equation}
e^{ib\qop}\ket{p} = \ket{p+b} ~.
\end{equation}

Finally, these relations tells us in the usual way that the transformation function between the coordinate and its conjugate momentum is given by 
\begin{equation}
\bracket{q}{p} = N e^{ipq}
\label{eq:qp}
\end{equation}
for some normalization constant $N$.%
\footnote{$N=1/\sqrt{2\pi}$ if we take $\bracket{q}{q'}=\delta(q-q')$.
} %
This follows from acting with $\bra{p}$ on Eq.(\ref{eq:Tq}) for an infinitesimal translation $a=-\delta q$ and integrating:
\begin{subequations}
\begin{align}
\bracket{q+\delta q}{p} &= \melt{q}{e^{i\delta q\, \pop}}{p} \\
 &\approx \melt{q}{\Id + i\delta q\, \pop}{p} \\
 &= \bracket{q}{p} + i\delta q\, p\bracket{q}{p}
\end{align}
\label{eq:qpinf}%
\end{subequations}
so that
\begin{equation}
\frac{\partial}{\partial q}\bracket{q}{p} = ip\,\bracket{q}{p} ~.
\label{eq:qpderiv}
\end{equation}
In this way the relationship (\ref{eq:qp}) between a continuous position coordinate and momentum, defined as the generator of translations in that coordinate, is given essentially by the Fourier transform. Interestingly, we will see that when viewed through the appropriate mathematical lens, the essential features of this web of relationships among coordinate, momentum, translations, the CCR, and the Fourier transform persist when the coordinate is instead discrete. A crucial exception to this assertion concerns the (non-)existence of an infinitesimal generator of translations for a discrete coordinate, and this turns out to be the key feature that leads to the existence of the polymer quantization.

\subsection{Weyl algebra}
\label{sec:weyl}

Polymer quantization begins from the same starting point as conventional quantum theory based on the canonical commutation relations $[\xop,\pop] = i\hbar\,\Id$ and $[\xop,\xop]= [\pop,\pop] = 0$, where henceforth we will use the notation $\qop=\xop$ for the coordinate. These relations can be expressed in a physically equivalent form due to Weyl that lends itself more readily to making mathematically rigorous statements about the algebra of operators on Hilbert space in a quantum theory. What to a physicist might at first seem like technicalities turn out to be the doorway to the genuinely distinct quantization that is polymer quantum mechanics. It is to Weyl's formulation of the CCR and the attendant technicalities and results that we now turn.

The starting point for Weyl's formulation are the momentum and coordinate translation operators, in this context traditionally denoted $\Uop$ and $\Vop$,
\begin{subequations}
\begin{align}
    \Uop(\alpha) &= e^{i\alpha \hat{x}/\hbar}   \label{eq:WeylopU}  \\ 
    \Vop(\beta)  &= e^{i\beta \hat{p}/\hbar} ~, \label{eq:WeylopV}
\end{align}
\label{eq:Weylops}%
\end{subequations}
where $\alpha$ and $\beta$ are real parameters with dimensions of momentum and length, respectively. The canonical commutation relations can be expressed in terms of the Weyl operators $\Uop$ and $\Vop$ using the Campbell-Baker-Hausdorff (CBH) expansion,
\begin{equation}
e^{\Aop}e^{\Bop} = e^{\Zop}  
\label{eq:CBHAB} 
\end{equation}
where  
\begin{equation}
\Zop = 
 \Aop + \Bop + \frac{1}{2}[\Aop,\Bop] + \frac{1}{3!}\frac{1}{2}\left([\Aop,[\Aop,\Bop]] + [\Bop,[\Bop,\Aop]]\right) 
      + \frac{1}{4!}[[[\Bop,\Aop],\Aop],\Bop]+ \dots   ~,
\label{eq:CBHZ}
\end{equation}
a result closely related to the Hadamard formula, Eq.(\ref{eq:Hadamard}).%
\footnote{This result is proved in many texts on Lie groups and Lie algebras, see e.g.,\cite{hallLie2e}. A handy summary of this and related expansions such as the Hadamard formula noted above can be found in \cite{ZachosCBH}.
} %

For two operators which commute, $\Zop=\Aop+\Bop$. For position and momentum, the commutator $[\xop,\pop] = i\hbar\,\Id$ is not zero, but does commute with $\xop$ and $\pop$, $[\xop,[\xop,\pop]]=[\pop,[\xop,\pop]]=0$, simplifying the CBH expansion for the products of the Weyl operators $\Uop$ and $\Vop$ considerably. A short calculation shows that the canonical commutation relations correspond formally to the Weyl relations 
\begin{subequations}
\begin{align}
    \Uop(\alpha)\Vop(\beta)      &=  e^{-i\alpha\beta/\hbar}\Vop(\beta)\Uop(\alpha) \label{eq:WeylUV}\\
    \Uop(\alpha_1)\Uop(\alpha_2) &= \Uop(\alpha_1+\alpha_2) \label{eq:WeylUU}\\
    \Vop(\beta_1)\Vop(\beta_2)   &= \Vop(\beta_1 + \beta_2) \label{eq:WeylVV} ~,
\end{align}
\label{eq:WeylRel}%
\end{subequations}
consistent with the interpretation of $\Uop$ and $\Vop$ as translation operators. The set of finite linear combinations of the unitary operators $\Uop(\alpha)$ and $\Vop(\beta)$ equipped with the product rules Eqs.~(\ref{eq:WeylRel}) form an algebra of operators  
called the \emph{Weyl algebra} \cite{moretti2e,Wald94,afwPQM}. (In this context, an ``algebra'' is a vector space equipped with a distributive product \cite{szekeres}.) Introduction of the Weyl algebra turns out to be useful because the Weyl operators $\Uop$ and $\Vop$ are better mathematically behaved than the infinitesimal generators $\xop$ and $\pop$,%
\footnote{This is because the unitary translation operators used in Weyl's formulation are bounded, while Schur's lemma tells us the generators $\xop$ and $\pop$ cannot both be \cite{RSI}. 
} %
and indeed may be well-defined even in circumstances where $\xop$ and $\pop$ may not. This is precisely the situation that opens the door for polymer quantum mechanics.

Representations of the Weyl algebra can be used to define quantum theories that embody the CCR, as we now explain.

\subsection{The Stone-von Neumann Theorem}
\label{sec:stone}

A \emph{representation} of an algebra $\mathcal{A}$ is a choice of a vector space $\Hilbert$ equipped with a mapping from the algebra to the set of linear operators on $\Hilbert$ such that the mapping is a homomorphism, which is to say that the algebraic structure is faithfully carried over to the space of linear operators \cite{szekeres}.%
\footnote{The idea of a \emph{homomorphism} in general is a map that preserves relevant algebraic structures. In the present context, a homomorphism is a map from the Weyl algebra to linear operators on $\Hilbert$ that preserves the Weyl relations. 
} %
We then say that $\Hilbert$ ``carries'' a representation of $\mathcal{A}$. The most familiar example is that of linear representations of groups, in which group elements are mapped to matrices on a vector space, and the group operation corresponds to the matrix product. Two representations on spaces $\Hilbert$ and $\Hilbert'$ are \emph{unitarily equivalent} when there is a unitary transformation $U$ such that $\ket{\psi'}=U\ket{\psi}$ and $A'=UAU^{\dagger}$ relating states and operators on $\Hilbert$ and $\Hilbert'$. Unitarily equivalent representations differ essentially by a basis change, and are therefore effectively the same representation. A representation is said to be \emph{irreducible} when the action of the algebra leaves no non-trivial subspace of $\Hilbert$ invariant. 
In this sense irreducible representations are as ``small'' as they can be and still faithfully represent the algebraic structure of the original algebra. 

In the case of quantizing a classical physical system, we wish to find representations of the Weyl algebra on a Hilbert space. The Hilbert space $\Hilbert$ that carries that representation is identified with the space of quantum states of the system, and the Weyl operators are identified with the translation operators for position and momentum, with operators for other physical quantities built from position and momentum as described in Sec.~\ref{sec:canquant}.

The representation most commonly employed in quantum mechanics is referred to as the ``Schr\"{o}dinger representation.''  In the Schr\"{o}dinger representation the Hilbert space $\Hilbert$ is $\Hsch = L^2(\Re,dx)$, the space of functions of a real position variable $x$ that are square-integrable in the standard (Lebesgue) measure $dx$ on the real line $\Re$ \cite{szekeres,hallQM}. The position and momentum operators act on position-space wave functions in the usual way as
\begin{subequations}
\begin{align}
\xop\, \psi(x) &\equiv \melt{x}{\xop}{\psi} \\ 
             &= x\,\psi(x) 
\label{eq:Schrepx} \\
\pop\, \psi(x) &\equiv \melt{x}{\pop}{\psi} \\ 
	         &= -i\hbar \frac{\partial}{\partial x} \psi(x) ~,  
\label{eq:Schrepp} 
\end{align}
\label{eq:Schrepxp}%
\end{subequations}
where the latter expression (compare Eq.(\ref{eq:qpderiv})) follows as usual from the interpretation of $\pop$ as the generator of infinitesmal translations in $x$ \cite{Sakurai3e}. Accordingly, the action of the Weyl operators in the Schr\"{o}dinger representation is
\begin{subequations}
\begin{align}
\Uop(\alpha) \psi(x) &\equiv \melt{x}{\Uop(\alpha)}{\psi} \\ 
                     &= e^{+i\alpha x/\hbar}\psi(x) 
\label{eq:SchrepU} \\
\Vop(\beta)  \psi(x) &\equiv \melt{x}{\Vop(\beta)}{\psi} \\ 
	                 &= \psi(x+\beta)   ~.
\label{eq:SchrepV} 
\end{align}
\label{eq:SchrepUV}%
\end{subequations}
Similarly, in the momentum representation%
\footnote{In the study of quantization, the choice of position or momentum representation is referred to as ``polarization" \cite{hallQM}.
} %
we have 
\begin{subequations}
\begin{align}
\Uop(\alpha) \tilde{\psi}(p) &\equiv \melt{p}{\Uop(\alpha)}{\psi} \\ 
                     &= \tilde{\psi}(p-\alpha) 
\label{eq:SchrepUp} \\
\Vop(\beta)  \tilde{\psi}(p) &\equiv \melt{p}{\Vop(\beta)}{\psi} \\ 
	                 &= e^{+i\beta p/\hbar}\tilde{\psi}(p) ~.  
\label{eq:SchrepVp} 
\end{align}
\label{eq:SchrepUVp}%
\end{subequations}

The \textbf{Stone von-Neumann theorem }
\cite{hallQM,prugovecki2e,blankexner2e,moretti2e,Wald94} 
states that all irreducible representations of the Weyl algebra are unitarily equivalent to the Schr\"odinger representation \emph{provided} that the Weyl operators are weakly continuous. 
Weak continuity is the condition that
\cite{prugovecki2e,Thiemann} 
\begin{equation}
    \lim_{\alpha \to 0} \melt{\varphi}{\Uop(\alpha)}{\psi} = \bracket{\varphi}{\psi} 
	\text{  and  } 
	\lim_{\beta \to 0}\melt{\varphi}{\Vop(\beta)}{\psi} = \bracket{\varphi}{\psi} 
\label{eq:weakcont}
\end{equation}
for all $\ket{\varphi},\ket{\psi}\in\Hilbert$. Roughly, this means that matrix elements of $\Uop(\alpha)$ and $\Vop(\beta)$ behave as though $\Uop(\alpha)$ and $\Vop(\beta)$ reduce continuously to the identity as $\alpha\to 0$ and $\beta\to 0$, a seemingly mild condition consistent with the interpretation of $\Uop(\alpha)$ and $\Vop(\beta)$ as translation operators in continuous momentum and position variables. 


Additonally, \emph{if} the Weyl operators are weakly continuous, the Stone-von Neumann theorem guarantees that there are operators $\xop$ and $\pop$ for which operators obeying the Weyl relations take the form of Eqs.(\ref{eq:Weylops}).%
\footnote{This result is usually stated as a separate theorem due to Stone.
} %
Given weak continuity, the Stone-von Neumann theorem thus guarantees that the ``usual'' quantization is essentially the \emph{only} quantization (up to questions of operator ordering of physical observables). 

The requirement of weak continuity is the door that provides an opening for identifying representations of the canonical commutation relations (the Weyl algebra) that are not unitarily equivalent to the Schr\"{o}dinger representation, and therefore to polymer quantum mechanics. In the polymer representation, the Weyl operators are \emph{not} (both) weakly continuous, and the resulting quantum theory is genuinely distinct from standard Schr\"{o}dinger quantum theory. Let us now describe how this arises.

\section{Polymer quantum mechanics}
\label{sec:pqm}

The essence of polymer quantization is that it is a representation of the Weyl algebra on a Hilbert space of states that is no longer $L^2(\Re, dx)$, but rather, a space of states on an underlying configuration space with a topology distinct from the standard topology on $\Re$ -- one in which space is effectively discrete. This allows a representation for which (at least one of) the Weyl operators are not weakly continuous, as we shall see, and thus turns out to be distinct from the Schr\"{o}dinger representation, yielding a distinct quantum theory.

To see how this works, recall that we are seeking to represent an algebra generated by unitary operators $\Uop(\alpha)$ and $\Vop(\beta)$ that obey the Weyl conditions Eqs.(\ref{eq:WeylRel}), a relation among the momentum and position translation operators that is physically equivalent to the canonical commutation relations among the infinitesimal \emph{generators} of translations, $\xop$ and $\pop$. In seeking an alternative to Schr\"{o}dinger quantization, however, if we let go of the requirement that the Weyl operators are (both) weakly continuous, the definitions Eqs.(\ref{eq:Weylops}) that \emph{motivated} the Weyl relations, and therefore the operators $\xop$ and $\pop$ themselves, are no longer a required part of the discussion. Indeed, the existence of a representation of the Weyl algebra for which one of the generators $\xop$ or $\pop$ is \emph{not} well-defined is central to the reason the polymer representation exists and is distinct from the Schr\"{o}dinger representation. 
This will have important consequences when we formulate the dynamics of the resulting quantum theory.

The starting point for polymer quantum mechanics, then, is to assume that the underlying configuration space of a quantum particle consists in all possible positions for the particle, \emph{i.e.~}the position $x$ of the particle may take on any real value, $x\in\Re$, just as it does in ordinary quantum mechanics. However, the \emph{topology} with which $\Re$ is endowed is not the standard one, but rather, the real line endowed with what is known as the ``discrete'' topology, in which every point in $\Re$ is an open set.%
\footnote{For later reference, note that every function whose domain is a set with the discrete topology is therefore automatically continuous, since by definition a function is continuous if the pre-image of an open set is an open set \cite{szekeres}.
} %
We will denote the real line with the discrete topology $\Red$. In this way, every position $x\in\Red$ is topologically isolated from its neighbors, making space effectively discrete without preemptively declaring space to be a lattice. 
This is consistent with the notion that in quantum gravity, space(-time) will become discrete. As we will see, the consequences of this assumption for the resulting quantum theory are significant.

Under the assumption that particle positions live in $\Red$, the corresponding particle \emph{momentum} space is what is known as the \emph{Bohr compactification} of the real line, denoted $\Reb$.  Let us see how this arises. 
(The reader interested in the 
technical details of the construction of the Bohr compactification and its relation to the quantum momentum space may consult Appendix \ref{app:bohr}.) 

To a physicist, ``momentum'' is the physical quantity that is conserved when a system possesses a translation invariance. In keeping with Noether's theorem, the generator of translations is identified with the momentum \cite{HellSah21} (and the relation with the CCR is thereby established, as described in Sec.\ref{sec:translate}).  As a direct consequence, the relationship between position and momentum is the Fourier transform, because the transformation function between $x$ and $p$ so identified is (proportional to) $\exp(ipx)$.  

Let us put this observation on a more formal footing. The real line may be regarded as a group, with addition as the group operation. Harmonic analysis teaches us that the natural domain of the Fourier transform on a group $G$ is what is called the \emph{dual group} $\hat{G}$. Thus, if the group $G$ is thought of as the quantum configuration space of some physical system, the dual group $\hat{G}$ is identified with the corresponding quantum momentum space. In the case of the real numbers, the group dual to $\Re$ with the standard topology is also $\Re$, and so in Schr\"{o}dinger quantum theory on the real line the momentum is also a real number, $p\in\Re$.

Now suppose the real line is equipped with the discrete topology instead of the usual one. The dual group to the discrete real line $\Red$ is the Bohr compactification $\Reb$ of $\Re$. 
In this way, we see that 
\emph{the polymer momentum space dual to a position space with the discrete topology is the Bohr compactification of the Schr\"{o}dinger momentum space.}
In fact, as shown in Appendix \ref{app:appdiscretedual}, this is a general feature of polymer quantum mechanics formulated on \emph{any} configuration space.

While the discrete real line $\Red$ is easy enough to visualize as a one-dimensional ``dust'' of positions, there seems to be no easy story to tell about what the Bohr compactification $\Reb$ ``looks like'' without knowing more about how it is constructed. While this will require a detour into mathematical territory unfamiliar to most physicists (for which see  Appendix \ref{app:bohr}), the underlying idea –– that momentum space is in a precise sense the dual of position space given by the Fourier transform –– is the same as in ordinary quantum mechanics.  That the resulting quantum theory is distinct 
from ordinary quantum theory flows \emph{entirely} from the assumption that the topology of position space is discrete.

Fortunately, the resulting quantum theory (polymer quantum mechanics) 
turns out to be surprisingly easy to work with, in spite of the lack of an easy picture of $\Reb$. This is because $\Re$ is dense in $\Reb$ and all of the complex structure of $\Reb$ is ``at infinity''.  For the most part, calculations on $\Reb$ therefore ``look like'' calculations on $\Re$, with the crucial exception that integration (and therefore the inner product) is defined differently on $\Reb$.  The discreteness of $\Red$ and the novel inner product on $\Reb$ will lead us to the central features of the polymer representation.

In the following subsection we will see how to construct integrals and inner products on $\Reb$. 
With this in hand, polymer quantum mechanics follows from seeking a representation of the Weyl algebra on a Hilbert space of states defined on $\Red$, 
which will lead naturally to a quantum theory defined on a discrete graph of positions. 
The Weyl operators can then be used to define Hamiltonian dynamics in PQM.

\subsection{Integrals and inner products on $\Reb$}
\label{sec:intinnerRb}

Most of the technical details of the definition of the Bohr compactification of $\Re$ described in the Appendix \ref{app:bohr} turn out not to play much of a direct role in calculations in polymer quantum mechanics. One agreeable exception is that integration on $\Reb$ (and therefore in our polymer momentum space) is defined in such a way that periodic functions become square-normalizable.

As described in the appendix, the construction of $\Reb$ is rooted in viewing the real numbers as an Abelian group under addition, and as described above, $\Reb$ is then the dual group to the discrete real numbers $\Red$. The natural measure on a (locally compact) Abelian group is called the \emph{Haar measure} \cite{moretti2e,Harmonic3e,RudinFourier,haarjoy}. This is the unique (up to overall scale) \emph{translation invariant} measure $\mu_H$ on the group, in the precise sense that the Haar measure of a (measurable) set is invariant under the action of the group operation on the set. The Haar measure thus weights the measure of (measurable sets of) group elements as uniformly as possible: the measure of a (measurable) subset translated anywhere in the group remains constant.
On discrete groups, the Haar measure is the counting measure $\mu_c$, which, as the name suggests, simply ``counts'' the number of points in the set being integrated over by adding over them, and is usually normalized to one on singletons.  On the real line (with the standard topology), the usual Lebesgue measure $dx$ is the Haar measure, because $dx= d(x+a)$ for any fixed $a$.

The Haar measure $d\mu_H$ on $\Reb$ is 
\begin{equation}
    \int_{\Reb}d\mu_H f(p)  = \lim_{L \to \infty} \frac{1}{2L} \int_{-L}^{L}dp\, f(p)  ~,
\label{eq:BohrInt}
\end{equation}
where $dp$  
is the standard Lebesgue measure on $\Re$ \cite{HewittRossI,Harmonic3e,bojo13,Thiemann}, %
and we have switched notation $x\rightarrow p \in\Reb$ from the appendix since $\Reb$ is going to be our momentum space.
Integrals of functions $f$ over $\Reb$ thus take the form of an asymptotic average of $f$ over $\Re$. 
An inner product for complex-valued functions on $\Reb$ can then be defined in the expected way as
\begin{equation}
    \bracket{f}{g} = \int_{\Reb}d\mu_H\, f(p)^* g(p) ~.
\label{eq:BohrIP}
\end{equation}
In this inner product, the periodic functions $e^{-ip\mu}$  
become square-normalizable, and are orthonormal with respect to $\mu$ 
on $\Reb$. To see that this is the case, in an obvious notation 
it is easy to check that
\begin{subequations}
\begin{align}
\bracket{\mu}{\nu}   
   &= \lim_{L \to \infty} \frac{1}{2L} \int_{-L}^{L}dp\, e^{ip\mu}e^{-ip\nu}  \\
   &= \delta_{\mu\nu} 
\end{align}
\label{eq:BohrIPexp}%
\end{subequations}
for all real $\mu$, $\nu$. 
The periodic functions of the form $e^{-ip\mu}$, $\mu\in\Re$ form an uncountable orthonormal basis of functions on $\Reb$ in the inner product given by the Bohr compactification's Haar measure \cite{Harmonic3e,bojo13}. 
When completed in this inner product, the resulting space is $L^2(\Reb,d\mu_H)$, the square-integrable functions on $\Reb$.  As will be seen shortly, this is the polymer Hilbert space $\Hpoly$.

\subsection{Polymer Hilbert space: position and momentum representations}
\label{sec:Hpoly}

In this section we will build the Hilbert space $\Hpoly$ of the polymer representation, in some ways paralleling the mathematical structure of loop quantum cosmology \cite{Ashtekar:2006wn,abl03,bohrpaper,bojo13,Thiemann}. We will follow most closely   the approach of Corichi \textit{et al.} in \cite{continuumlimit} (building on \cite{afwPQM,physhilbspace}), 
but provide additional detail relative to these references.  
One significant difference is that we will focus more on the position representation than the previous literature, because that will turn out to be more convenient when considering compact configuration spaces as we do in the sequel.

The essential physical idea is to model space by the discrete real line $\Red$. As we have seen, the quantum momentum space is then the Bohr compactification of the real line $\Reb$.  After suitably defining corresponding quantum position and momentum states, we will proceed to describe the Hilbert space $\Hpoly$ built from these states in both momentum and position representations and see that it carries a representation of the Weyl algebra that does not meet the conditions of the Stone-von Neumann theorem, and is therefore not unitarily equivalent to the Schr\"{o}dinger representation; this is called the ``polymer representation''.  From there, we show how quantum theories may be formulated on subspaces $\Hgraphmuo \subset \Hpoly$ defined on discrete, regular graphs of positions.

Note that these constructions are \emph{choices}, not conclusions. What remains to show is that these choices yield a well-defined quantum theory, that is to say, a well-defined Hilbert space and a representation of the Weyl algebra that has ordinary quantum mechanics (the Schr\"{o}dinger representation) as its classical limit.
Let us begin.

Modeling space as the discrete real line $\Red$ leads us to introduce a set of (what will be) position basis states $\{\ket{\mu}\}$, where the label $\mu$ can assume any real value. These basis states will reflect the topology of the discrete real line by endowing them with the discrete ``polymer inner product'' 
\begin{equation}
\bracket{\mu}{\nu} = \delta_{\mu\nu} ~.
\label{eq:polymerip}
\end{equation}
The completeness relation for these states will be expressed as
\begin{subequations}
\begin{align}
\mathbb{I} &= \sum_{\mu\in\Re}\, \ketbra{\mu}{\mu} \label{eq:countingsum} \\
    &\equiv \int_{\Red} d\mu_c\,  \ketbra{\mu}{\mu} ~,    \label{eq:countingintegral}
\end{align}
\label{eq:qcomplete}%
\end{subequations}
where as indicated the sum over all real numbers should be regarded as an integral over the real line using the counting measure $\mu_c$ on $\Red$, since the counting measure is the Haar measure on discrete groups. %
(Note the counting measure is taken to be dimensionless even if the position $\mu$ carries units, consistent with the common notation expressed in Eq.~(\ref{eq:countingsum}).)

We can see immediately the seeds for why a quantum theory built on $\Hpoly$ will lead to a representation of the Weyl algebra that violates the weak continuity requirement of the Stone-von Neumann theorem, because
\begin{subequations}
\begin{align}
\lim_{\epsilon \to 0}\bracket{\mu}{\mu+\epsilon} 
     &= 0  \\
	 &\neq \bracket{\mu}{\mu}  ~.
\end{align}
\label{eq:iplimit}%
\end{subequations}

To build the Hilbert space $\Hpoly$ from these discrete position states, following Corichi \textit{et al.}~\cite{continuumlimit} we introduce so-called ``cylinder states''%
\footnote{This rather cryptic naming has its origins in probability theory. The rough idea seems to be that these states are defined by fixing a finite (or sometimes merely countable) number of values of $\mu$ out of the uncountably infinite number of possible values of $\mu\in\Re$, leaving the rest unspecified, so these states can be pictured rather fancifully as living on a ``cylinder'' in the space of all possible values of $\mu$ that might have been selected.
} %
that correspond to finite linear combinations of the kets $\ket{\mu}$:
\begin{equation}
    \ket{\psi_{\mathrm{Cyl}}}=\sum_{i=1}^N a_i \ket{\mu_i} ~,
\label{eq:cylstatepsi}
\end{equation}
where $N$ may take any (for now) finite value. Consider another cylinder state given by
\begin{equation}
    \ket{\phi_{\mathrm{Cyl}}}=\sum_{i=1}^M b_i \ket{\nu_i} ~,
\label{eq:cylstatephi}
\end{equation}
noting that the set of points labeled by the $\{\nu_i\}$ will in general be different from the  $\{\mu_i\}$. The inner product between these states is clearly
\begin{equation}
    \bracket{\psi_{\mathrm{Cyl}}}{\phi_{\mathrm{Cyl}}} = \sum_{k} a_k^*\, b_k  ~,
\label{eq:cylip}
\end{equation}
where the index $k$ runs over the intersection $\{\mu_i\}\cap\{\nu_j\}$ between the sets of position labels with non-zero coefficients in the two states.  This is well-defined because it is always a finite sum.
The Hilbert space $\Hpoly$ is the Cauchy-completion of the space of cylinder states in this inner product.%
\footnote{%
``Cauchy completion'' of an inner product space is the act of forming a new space by adding 
limits of all Cauchy sequences of vectors in the space to the space itself to yield a Hilbert space \cite{szekeres}.
In \cite{afwPQM}, $\Hpoly$ is defined by allowing cylinder states that are sums over \emph{countable} sequences of points, not just finite sums, with the coefficients satisfying $\sum_i |a_i|^2 < \infty$. In the present definition due to \cite{continuumlimit}, these convergent infinite sequences are ``added in'' in the process of Cauchy-completing the space. See also our discussion of the AFW conditions in Sec.\ref{sec:posgraph} below.
} %

$\Hpoly$ is a large, non-separable Hilbert space (that is to say, it contains no \emph{countable} dense subsets \cite{szekeres}) because the orthonormal basis $\{\ket{\mu}\}$ is uncountable. Shortly, though, we will decompose $\Hpoly$ into subspaces which are separable, which is the usual case encountered in quantum theory.

Next, we introduce the basis of states $\{\ket{p}\}$ defined by the transformation function 
\begin{subequations}
\begin{align}
  \bracket{p}{\mu} &= e^{-ip\mu/\hbar}  \\
        &\equiv \tilde{\phi}_{\mu}(p)  ~,
\end{align}
\label{eq:momtrans}%
\end{subequations}
consistent with the interpretation of these states as representing the momentum conjugate to the position coordinate.%
\footnote{Note that we choose the sign in the exponent opposite to the convention of \cite{continuumlimit} in order to be consistent with this interpretation of the physical meaning of $\mu$ and $p$. As emphasized in \cite{continuumlimit}, the mathematics does not care which is which.
} %
As such, since position is discrete, $p$ takes values in $\Reb$, the Bohr compactification of the real numbers $\Re$ \cite{afwPQM,continuumlimit}, the dual group to $\Red$. Indeed, states in $\ket{\psi}\in\Hpoly$ in this representation 
(or ``polarization'', in the language of quantization \cite{hallQM}) %
take the form of countable sums of plane waves 
\begin{subequations}
\begin{align}
\tilde{\psi}(p)  
     &\equiv \bracket{p}{\psi}   \\
     &= \sum_{j=1}^{\infty} a_j e^{-ip \mu_j/\hbar}  
\end{align}
\label{eq:momstates}%
\end{subequations}
that are normalizable in the inner product Eq.(\ref{eq:cylip}), so that $\sum_{j}|a_j|^2 < \infty$ \cite{afwPQM}.%
\footnote{Readers of Appendix \ref{app:bohr} will recognize that cylinder states in the momentum representation include Bohr's almost-periodic functions (as the space of continuous functions on $\Reb$, $AP(\Re)$ is dense in $L^2(\Reb,d\mu_H)=\Hpoly$),  
and that the Bohr compactification of $\Re$ is the set of maps ($\bra{p}$) from these functions ($\ket{\psi}$) to the complex numbers. Thus $p\in\Reb$.  Put another way, again in the language of the appendix, the states $\ket{p}$ introduced in Eq.(\ref{eq:momtrans}) are the \emph{characters} on $\Red$ in terms of which the Fourier transform on $\Red$ is defined. In \cite{afwPQM,continuumlimit} essentially the same argument is made instead considering the spectrum of the Abelian $C^*$-algebra formed by the Weyl operator $\Vop(\beta)$ on $\Hpoly$ defined in Sec.\ref{sec:HpolyWeyl}.
} %
We learned in Sec.\ref{sec:intinnerRb} that the natural inner product for functions on $\Reb$ is given by Eq.(\ref{eq:BohrIP}), and so we have
\begin{subequations}
\begin{align}
\bracket{\psi}{\phi} 
    &= \int_{\Reb} d\mu_H\, \tilde{\psi}(p)^*\tilde{\phi}(p)   \\
	&= \lim_{L\to \infty}\frac{1}{2L}\int_{-L}^{L}dp\, \tilde{\psi}(p)^*\tilde{\phi}(p)
\end{align}
\label{eq:themysterymachine}%
\end{subequations}
(The Haar measure $\mu_H$ is not to be confused with the position labels on the states $\ket{\mu}$.) Accordingly, the completeness relation for the states $\ket{p}$ may be expressed
\begin{subequations}
\begin{align}
\Id &= \int_{\Reb} d\mu_H\, \ketbra{p}{p}   \\
	&= \lim_{L\to \infty}\frac{1}{2L}\int_{-L}^{L}dp\, \ketbra{p}{p}     ~.
\end{align}
\label{eq:pcomplete}%
\end{subequations}
We note that this is consistent with Eqs.(\ref{eq:BohrIPexp}) and the polymer inner product, Eq.(\ref{eq:polymerip}):
\begin{subequations}
\begin{align}
\bracket{\mu}{\nu} &= \melt{\mu}{\Id}{\nu}  \\
   &= \int_{\Reb} d\mu_H\, \bracket{\mu}{p}\bracket{p}{\nu} \\ 
   &= \lim_{L \to \infty} \frac{1}{2L} \int_{-L}^{L}dp\, e^{+ip\mu /\hbar}e^{-ip\nu /\hbar}  \\
   &= \delta_{\mu,\nu}  ~.
\end{align}
\label{eq:Iconsistent}%
\end{subequations}
The Hilbert space $\Hpoly$ as represented by momentum-space wave functions is (isomorphic to) the space of square-normalizable functions on $\Reb$ under its Haar measure $d\mu_H$, $\Hpoly \isrep L^2(\Reb,d\mu_H)$ \cite{afwPQM,continuumlimit}.

We next consider the position representation. Given states $\ket{\psi}=\sum_{i=1}^{\infty} a_{\mu_i} \ket{\mu_i}$ for some countable sequence of points $\{\mu_i\}$ and $\ket{\phi}=\sum_{i=j}^{\infty} b_{\nu_j} \ket{\nu_j}$ for some other sequence $\{\nu_j\}$,
position ``wave functions'' are the coefficients
\begin{subequations}
\begin{align}
\psi(x) &\equiv \bracket{x}{\psi}   \\
	 &= \sum_{i=1}^{\infty} a_{\mu_i} \bracket{x}{\mu_i} \\
     &= \sum_{i=1}^{\infty} a_{\mu_i} \delta_{x,\mu_i}    ~,
\end{align}
\label{eq:xQWF}%
\end{subequations}
which is zero unless $x$ coincides with one of the $\mu_i$ defining $\ket{\psi}$.
Accordingly, the inner product in the position representation is (using the completeness relation, Eq.(\ref{eq:qcomplete})) 
\begin{subequations}
\begin{align}
\bracket{\psi}{\phi}  &= \int_{\Red}dx_c \bracket{\psi}{x}\bracket{x}{\psi}   \\
	 &= \sum_{x\in\Re} \psi(x)^*\phi(x)      \\
     &= \sum_{x\in\Re} \sum_{i,j} a_{\mu_i}^* \delta_{x,\mu_i}  b_{\nu_j} \delta_{x,\nu_j} \\
	 &= \sum_{i,j} a_{\mu_i}^*  b_{\nu_j} \delta_{\mu_i,\nu_j}   ~,  
\end{align}
\label{eq:posip}%
\end{subequations}
as expected from Eq.(\ref{eq:cylip}). This is of course zero unless the sequences $\{\mu_i\}$  and $\{\nu_j\}$ overlap.  $\Hpoly$ as represented by position-space wave functions is thus (also isomorphic to) the space of square-``integrable'' functions on the discrete real line in the counting measure $\mu_c$, $\Hpoly \isrep L^2(\Red,d\mu_c)$ \cite{afwPQM,continuumlimit}.

What is the status of the position and momentum operators on $\Hpoly$?  We will see that, as we have constructed it, there is a well-defined position operator on $\Hpoly$, \emph{but there is no corresponding momentum operator}. The Weyl operators (the position and momentum translation operators), however, are both well-defined, and this will turn out to be sufficient to do quantum mechanics.

Define the position operator $\xop$ in the expected way as the label operator
\begin{equation}
 \xop\ket{\mu} = \mu\ket{\mu} ~,
\label{eq:xopdef}
\end{equation}
which is clearly symmetric, $\melt{\mu}{\xop}{\nu} = \melt{\nu}{\xop}{\mu}^*$. The basic position functions are simply
\begin{subequations}
\begin{align}
\bracket{x}{\mu} &= \delta_{x,\mu}  \\
	 &\equiv \phi_{\mu}(x)          ~,
\end{align}
\label{eq:posefn}%
\end{subequations}
which are normalized, unlike the Dirac delta functions one encounters in the Schr\"{o}dinger representation.  That position states are represented by Kronecker deltas aligns with our choice of a discrete spatial topology.  
The action of the position operator on position-space wave functions is of course multiplicative,
\begin{subequations}
\begin{align}
\xop\psi(x)  &\equiv \melt{x}{\xop}{\psi}  \\
	 &= x\,\bracket{x}{\psi} \\     
	 &= x\,\psi(x)     ~.
\end{align}
\label{eq:xoppos}%
\end{subequations}
%
On the other hand, in the momentum representation we have
\begin{subequations}
\begin{align}
\xop\tilde{\psi}(p) &\equiv \melt{p}{\xop}{\psi}   \\
	 &= \sum_{i} a_{\mu_i} \melt{p}{\xop}{\mu_i}   \\
	 &= \sum_{i} a_{\mu_i} \mu_i \bracket{p}{\mu_i}   \\
	 &= \sum_{i} a_{\mu_i} \mu_i \, e^{-ip\mu_i/\hbar}    \\
	 &= +i\hbar\frac{\partial}{\partial p} \sum_{i} a_{\mu_i}  e^{-ip\mu_i/\hbar}    \\
	 &= +i\hbar\frac{\partial}{\partial p} \tilde{\psi}(p)   ~,
\end{align}
\label{eq:xopmom}%
\end{subequations}
as might have been expected.

What about the momentum operator? To answer this question, we turn to the representation of the Weyl algebra on $\Hpoly$.

\subsection{Polymer Hilbert space: Weyl operators}
\label{sec:HpolyWeyl}


Now let us consider the Weyl operators $\Uop(\alpha)$ and $\Vop(\beta)$, the translation operators in momentum and position, defined by%
\begin{subequations}
\begin{align}
\Uop(\alpha)\ket{p} &= \ket{p+\alpha}       \label{eq:WeyldefPQMU}\\
\Vop(\beta)\ket{\mu} &= \ket{\mu-\beta}  ~, \label{eq:WeyldefPQMV}
\end{align}
\label{eq:WeyldefPQM}%
\end{subequations}
as would be expected from the behavior of the Weyl operators in the Schr\"{o}dinger representation, Eqs.(\ref{eq:SchrepUV}-\ref{eq:SchrepUVp}). It can be easily checked that the adjoints of these operators are $\Uop(\alpha)^{\dagger}=\Uop(-\alpha)$ and  $\Vop(\beta)^{\dagger}=\Vop(-\beta)$, and these Weyl operators are therefore unitary.

We see immediately that $\Vop(\beta)$ fails to be weakly continuous, because 
$\melt{\mu}{\Vop(\epsilon)}{\mu}=\bracket{\mu}{\mu-\epsilon}=\delta_{\mu,\mu-\epsilon}=0$ for any value of $\epsilon>0$, so $\lim_{\beta\rightarrow 0}\melt{\mu}{\Vop(\beta)}{\mu}\neq \bracket{\mu}{\mu}=1$.  
By Stone's theorem, there is no Hermitian operator $\pop$ on $\Hpoly$ that can be exponentiated to give $\Vop(\beta)$.  This is precisely what we expect in a theory with a fundamentally discrete representation of space, because the momentum operator generates infinitesimal spatial translations \cite{afwPQM}. 
Nonetheless, the Weyl operators as defined here obey the relations one might expect if a momentum operator \emph{did} exist. As we shall see, $\Hpoly$ also carries a representation of the Weyl relations, Eqs.(\ref{eq:WeylRel}).

The action of $\Uop(\alpha)$ on position kets and $\Vop(\beta)$ on momentum kets can be found using the completeness relations, Eqs.(\ref{eq:pcomplete}) and (\ref{eq:qcomplete}):
\begin{subequations}
\begin{align}
\Uop(\alpha)\ket{\mu} &= \Uop(\alpha)\cdot\Id\cdot\ket{\mu}  \\
     &= \int d\mu_H\, \Uop(\alpha)\ket{p}\bracket{p}{\mu} \\
     &= \int d\mu_H\, \ket{p+\alpha}e^{-ip\mu/\hbar} \\
     &= \int d\mu_H\, \ket{p}\,e^{-ip\mu/\hbar}e^{+i\alpha\mu/\hbar} \\
     &= e^{+i\alpha\mu/\hbar} \int d\mu_H\, \ket{p}\bracket{p}{\mu} \\
	 &= e^{+i\alpha\mu/\hbar}\ket{\mu} ~,
\end{align}
\label{eq:Uopmu}%
\end{subequations}
using the translation invariance of the Haar measure. Incidentally, this shows that $\Uop(\alpha)$, in contrast to $\Vop(\beta)$, \emph{is} weakly continuous 
($\lim_{\alpha\rightarrow 0}\melt{\mu}{\Uop(\alpha)}{\nu}=\lim_{\alpha\rightarrow 0}e^{i\alpha\nu}\bracket{\mu}{\nu}=\bracket{\mu}{\nu}$)
and is as expected the exponential $\Uop(\alpha)=\exp(i\alpha\xop/\hbar)$ of the position operator.

Similarly,
\begin{subequations}
\begin{align}
\Vop(\beta)\ket{p} &= \Vop(\beta)\cdot\Id\cdot\ket{p}  \\
     &= \sum_{x}\, \Vop(\beta)\ket{x}\bracket{x}{p} \\
     &= \sum_{x}\, \ket{x-\beta}e^{+ipx/\hbar} \\
     &= \sum_{x}\, \ket{x}e^{+ipx/\hbar}e^{+ip\beta/\hbar} \\
     &= e^{+ip\beta/\hbar} \sum_{x}\, \ket{x}\bracket{x}{p} \\
	 &= e^{+ip\beta/\hbar}\ket{p} ~,
\end{align}
\label{eq:Vopp}%
\end{subequations}
just as one would expect if $\Vop(\beta)$ \emph{were} the exponential of a momentum operator.

These actions can then be used to check that the action of these operators on wave functions in the position and momentum representations expressed in Eqs.(\ref{eq:SchrepUV}-\ref{eq:SchrepUVp}) hold in polymer quantum mechanics as well.
%
Finally, it is straightforward to check using Eqs.(\ref{eq:WeyldefPQM}-\ref{eq:Vopp}) that the Weyl operators on $\Hpoly$ satisfy the Weyl relations, Eq.(\ref{eq:WeylRel}), with no need for infinitesimal generators.
%
Thus, $\Hpoly$ carries an (irreducible) representation of the Weyl algebra that is not weakly continuous and therefore not unitarily equivalent to the Schr\"{o}dinger representation on $L^2(\Re,dx)$ –– the polymer representation \cite{afwPQM}.

With these tools in place, we now wish to learn how to construct model quantum theories on $\Hpoly$. It turns out that in order to do this, it is necessary to imbue the polymer framework with some additional structure.

\subsection{Quantum mechanics on position graphs}
\label{sec:posgraph}

We have already noted that the $\Hpoly$ is a large, non-separable Hilbert space. In order to construct quantum theories in this space that (may) have Schr\"{o}dinger quantum mechanics as their continuum limit, it is helpful to decompose $\Hpoly$ into separable subspaces, and build quantum models in one of those subspaces. Indeed, one can see from the position representation $\Hpoly\isrep L^2(\Red,dx_c)$ that the support of position-space wave functions on $\Red$ must be tightly constrained. Consider something as simple as the constant function $f(x)=1$. Its integral on an interval $[a,b]$ under the counting measure takes the form $\sum_{x\in[a,b]}\, 1$ over \textit{all of the uncountably many points} in that interval. Clearly no function that takes non-zero values on any non-empty interval 
in $\Re$ can be normalizable in the counting measure, and therefore belong to $\Hpoly$.  Interestingly, this tells us that the intersection between $\Hpoly$ and states in $\Hsch=L^2(\Re,dx)$ in the Schr\"{o}dinger representation is only the zero state \cite{afwPQM}!

The approach taken by Ashtekar, Fairhust, and Willis in \cite{afwPQM} is to define a \emph{graph} $\gamma$ that consists in a countable set of positions $\{x_i \}\subset\Re$ subject to two requirements we shall refer to as the Ashtekar-Fairhurst-Willis (AFW) conditions:
\begin{enumerate}[nosep, label=\arabic*.]
\item The points $x_i \in \gamma$ do not contain sequences with accumulation points (in the usual topology on $\Re$).
\item For intervals of length above some threshold $\ell_\gamma$, there exists a maximum density of points $\rho_\gamma$ on the interval.
\end{enumerate}
(These conditions were imposed to guarantee convergence of certain series they employ in subsequent developments.)
They then define $\mathrm{Cyl}_{\gamma}$ to be the set of normalizable cylinder states defined on $\gamma$ (subject to a fall-off condition, again for technical reasons), and $\mathrm{Cyl}$ to be union of the $\mathrm{Cyl}_{\gamma}$ for all possible graphs.  In their approach, $\Hpoly$ is the Cauchy-completion of $\mathrm{Cyl}$ in the polymer inner product, Eq.(\ref{eq:polymerip}).

It is worth observing that the first of the AFW conditions is sufficient to conclude that the topology induced on each such graph $\gamma$ is the discrete topology; all of its points are isolated. 
These graphs are the analog in PQM of the graphs of quantum geometry on which the polymer-like states of loop quantum gravity are defined \cite{afwPQM}.

\begin{figure}
\includegraphics[width=0.75\textwidth]{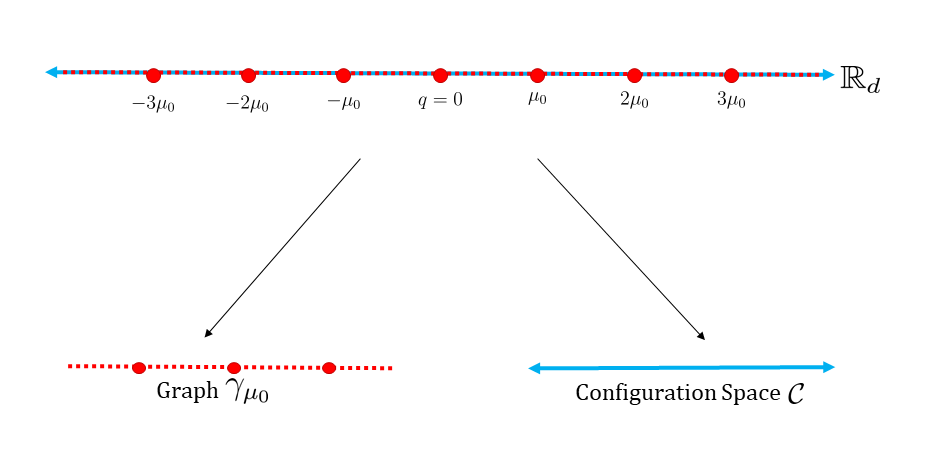}
\caption{Position space for a point particle in one-dimensional polymer quantum mechanics. The quantum configuration space $\mathcal{C}=\Red$ is the real line with the discrete topology. In order to admit a well-defined dynamics, states are restricted to support on a graph $\gamma_{\mu_0}$ that is a regular lattice on $\mathcal{C}$ with points equally spaced at distance $\mu_0$. 
}  
\label{fig:graphgamma}
\end{figure}

In order to construct viable quantum theories on $\Hpoly$, however, the choice of graphs must be restricted even further. If the Weyl operator $\Vop(\beta)$ is to remain well-defined when its action is restricted to states on a single graph, it is necessary to ensure that $V(\lambda) \ket{x_j}=\ket{x_j + \lambda}$ is another state on the same graph for every $x_j\in\gamma$. This can be achieved by introducing a fixed length scale $\mu_0$, and restricting attention to the regular lattice $\gamma_{\mu_0}$ defined by
\begin{equation}
    \gamma_{\mu_0} \equiv \{ x_k = k\mu_0,  k\in \Int \}.
\end{equation}
Such a graph is illustrated in Figure \ref{fig:graphgamma}. The space of states in $\Hpoly$ whose support is restricted to such a regular graph is a separable Hilbert space we will denote $\Hgraphmuo$. 
(This space is separable because it has a countable basis $\{\ket{k\mu_0},k\in\Int\}$.)
The action of $\Vop(\beta)$ on $\Hgraphmuo$ will be similarly restricted to lattice translations $\Vop(k\mu_0)$ of minimum step size $\mu_0$. While this might seem highly restrictive, it turns out to be sufficient to define interesting quantum theories that have a reasonable continuum limit.

Finally, it is useful to observe that the momentum-space representation $\tilde{\psi}(p)$ of quantum states $\ket{\psi}\in\Hgraphmuo$ are actually \emph{periodic} functions of $p$, with fundamental frequency $\mu_0/2\pi\hbar$  (cf.~Eq.(\ref{eq:momstates})). 
On such states, the polymer inner product in momentum space, Eq.(\ref{eq:themysterymachine}), is equivalent to the inner product \cite{continuumlimit}
\begin{equation}
\bracket{\psi}{\phi}_{\mu_0} = 
  \frac{\mu_o}{2\pi\hbar}\int_{-\pi\hbar/\mu_0}^{\pi\hbar/\mu_0} dp\, \tilde{\psi}(p)^*\tilde{\phi}(p)  ~,
\label{eq:polyiplattice}
\end{equation}
which can be confirmed by checking that $\bracket{j\mu_0}{k\mu_0}_{\mu_0} = \delta_{j,k}$, 
in agreement with Eq.~(\ref{eq:BohrIPexp}). 
In this sense, the compactness of the momentum space $\Reb$ is made manifest in the inner product for states on $\Hgraphmuo$.


\subsection{Hamiltonian dynamics on a discrete graph}
\label{sec:dynamics}

We have seen that in the polymer quantization with a discrete position space, the momentum operator is no longer well defined. How then is Hamiltonian dynamics to be formulated in this theory?  With the intention of constructing a quantum theory that might have the Schr\"{o}dinger representation as some kind of continuum limit, the usual solution is to approximate the momentum operator in terms of operators which \emph{are} well defined, namely the Weyl operator $\Vop(\beta)$ \cite{afwPQM}.  Inspired by lattice gauge theory, since $\Vop(\beta) = e^{i\pop\beta /\hbar}$ ``in spirit'' 
-- a sentiment sometimes expressed as $\Vop(\beta)=\widehat{\exp(ip\beta/\hbar)}$ -- a self-adjoint approximate momentum operator may be defined as
\begin{subequations}
\begin{align}
\pop_{\mu_0} &\equiv -\frac{i\hbar}{2\mu_0}(\Vop(\mu_0)-\Vop(-\mu_0)) \\
     &= \frac{\hbar}{2\mu_0i}\left(\widehat{e^{ip\mu_0/\hbar}} - \widehat{e^{-ip\mu_0/\hbar}} \right) \label{eq:momdefpqmV}\\
     &= \frac{\hbar}{\mu_0}\widehat{\sin\left(\frac{p\mu_0}{\hbar}\right)}   ~.
\label{eq:momdefpqm}
\end{align} 
\end{subequations}
One sees that (again, in spirit) this is approximately $p$ when $p \ll \hbar/\mu_0$; the physics is only sensitive to the presence of a fundamental minimum length scale $\mu_0$ for momenta large enough to probe that scale.  Aspects of this observation have been carefully explored in earlier work \cite{afwPQM,physhilbspace,continuumlimit}.  We will show explicitly how this manifests in the examples developed in the remainder of the paper.

In order to construct the kinetic term $\pop^2/2m$ in a non-relativistic Hamiltonian, one appproach would be simply to square the operator $\pop_{\mu_0}$. That choice, however, has the inconvenient feature that it involves $\Vop(\pm\mu_o)^2 = \Vop(\pm 2\mu_o)$, and thus translations by two steps on the lattice $\gamma_{\mu_0}$.  Instead, the usual choice is to define a self-adjoint approximate momentum-squared operator by  
\begin{subequations}
\begin{align}
\widehat{p^2_{\mu_0}} &\equiv \frac{\hbar^2}{\mu_0^2}(2\Id -\Vop(\mu_0)-\Vop(-\mu_0)) \\
     &= \frac{\hbar^2}{\mu_0^2}\left(2\Id -\widehat{e^{ip\mu_0/\hbar}} - \widehat{e^{-ip\mu_0/\hbar}} \right) \label{eq:mom2defpqmV}\\
     &= \frac{2\hbar^2}{\mu_0^2}\left(\Id -\widehat{\cos\left(\frac{p\mu_0}{\hbar}\right)}\right)   ~.
\label{eq:mom2defpqm}
\end{align}
\end{subequations}
All the tools are now in place to do quantum mechanics in the Hilbert space $\Hgraphmuo$. A self-adjoint single-particle Hamiltonian on $\Hgraphmuo$ can be defined as
\begin{equation}
\hat{H}_{\mu_0} = \frac{\widehat{p^2_{\mu_0}}}{2M} + \widehat{V(x)} ~.
\end{equation}
The dynamics defined by $\hat{H}_{\mu_0}$ thus leaves $\Hgraphmuo$ invariant for any lattice $\gamma_{\mu_0}$; these are the analogues in PQM of the ``superselection sectors" of loop quantum cosmology \cite{liv-rev,ashsingh11}. 
This is the construction we will examine in the examples of a particle on a ring 
and a particle in a box 
to follow, after we have discussed the restriction to compact configuration spaces.  
The simple harmonic oscillator and free particle ``polymerized'' in this manner have been described previously in \cite{afwPQM,continuumlimit}.

Finally, an amusing consistency check is to compare the polymer-quantized commutator $[\xop,\hat{H}_{\mu_0}]$ to its Schr\"odinger-quantized counterpart. Using Eq.~(\ref{eq:WeyldefPQMV}), a short calculation shows that $[\xop,\Vop(\mu_0)] = -\mu_0\Vop(\mu_0)$ on the lattice $\gamma_{\mu_0}$ \cite{afwPQM}, from which, using Eq.~(\ref{eq:mom2defpqmV}), it quickly follows that
\begin{equation}
    [\xop, \hat{H}_{\mu_0}]=\frac{\hbar^2}{2M\mu_0} \left( V(\mu_0) - V(-\mu_0)\right)  ~.
\end{equation}
From Eq.~(\ref{eq:momdefpqmV}) we see that this is the same as
\begin{equation}
    [\xop, \hat{H}_{\mu_0}]=\frac{i\hbar}{M}\, \pop_{\mu_0}  ~,
\end{equation}
in complete agreement with the standard result. 

While the first part of this paper has emphasized a pedagogical presentation of the essentials of polymer quantum mechanics, the results presented in the remainder are new. 

\section{Polymer quantum mechanics on compact spaces}
\label{sec:pqmcompact}

The presentation of polymer quantum mechanics so far has taken the classical configuration space to be the entire real line $\Re$.  In the examples we explore in the sequel, the classical configuration space is instead compact -- a circle and a closed interval in the case of a particle on a ring or a particle in a box, respectively. How does this restriction impact the formulation of PQM on such spaces?  

Recall the general strategy for polymer quantization as we have presented it begins by endowing the classical configuration space with the discrete topology. The corresponding polymer momentum space is then the Bohr compactification of the Schr\"{o}dinger momentum space, and is therefore compact.  After constructing the polymer Hilbert space $\Hpoly$, viable quantum theories may be formulated on separable subspaces $\Hgraphmuo$ of $\Hpoly$ defined by regular position graphs.

Now suppose the configuration space is compact. What are the consequences of this? As discussed in detail in Appendix \ref{app:bohrcompact}, the group dual to a compact group is discrete. \textit{A compact configuration space thus has a discrete momentum space, so the momentum is quantized in the Schr\"{o}dinger representation.}  (In the case of a particle on a ring, this is the particle's angular momentum; see Sec.~\ref{sec:ringschroquant}.) This illustrates that the quantization of the momentum on compact configuration spaces is a kinematical feature of the theory, due to the relationship between the structure of the configuration space and translations on it.

The situation is different when we polymerize. Unless the configuration space is finite, \textit{a discretized configuration space is no longer compact}. (Infinite spaces with the discrete topology aren't compact since an open cover by singletons has no finite subcover.) Hence, the corresponding momentum space is no longer strictly discrete! (This lack of discreteness is, however, ``at infinity".) Thus, a polymer-quantized compact configuration space has a momentum space that is compact, since it is the Bohr compactification of the Schr\"{o}dinger momentum space, but not strictly discrete. We will see how this works in detail for the particle on a ring in Sec.~\ref{sec:ringpolyquant}.%
\footnote{The treatment of the particle in a box is reserved for Appendix \ref{app:pqmbox} because, while polymer quantum mechanics may be formulated for it in much the same way as for the particle on a line or a ring, and may be solved using the same methods, the technical setting is different because unitary Weyl operators cannot be defined for the infinite well \cite{hallQM}.
} %


That the polymer-quantization of a compact configuration space is not in fact compact still has significant consequences, however. As we shall see shortly, the compactness of the Schr\"{o}dinger configuration space implies that when we pass to quantum theories defined on regular position graphs $\gamma_{\mu_0}$, these graphs must in fact not only be regular, but \emph{finite}. This fact means that the dynamics of polymer quantum theories defined on such graphs becomes amenable to explicit solution by discrete methods.

\subsection{Position graphs on compact spaces}
\label{sec:posgraphc}

Compact sets in $\Re^n$ are closed and bounded. More generally, a topological space is compact if every cover of the space by open sets (in the given topology) has a subcover that requires only a finite number of open sets. 
The Heine-Borel theorem states that for subsets of $\Re^n$, this definition is equivalent to the more familiar notion \cite{szekeres,basictopology}. 

Another important property of compact sets is given by the Bolzano-Weierstrass theorem, which says that a subset of a compact space with an infinite number of points must have a limit point \cite{basictopology}. 
This has important implications for polymer quantum mechanics. The first of the AFW conditions restrict the choice of graphs to subsets of the configuration space \textit{without limit points} (in the usual topology on $\Re$). Therefore, the Bolzano-Weierstrass property implies that if the Schr\"odinger configuration space is compact, only graphs with \textit{finite} numbers of points are admissible, whether or not the graphs are regular. Therefore, for quantum systems on compact configuration spaces, all graphs consistent with the AFW conditions carry an extra parameter $N$ which corresponds to the number of points on the graph. This result is consistent with what is already obvious for regular graphs $\gamma_{\mu_0}$ on $\Re$, that such a graph on a bounded interval is limited to a finite number of points. The Bolzano-Weierstrass property guarantees that this result holds for graphs satisfying the AFW conditions on an arbitrary compact configuration space.

This feature of the admissible graphs on compact configuration spaces will provide the basis for analyzing the quantum mechanics of a polymer quantum particle confined to motion on a ring or in a box in the sequel.

\section{Polymer Particle on a Ring}
\label{sec:pqmring}

Let us see how all of this works for the explicit example of a particle of mass $M$ confined to motion on a ring of radius $R$. The Hamiltonian for such a particle is simply the free particle Hamiltonian 
\begin{subequations}
\begin{align}
H &= \frac{L_z^2}{2I} \\
  &= \frac{p^2}{2M} ~,
\end{align}   
\label{eq:Hring}%
\end{subequations}
which we choose to express in terms of the linear momentum $p$
using $L_z=Rp$ and $I=MR^2$
in order to directly compare with the particle on a line considered previously. All physical quantities are of course subject to the periodicity condition $\psi(\varphi)=\psi(\varphi+2\pi)$.

\subsection{Schr\"{o}dinger quantization}
\label{sec:ringschroquant}

The Schr\"{o}dinger quantization of particle on a ring is a familiar textbook example of quantization of a physical system with a configuration space that has non-trivial spatial structure \cite{McIntyre}.  Let us briefly examine it from the perspective of representations of the Weyl algebra being explored in this paper.

In this case the configuration space (the ring) is isomorphic to $\CircT$, the circle group. From Appendix \ref{app:bohrcompact} we know that the dual group to $\CircT$, the quantum momentum space, is isomorphic to the integers, $\CircThat = \Int$. This is precisely what we expect from more elementary considerations: the momentum in question is really the \emph{angular} momentum, the momentum conjugate to the position coordinate on the circle, which for a particle on a ring is quantized. As with the particle on a line, we can define ``position" eigenstates $\ket{\varphi}$ ($\equiv\ket{\varphi+2\pi n}$) 
and a dual basis of momentum states $\bra{m}$ defined by the Fourier transform (the characters on $\CircT$, in the language of Appendix \ref{app:bohr}), $\bracket{m}{\varphi}=e^{-im\varphi}/\sqrt{2\pi}=e^{-im\hbar\varphi/\hbar}/\sqrt{2\pi}\equiv \Phi_m(\varphi)^*$ for $\varphi\in\CircT$ and $m\in\Int$, using the conventional normalization $\int_0^{2\pi}d\varphi\,  |\Phi_m(\varphi)|^2=1$.  As is well known, the angular momentum eigenvalues are $L_z=m\hbar$, $m\in\Int$, while the doubly-degenerate energies $E_m$ for the Hamiltonian Eq.~(\ref{eq:Hring}) are $E_m=(m\hbar)^2/2MR^2$.

It may be quickly checked that the resolution of the identity
\begin{equation}
\int_0^{2\pi}d\varphi\, \ketbra{\varphi}{\varphi} = \Id    
\label{eq:phicomplete}
\end{equation}
is consistent with the inner product
\begin{equation}
\bracket{m}{n} = \delta_{mn} ~,
\label{eq:mprod}
\end{equation}
while the identity expressed in terms of the (angular) momentum states
\begin{equation}
\sum_{m\in\Int}\,\ketbra{m}{m} = \Id
\label{eq:mcomplete}
\end{equation}
is consistent with the expected inner product in position space,
\begin{subequations}
\begin{align}
\bracket{\varphi}{\varphi'} &= \melt{\varphi}{\Id}{\varphi'} \\
  &= \frac{1}{2\pi} \sum_{m\in\Int}\, e^{im(\varphi-\varphi')} \\
  &= \sum_{m\in\Int}\, \delta(\varphi-\varphi'-2\pi m) \\
  &= \delta(\varphi-\varphi') ~.
\end{align}
\label{eq:phiprod}%
\end{subequations}
Here the equality between the second and third lines is known as the \emph{Poisson summation formula}, 
and the last line follows because only one term in the sum will ever contribute for $\varphi, \varphi'\in [0,2\pi)$.
The space of Schr\"odinger-quantized position wave functions is $\mathcal{H}_{\mathrm{Schr}}^{\varphi}=L^2(\CircT,d\varphi)$, and the (angular) momentum wave functions live in $\mathcal{H}_{\mathrm{Schr}}^{L_z}=L^2(\Int,d\mu_c)$.

As in the case of a particle on a line, we may define unitary Weyl operators $\Uopr(n)$ and $\Vopr(\phi)$, the translation operators in momentum and position, by
\begin{subequations}
\begin{align}
\Uopr(n)\ket{m} &= \ket{m+n}  \\
\Vopr(\phi)\ket{\varphi} &= \ket{\varphi-\phi}  ~.  \label{eq:WeyldefPQMringpos}
\end{align}
\label{eq:WeyldefPQMring}%
\end{subequations}
Just as in the linear case, it is easy to check that correspondingly we have
\begin{subequations}
\begin{align}
\Uopr(n)\ket{\varphi} &= e^{in\varphi}\ket{\varphi}  \\
\Vopr(\phi)\ket{m} &= e^{im\phi}\ket{m}  ~,  \label{eq:WeylOpEigenrmom}
\end{align}
\label{eq:WeylOpEigenr}%
\end{subequations}
so that (as expected) these states are translation eigenfunctions, whether or not infinitesimal generators for those translations exist.
The Weyl algebra is similarly given by
\begin{subequations}
\begin{align}
\Uopr(m)\Vopr(\phi) &= e^{-im\phi}\,\Vopr(\phi)\Uopr(m) \\
\Uopr(m)\Uopr(n) &= \Uopr(m+n) \\
\Vopr(\phi)\Vopr(\varphi) &= \Vopr(\phi+\varphi) ~.
\end{align}
\label{eq:WeylRelr}%
\end{subequations}

It is worth commenting that the Stone-von Neumann theorem does not actually apply to the particle on a ring because of the discreteness of the momentum, so $\Uopr(n)$ fails to be weakly continuous.  One manifestation of this is that the Weyl algebra we have defined 
in Eqs.~(\ref{eq:WeylRelr}) is not unique; there are unitarily inequivalent representations parameterized by a factor of $e^{i\theta}$ for $\theta\in [0,2\pi)$.%
\footnote{The most general boundary condition compatible with the periodicity of the probability density, $|\psi(\varphi+2\pi)|^2=|\psi(\varphi)|^2$, is $\psi(\varphi+2\pi)=e^{i\theta}\psi(\varphi)$ \cite{vanenk25}. In this paper we confine our attention to $\theta=0$. For a more refined treatment of the Schr\"odinger particle on a ring that discusses this quantization ambiguity, see e.g.~example 6.1 in Ref.~\cite{bojo10}.
} %
In spite of this, these representations do not yield quantizations that reflect the discreteness expected in quantum gravity.  For that, we turn once again to polymer quantization.

\subsection{Polymer quantization}
\label{sec:ringpolyquant}

Let us now work out the details for a \emph{polymer} particle on a ring. The polymer configuration space is taken to be the discretized ring $\CircT_d$. As described at the beginning of Sec.~\ref{sec:pqm} (and worked out in detail in Appendix \ref{app:appdiscretedual}), the polymer momentum space is in general the Bohr compactification of the Schr\"odinger momentum space.  Thus, the polymer momentum space for a particle on a ring is $\IntB$, the Bohr compactification of the integers.

The Haar measure on $\IntB$ is given by 
\begin{equation}
\int_{\IntB}d\mu_H\, f(n)  = \lim_{M\to\infty} \frac{1}{2M+1}\sum_{n=-M}^{M} f(n) ~,
\label{eq:HaarZ}
\end{equation}
a result closely analogous to Eq.(\ref{eq:BohrInt}) for $\Reb$. (The Haar measure on compact groups is given in general by a finite, translation invariant mean of this type 
\cite{StackHaarBohr,haarjoy}.) Otherwise, computations in $\IntB$ proceed as if it were simply the integers, because the additional elements added to $\Int$ to make $\IntB$ are ``at infinity".

At this point, we can proceed with quantization in the same manner as the linear case. We define an inner product on the discrete ring $\CircT_d$ by
\begin{equation}
\bracket{\varphi}{\varphi'}=\delta_{\varphi,\varphi'} ~.
\label{eq:polymeripr}
\end{equation}
With polymer momentum states given by the transformation function (character, in the terminology of Appendix \ref{app:bohr})
\begin{equation}
\bracket{m}{\varphi}=e^{-im\varphi} ~,
\label{eq:momtransr}
\end{equation}
we have the resolution of the identity
\begin{subequations}
    \begin{align}
\Id &= \int_{\IntB} d\mu_H\, \ketbra{m}{m} \\
    &= \lim_{M\rightarrow\infty} \frac{1}{2M+1}\sum_{m=-M}^{M}\ketbra{m}{m}  ~.
\end{align}
\label{eq:mcompleter}
\end{subequations}
It may be confirmed with just a little effort%
\footnote{For $\varphi=\varphi'$, the result is immediate.  For $\varphi\neq\varphi'$, each partial sum is a geometric series, and the triangle inequality may be used to bound the sum from above by a factor that does not depend on $N$.
} %
that this correctly reproduces the discrete inner product, Eq.~(\ref{eq:polymeripr}). 

As in the Schr\"odinger case, unitary polymer Weyl operators may be defined by Eqs.~(\ref{eq:WeyldefPQMring}), and in the same way as the particle on a line they also satisfy Eqs.~(\ref{eq:WeylOpEigenr}), and hence the Weyl relations, Eqs.~(\ref{eq:WeylRelr}).

The definition of the polymer Hilbert space $\Hpoly$ follows in the same way by constructing cylinder states and completing in the inner product of Eq.~(\ref{eq:polymeripr}).  The resulting space of polymer-quantized position wave functions is $\Hpoly^{\varphi}=L^2(\CircT_d,d\mu_c)$, and the (angular) momentum wave functions live in $\mathcal{H}_{\mathrm{poly}}^{L_z}=L^2(\IntB,d\mu_H)$.
$\Hpoly$ is again non-separable. 

The final step is to define regular graphs on $\Hpoly^{\varphi}$ and construct quantum theories on the corresponding separable subspaces.

\subsection{Position graphs on a ring}
\label{sec:posgraphring}

The configuration space for a particle on a ring is (isomorphic to) the circle $\CircT$, so all functions on it must satisfy periodic boundary conditions.  Equivalently, when convenient one may regard the ring as an equivalence class of positions $x\mod 2\pi R$ on the real line. 
In the following analysis we will identify points $\varphi$ on the ring with their position $x= \varphi R$ along the circumference of the ring, though we will not concern ourselves with the location of the origin; only the periodicity of the ring's geometry is relevant.

\begin{figure}
\centering
\includegraphics[height=2 in]{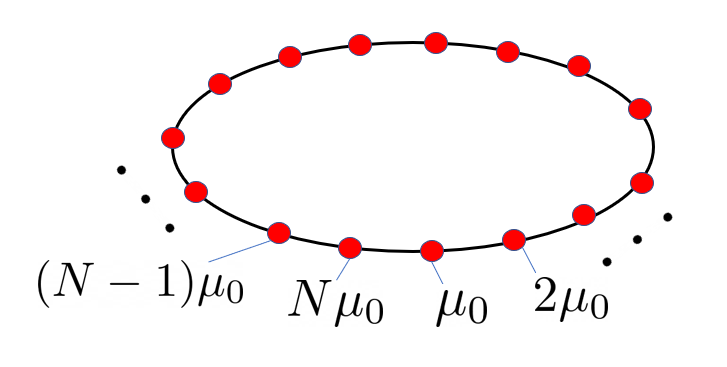}
\caption{A regular graph $\gamma_{\mu_0}$ of positions on the polymer quantized ring. The graph  consists of $N$ points distributed along the ring with uniform spacing $\mu_0 = 2\pi R/N$. The corresponding set of orthonormal position kets can be indexed from 1 to $N$, where  $\ket{x_k}\equiv\ket{x_k+N\mu_0}$. These kets form the basis for the finite-dimensional polymer graph Hilbert space $\mathcal{H}_{\gamma_{\mu_0}} = L^2(\Int_N , d\mu_c )$.}
\label{fig:ringpositions}
\end{figure}

To construct quantum theories on the discretized ring $\CircT_d$ we restrict support of states in $\Hpoly$ to regular graphs $\gamma_{\mu_0}$ with a finite number of points $N$, as discussed in Sec.~\ref{sec:posgraphc}. In order to ensure that the Weyl operator $\Vopr(N\mu_0)$ leaves the ring invariant, $\mu_0$ must divide the circumference of the ring, so $N\mu_0=2\pi R$.%
Since the origin of the coordinates is not relevant, we choose the labeling convention $x_k = k\mu_0$ with $k\in\{1,\ldots,N\}$, so that $x_{k+nN}=x_k$ for $n\in\Int$. A graph with $N$ points is thus $\gamma_{\mu_0}(N) = \{x_1,\ldots,x_N\}$.
Figure \ref{fig:ringpositions} illustrates such a graph of positions on a ring.
The corresponding set of orthonormal position states $\{\ket{x_k}, k=1,\ldots,N\}$ is then the basis for the \emph{finite dimensional} graph Hilbert space $\mathcal{H}_{\gamma_{\mu_0}} = L^2(\Int_N , d\mu_c )$.  (Here $\Int_N \equiv \Int\mod N$ are the integers modulo $N$.)
Cylinder states in $\ket{\psi}\in\mathcal{H}_{\gamma_{\mu_0}}$ are then
\begin{equation}
\ket{\psi} = \sum_{k=1}^N c_k\,\ket{x_k}
\label{eq:psir}
\end{equation}
subject to the normalization condition
\begin{equation}
    \sum_{k=1}^N |c_k|^2 = 1 
\label{eq:psirnorm}
\end{equation}
using the orthonormality of the position kets,
\begin{equation}
\bracket{x_j}{x_k} = \delta_{jk} ~.
\label{eq:xON}
\end{equation}
Because $\ket{x_{k+nN}}=\ket{x_k}$ for any $n\in\Int$, we also have $c_k=c_{k+nN}$.

In momentum space, using $\varphi_k = x_k/R = 2\pi k/N$, Eq.~(\ref{eq:psir}) reads
\begin{subequations}
\begin{align}
\psi(m) &= \sum_{k=1}^N c_k\bracket{m}{x_k} \\
 &= \sum_{k=1}^N c_k\, e^{-im\varphi_k} \\
 &= \sum_{k=1}^N c_k\, e^{-2\pi i mk/N} ~.
\label{eq:psirmom}
\end{align}
\end{subequations}

\subsection{Operator solution}
\label{sec:ringop}

The textbook analysis of the particle on a ring involves solving the time-independent Schr\"odinger equation for a free particle and then imposing periodic boundary conditions that capture the ring's geometry \cite{McIntyre}. We will employ a similar strategy for the polymer particle on a ring. Thus, we are to solve the Schr\"odinger equation on a finite graph of $N$ points on the ring in order to find the polymer energy eigenvalues and eigenstates for this system. (While not in the context of polymer quantum mechanics, there has been recent work analyzing similar finite discrete ring systems in quantum mechanics such as \cite{chung2023quantum}.) 

Because the particle on a ring is ``locally" like a free particle on a line, we can import the machinery developed for polymer quantum dynamics developed in section \ref{sec:dynamics} to analyze the polymer quantum mechanics of a particle on a ring. The fact that the graph on which the theory is defined is finite means that the problem lends itself to explicit closed form solution.

In the remaining sections of the paper, we solve the time-independent Schr\"odinger equation on the ring in two ways. First, we employ the properties of the Weyl operators used in defining the polymer Hamiltonian to find the ring eigenstates and energies in the graph Hilbert space $\Hgraphmuo$, and then find the solution a different way using a recurrence relation. We will then show how a process of renormalization can be used to recover the results of the Schr\"odinger quantization in the continuum limit $N\rightarrow\infty$, similar to the procedure developed in \cite{physhilbspace}. Finally, we will apply these results to explore the time dependence of quantum states in this polymer-quantized system.

From Eq.~(\ref{eq:Hring}), with $\widehat{p^2_{\mu_0}}$ given by Eq.~(\ref{eq:mom2defpqm}) we have the polymer Hamiltonian for a particle on a ring
\begin{equation} 
    \hat{H}_{\mathrm{r}}= \frac{\hbar^2}{2M\mu_0^2}\left[2\Id-\Vopr(\mu_0)-\Vopr(-\mu_0)\right]  ~,
\label{eq:Hringpoly}
\end{equation}
in terms of which the polymerized time-independent Schr\"{o}dinger equation for the particle on a ring is that of a free particle on the graph $\gamma_{\mu_0}$, 
\begin{equation}
\frac{\widehat{p^2_{\mu_0}}}{2M} \ket{\psi} = E\ket{\psi} ~,  
\end{equation}
so that
\begin{equation} 
    \frac{\hbar^2}{2M\mu_0^2}\left[2\Id-\Vopr(\mu_0)-\Vopr(-\mu_0)\right] \ket{\psi} = E\ket{\psi} ~.
\label{eq:ringTISE}
\end{equation}
Because the translation operators $\Vopr(n\mu_0)$ commute for any $n\in\Int$, this difference equation is easily solved by finding the eigenfunctions of the translation operators.

The action of $\Vopr(\mu_0)$ on a generic cylinder state, Eq.~(\ref{eq:psir}), is (cf.~Eq.~(\ref{eq:WeyldefPQMringpos}))
\begin{align}
 \Vopr(\pm\mu_0) \ket{\psi} &= \sum_{k=1}^N c_k \ket{x_{k\mp 1}} \\
  &= \sum_{k=1}^N c_{k\pm 1} \ket{x_{k}} ~,
\end{align}
where we recall that $c_{k+nN}=c_k$ for $n\in\Int$, so that $c_0=c_N$ and $c_{N+1}=c_1$.
We can build up arbitrary translations on the ring as $\Vopr(n\mu_0)=\Vopr(\mu_0)^n$, and in particular 
\begin{equation}
    \Vopr(\pm\mu_0)^N \ket{x_k} = \ket{x_{k\mp N}}= \ket{x_k} ~,
\end{equation}
and thus $\Vopr(\pm\mu_0)^N = \Id$. Since the Weyl operators are unitary, the eigenvalues of $\Vopr$ must have norm 1, so that with
\begin{equation}
  \Vopr(\mu_0)\ket{v_m} = v_m \ket{v_m} ~,  
\label{eq:Voprefneqn}
\end{equation}
we have $v_m=e^{i\delta_m}$. $\Vopr(\mu_0)^N = \Id$ then tells us that $v_m^N =1$, giving the possible eigenvalues
\begin{equation}
    v_m = e^{2\pi im/N},\ m = 1, \dots,N  ~.
\end{equation}
(In passing, we note that while this labeling of the eigenvalues is convenient for present purposes, $m$ could in principle be chosen from any set of $N$ consecutive integers.)
To find the eigenvectors $\ket{v_m}$, write
\begin{equation}
    \ket{v_m} = \sum_{k=1}^N c_{m,k} \ket{x_k} 
\end{equation} 
and apply $\Vopr(\mu_0)$, yielding
\begin{equation}
    \Vopr(\mu_0) \ket{v_m} = \sum_{k=1}^N c_{m,k} \ket{x_{k-1}} = \sum_{k=1}^N e^{2\pi i m/N} c_{m,k} \ket{x_k},
\end{equation}
indicating $c_{m,k+1} = e^{+2\pi i m/N} c_{m,k}$. Fixing $c_{m,1}$, each succeeding coefficient acquires a factor of $e^{+2\pi i m/N}$. Thus, 
\begin{equation}
    c_{m,k} = e^{+2\pi im(k-1)/N} c_{m,1}  ~.
\end{equation}
Therefore, the eigenstates of the translation operator $\Vopr(\mu_0)$ in the position basis are
\begin{equation} 
\ket{v_m}= c_1\sum_{k=1}^N \exp \left(\frac{2\pi im(k-1)}{N}\right) \ket{x_k} ~,
\label{eq:messyket}
\end{equation}
where $c_1$ is fixed by normalization. Absorbing a factor of $e^{-2\pi im/N}$ into $c_1$ and normalizing, we have finally 
\begin{subequations}
\begin{align}
\ket{v_m} &= \frac{1}{\sqrt{N}}\sum_{k=1}^N e^{\frac{2\pi imk}{N}}\ket{x_k} \\ 
   &= \frac{1}{\sqrt{N}}\sum_{k=1}^N e^{im\varphi_k}\ket{x_k} ~,
\end{align}
\label{eq:ringefns}%
\end{subequations}
with eigenvalue 
$v_m = e^{2\pi im/N}=e^{i\varphi_m}$. 

Because $\Vopr(-\mu_0)=\Vopr(\mu_0)^{\dagger}=\Vopr(\mu_0)^{-1}$, $\Vopr(-\mu_0)$ has the same eigenstates as $\Vopr(\mu_0)$, but its eigenvalues are the complex conjugates $v_n^*$. The $\ket{v_m}$ are therefore also the eigenstates of the polymer Hamiltonian,
\begin{equation}
    \frac{\hbar^2}{2M\mu_0^2}\left[2\Id-\Vopr(\mu_0)-\Vopr(-\mu_0)\right] \ket{v_m} = E_m\ket{v_m} ~,
\label{eq:ringTISEv}
\end{equation}
giving the energy eigenvalues
\begin{subequations}
\begin{align}
E_m &= \frac{\hbar^2}{2M\mu_0^2}\left(2-v_m-v_m^*\right) \\
    &= \frac{\hbar^2}{2M\mu_0^2}\left(2-e^{2\pi im/N}-e^{-2\pi im/N}\right) \\
    &= \frac{\hbar^2}{M\mu_0^2}\left(1 - \cos\left(\frac{2\pi m}{N} \right) \right) ~.
\end{align}
\label{eq:ringevals}%
\end{subequations}
It is interesting to note that these eigenvalues are bounded both below (by $E_N=0$, in the present labeling) \emph{and} above, unlike the Schr\"odinger-quantized ring. The upper-bound $E_{\frac{N}{2}}= \frac{2\hbar^2}{M\mu_0^2}$ may be regarded as a UV-cutoff that is due to the presence of the minimum length scale $\mu_0$. (The existence of a cutoff in this system is inevitable due to the finite dimensionality of the graph Hilbert space.) Note that except for the ground state $E_N$ and the upper-bound $E_{\frac{N}{2}}$ (which is attained only on lattices with an even number of points), the energies are two-fold degenerate, $E_{m} = E_{N-m}$, as would be expected because they represent counter-propagating modes on the ring. 
We will discuss the continuum limit of this expression for the polymer-quantized energies in Sec.~\ref{sec:ringcont} below.

Finally, for completeness we observe that the translation eigenstates we have constructed are just the momentum eigenstates $\ket{m}$ previously introduced restricted to the lattice $\gamma_{\mu_0}$. Indeed, even though we have expressed our position kets and translation operators in terms of the linear position $x_k=k\mu_0$ on the ring, we could equally well have written everything in terms of the corresponding angular positions $\varphi_k$; 
the lattice step $\mu_0$ corresponds to the fundamental angular step $\varphi_1 = \mu_0/R=2\pi/N$. In those terms, $\Vopr(\mu_0)\equiv \Vopr(\varphi_1)$ and $v_m=e^{im\varphi_1}$, so that the eigenvalue equation Eq.~(\ref{eq:Voprefneqn}) is consistent with Eq.~(\ref{eq:WeylOpEigenrmom}), and (identifying $\ket{\varphi_k}\equiv\ket{x_k}$ for position eigenstates on the graph $\gamma_{\mu_0}$%
\footnote{This is in contrast to the corresponding basis states defined on the continuous ring, which satisfy $\ket{\varphi} = \sqrt{R}\ket{x}$.
}%
)
\begin{equation}
    \bracket{\varphi_k}{v_m} = \frac{1}{\sqrt{N}}e^{im\varphi_k} ~,
\end{equation}
consistent with Eq.~(\ref{eq:momtransr}) up to the normalization appropriate to the restriction to the lattice. 
We will not, however, be using the momentum representation further in this paper.

\subsection{Solution by recurrence}
\label{sec:ringrecur}

We will now find the eigenstates and energies of the particle on a ring by an alternate method, using the polymerized time-independent Schr\"odinger equation to find a recurrence relation satisfied by the coefficients of the eigenfunctions. Inserting $\ket{\psi} = \sum_k c_k \ket{x_k}$ into Eq.~(\ref{eq:ringTISE}), we have
\begin{subequations}
\begin{align}
    \alpha\lambda \ket{\psi} &= \alpha [2\mathbb{I} - V(\mu_0) - V(-\mu_0)]\,\sum_{k=1}^N c_k \ket{x_k}  \\
    \lambda \ket{\psi} &= \sum_{k=1}^N\, [2c_k - c_{k-1} - c_{k+1}]\ket{x_k}  ~,
\end{align}
\end{subequations}
where we have defined $\alpha \equiv \frac{\hbar^2}{2M\mu_0^2}$ and $\lambda \equiv E/\alpha$. Acting on both sides with  $\bra{x_k}$ yields the recurrence relation
\begin{equation} \label{eq:ringrecurr}
    c_{k+1} = (2-\lambda)c_{k} - c_{k-1} ~.
\end{equation}
We will take $c_1$ and $c_2$ as given (to be determined by boundary conditions and normalization), use Eq.~(\ref{eq:ringrecurr}) for $k>2$, and require that $c_{k+N} = c_k$ for all $k$ to enforce the periodicity of the ring.

Eq.~(\ref{eq:ringrecurr}) is a linear, second-order recurrence relation with constant coefficients. This type of recurrence relation has a general solution in terms of the roots of a characteristic polynomial \cite{differenceEqns}. In this case, the characteristic polynomial is
\begin{equation} \label{eq:ringcharpoly}
    t^2 + (\lambda - 2)t +1 = 0
\end{equation}
with roots
\begin{align}
    t_{\pm} &= \frac{2 - \lambda \pm \sqrt{\lambda^2 - 4\lambda}}{2} ~.
\label{eq:ringroots}
\end{align}
If there are 2 distinct real roots $t_+$ and $t_-$, the solution to the recurrence relation is
\begin{equation}
    c_k = A t_{+}{}^k + B t_{-}{}^k  ~.
\end{equation}
This will occur when $\lambda^2 - 4\lambda > 0$, which corresponds to $\lambda < 0$ or $\lambda > 4$. Since the solutions are exponentials, the only way to satisfy the periodic boundary conditions is if $t_{+} = 1$, $t_{-} = -1$. However, these values are not possible roots to the characteristic polynomial for $\lambda < 0$ or $\lambda > 4$, so these solutions cannot satisfy the boundary conditions. This restricts the possible values of $\lambda$ to the range $0 \leq \lambda \leq 4$, which corresponds to the restriction on the allowed energies that
\begin{equation}
    0 \leq E \leq \frac{2\hbar^2}{M\mu_0^2}  ~.
\end{equation}

If there is only one distinct root $t_*$ of the characteristic polynomial, instead we get solutions of the form
\begin{equation}
    c_k = At_*{}^k + Bkt_*{}^k  ~.
\end{equation}
This occurs when $\lambda = 0$ and $t_* = 1$ or $\lambda = 4$ and $t_* = -1$. Since our solutions must be periodic and the second term grows with $k$, we must have $B = 0$.  This gives us our first eigenstate
\begin{equation}
    E = \alpha \lambda = 0, \quad \ket{\psi} = A \sum_k \ket{x_k} ~.
\end{equation}
The remaining solution $c_k = At_*{}^k$ is periodic with $t_* = 1$. It is only periodic with $t_* = -1$ if $N$ is even. In this case, we also have
\begin{equation}
    E = \alpha \lambda = \frac{2\hbar^2}{M\mu_0^2}, \quad \ket{\psi} = A \sum_k (-1)^k \ket{x_k}  ~.
\end{equation}

The last possibility is that the characteristic equation has 2 complex roots, $t_{\pm} = r e^{\pm i\tau}$. In this case we have solutions of the form
\begin{equation} \label{eq:ringcomplexroot}
    c_k = r^k\left(A\cos(k\tau) + B\sin(k\tau)\right)
\end{equation}
Relating $r$ and $\tau$ back to $\lambda$,
\begin{subequations}
\begin{align}
    \mathcal{R}e(t_+) &= \frac{1}{2}(2 - \lambda)\\
    \mathcal{I}m(t_+) &= \frac{1}{2}\sqrt{\lambda(4-\lambda)} ~,
\end{align}
\end{subequations}
giving
\begin{subequations}
\begin{align}
    \nonumber r^2 &= |\mathcal{R}e(t_+)|^2 + |\mathcal{I}m(t_+)|^2\\
    &= \frac{1}{4}\left((2-\lambda)^2 + 4\lambda - \lambda^2 \right)\\
    &= 1 ~.
\end{align}
\end{subequations}
We can similarly solve for $\tau$,
\begin{equation} \label{eq:ThetaCondition}
    \cos(\tau) = \frac{\mathcal{R}e(t_+)}{r} = \frac{2 - \lambda}{2} ~,
\end{equation}
so that from Eq.~(\ref{eq:ringcomplexroot}) we find
\begin{equation} \label{eq:ringcomplexrootsimp}
    c_k = A\cos(k\tau) + B\sin(k\tau).
\end{equation}
We can make this solution $N-$periodic by requiring  $ N\tau = 2\pi m$ with $m\in\Int$. With this condition, the allowed energies are determined by Eq.~(\ref{eq:ThetaCondition}) and the eigenstates by Eq.~(\ref{eq:ringcomplexrootsimp}), resulting finally in
\begin{equation}
    E_m = \alpha \lambda = \frac{\hbar^2}{M\mu_0^2}\left(1 - \cos\left(\frac{2\pi m}{N}\right)\right)\\
    \label{eq:ringrecursoln1}
\end{equation}
\begin{equation}
    \ket{\psi_m} = A\sum_{k=1}^N \cos\left(\frac{2\pi k m}{N}\right)\ket{x_k} + B\sum_{k=1}^N \sin\left(\frac{2\pi k m}{N}\right)\ket{x_k} 
    ~.\label{eq:ringrecursoln2}
\end{equation}
The energy eigenvalues in Eq.~(\ref{eq:ringrecursoln1}) are of course identical with those found in Eq.~(\ref{eq:ringevals}). To complete the solution, we need to find the coefficients $A$ and $B$ for the eigenstates. Begin by rewriting Eq.~(\ref{eq:ringrecursoln2}) as
\begin{equation}
    \ket{\psi_m} = \frac{1}{2}\sum_{k=1}^{N} \beta  e^{2\pi i k m/N}  \ket{x_k}+\frac{1}{2}\sum_{k=1}^N \beta^*e^{-2\pi i k m/N}  \ket{x_k} ~,
\end{equation}
where $\beta \equiv A+iB$. There are three cases to consider. For the first case, take $\frac{N}{2} \neq m\neq N$. Normalizing, we find $|\beta|=\sqrt{\frac{2}{N}}$, so $\beta = e^{i\rho} \sqrt{\frac{2}{N}}$ for $\rho \in \Re$. The eigenstates in this subset are thus
\begin{subequations}
\begin{align}
    \ket{\psi_m} &= \frac{1}{\sqrt{2N}}\left(\sum_{k=1}^{N}  e^{(2\pi i k m/N + i\rho)}  \ket{x_k}+\sum_{k=1}^N e^{-(2\pi i k m/N + i\rho)}  \ket{x_k} \right) \\
    &= \sqrt{\frac{2}{N}}\sum_{k=1}^N \cos\left( \frac{2\pi km}{N} + \rho \right) \ket{x_k} ~.
\end{align}
\label{eq:ringrecursefns}%
\end{subequations}
Recalling the two-fold degeneracy between $E_{m}$ and $E_{N-m}$ of the energy eigenvalues in Eq.~(\ref{eq:ringrecursoln1}), we fix $\rho$ by demanding that these pairs of degenerate eigenvectors are orthogonal. In particular,
\begin{subequations}
\begin{align}
    0 &=\braket{\psi_{N-m}}{\psi_m} \\
      &= \frac{1}{2N} \sum_{k=1}^N \left( e^{(2\pi i k m/N - i\rho)}  +e^{(-2\pi i k m/N + i\rho)} \right)\left( e^{(2\pi i k m/N + i\rho)}+e^{-(2\pi i k m/N + i\rho)}\right) \\
    &= \frac{1}{2N}\sum_{k=1}^N \left( e^{4\pi ikm/N} + e^{-4\pi ikm/N} + e^{2i\rho}+e^{-2i\rho} \right) \\
    &=  \cos(2\rho)  ~,
\end{align} 
\end{subequations}
so that any
\begin{equation}
    \rho = (2\nu+1)\frac{\pi}{4} ~, \ \ \nu\in\Int 
\end{equation}
will do. The resulting real eigenfunctions are simply a different choice of eigenbasis in each degenerate subspace than that of Eq.~(\ref{eq:ringefns}). 

Notice, however, that the eigenstates $ \{\ket{v_m} \} $ found in Sec.~\ref{sec:ringop} were explicitly constructed as the eigenvectors of the translation operators, which do not share the degeneracy of the energy eigenstates. As such, while the states $\ket{\psi_m}$ are eigenstates of the Hamiltonian, they are not eigenstates of discrete translations. In fact, these states are related by
\begin{equation}
    \ket{\psi_m} = \frac{1}{\sqrt{2}}\left( e^{i\rho} \ket{v_m} +e^{-i\rho} \ket{v_{N-m}} \right)
\end{equation} 
for $\rho \in \{\dots,-\frac{\pi}{4}, \frac{\pi}{4},\frac{3\pi}{4},\dots \}$.

However, this does not hold for the $m=N$ or $m=N/2$ cases, which do not have degenerate eigenvalues. In these cases, the eigenstates $\{\ket{\psi_m} \} $ and $\{\ket{v_m} \} $ must coincide. First, consider the ground state $m=N$. In this case, Eq.~(\ref{eq:ringrecursoln2}) reduces to
\begin{equation}
    \ket{\psi_N} = A \sum_{k=1}^N \cos\left({2\pi k } \right) \ket{x_k} = A\sum_{k=1}^N \ket{x_k} ~, 
\end{equation}
so that $A=\frac{1}{\sqrt{N}}$, and indeed $\ket{\psi_N} = \ket{v_N}$.  Similarly, consider the maximum energy eigenstate with $m=\frac{N}{2}$, which only occurs in the case for a ring with an even number of points. Here, Eq.~(\ref{eq:ringrecursoln2}) becomes
\begin{equation}
    \ket{\psi_{\frac{N}{2}}} = A\sum_{k=1}^N \cos\left({\pi k }\right)\ket{x_k} = A \sum_{k=1}^N (-1)^k\ket{x_k}  ~,
\end{equation}
and once again $\ket{\psi_{\frac{N}{2}}} = \ket{v_{\frac{N}{2}}}$.

Finally, then, the recursion method naturally leads to a basis of eigenstates of the Hamiltonian given by
\begin{equation}
  \ket{\psi_m} = \frac{1}{\sqrt{N}}  
  \begin{cases}
        \displaystyle\sum_{k=1}^N \ket{x_k} &\,\text{ if }\, \, m=N\\
        
        \displaystyle\sum_{k=1}^N {\sqrt{2}}\cos\left( \frac{2\pi km}{N} + \frac{(2\nu+1)\pi}{4} \right) \ket{x_k} &\,\text{ if } \,\, m\neq N \text{ or } \frac{N}{2}\,,\, \nu\in \Int  \\
        
        \displaystyle\sum_{k=1}^N (-1)^k \ket{x_k} &\,\text{ if } \,\, m=\frac{N}{2}
    \end{cases} ~.%
\end{equation}%
Interestingly, this leads to a basis consisting of entirely real wavefunctions.

\subsection{Continuum limit}
\label{sec:ringcont}

We turn now to the continuum limit of the eigenvalues and eigenstates of the polymer particle on a ring. We will show that they approach the standard Schr\"odinger-quantized results in a suitable limit, consistent with previous work on the continuum limit in other polymer-quantized systems such as the harmonic oscillator \cite{continuumlimit,physhilbspace,gravitationalwave}. 

\begin{figure}[h]
    \centering
    \includegraphics[width=0.75\textwidth]{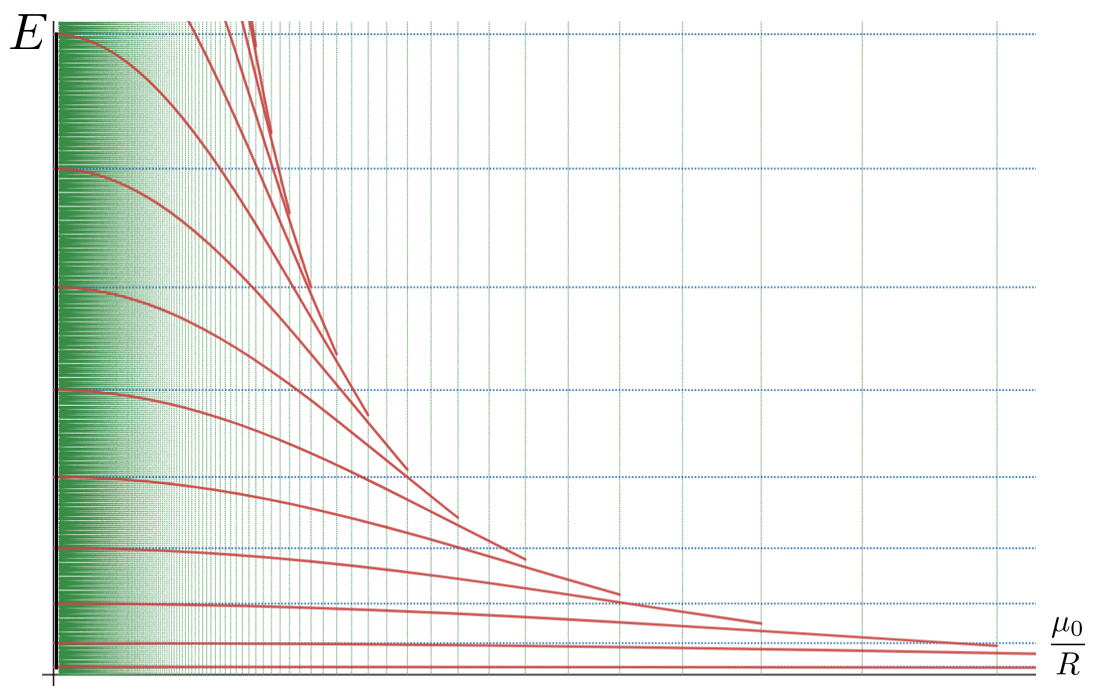}
    \caption{The spectrum of energy levels of the polymer-quantized particle on a ring as a function of $\mu_0/R$. The red curves show the behavior of these energy levels.  The green vertical lines are the allowed ratios of $\mu_0/R = 2\pi/N$ for $N\in\mathbb{N}$; only the points where the red energy curves intersect the green lines correspond to valid energy levels for the discretized ring. For a given value of $\mu_0/R$, only levels $E_m$ with $m\leq \frac{\pi R}{\mu_0}$  are allowed, reflecting the UV cutoff imposed by the finite lattice spacing $\mu_0$. Thus, as $\mu_0/R$ decreases, the polymer-quantized ring accommodates an increasing number of energy levels. The horizontal blue dotted lines are the energy levels of the corresponding Schr\"odinger-quantized system. As $\mu_0/R$ goes to zero, the allowable values of $\mu_0/R$ approach a continuum and the polymer-particle energies approach the standard values.}
\label{fig:continuum}
\end{figure}

In the case of the polymerized ring, we recall that the graph Hilbert space $\Hgraphmuo$ consists in cylinder states defined on finite regular graphs $\gamma_{\mu_0}$ such that the graph spacing $\mu_0$ divides the circumference of the ring, so that the number $N$ of graph points satisfies
\begin{equation}\label{eq:NtoR}
    N=\frac{2\pi R}{\mu_0} ~.
\end{equation}
The continuum limit is therefore achieved as $N\rightarrow\infty$ and $\mu_0\rightarrow 0$ with their product held fixed, so that (as expected) $\mu_0/R\rightarrow 0$. (This is not a continuous limit, however. Because of Eq.~(\ref{eq:NtoR}), the ratio $\mu_0/R$ may only take on discrete values $2\pi/N$ for $N\in\mathbb{N}$. These values nonetheless approach a continuum as $N\rightarrow\infty$ because the spacing $(\frac{2\pi}{N} - \frac{2\pi}{N-1})$ between adjacent points approaches zero.) In that limit, the energy eigenvalues given in Eq.~(\ref{eq:ringevals} become
\begin{subequations}
\begin{align}
E_m 
    &= \frac{\hbar^2}{M\mu_0^2}\left(1-\cos\left(\frac{m\mu_0}{R}\right)\right) \\
    &= \frac{\hbar^2}{M\mu_0^2} \left( \frac{1}{2!}\left( \frac{m\mu_0}{R}\right)^2 -\frac{1}{4!}\left( \frac{m\mu_0}{R} \right)^4 + \frac{1}{6!}\left( \frac{m\mu_0}{R} \right)^6 - \dots  \right) \\
    & \overset{\mathclap{\mu_0/R\rightarrow 0}}{=}\qquad \frac{(m\hbar)^2}{2MR^2} ~,
\end{align}
\end{subequations}
in agreement with the Schr\"odinger-quantized eigenvalues. This agreement is illustrated in Fig.~\ref{fig:continuum}.

Note that the UV cutoff $E_m \leq E_{\frac{N}{2}}$ at $m=\frac{N}{2}$ means that the the number of allowed polymer energies grows as $N$ increases, and correspondingly the allowed energies truncate for $\mu_0/R$ below $2\pi/N$ (i.e. $m\leq \frac{N}{2} = \frac{\pi R}{\mu_0}$). 
Put another way, the number of allowed energies increases as $\mu_0/R$ decreases, and the UV cutoff on the spectrum is lifted as $\mu_0/R$ goes to zero. This behavior can be clearly seen in the figure.

Finally, we investigate the continuum limit of the polymer eigenfunctions. In order to do this, we must accommodate the fact that these functions are defined on a different domain than that of the Schr\"odinger theory, the discrete lattice $\gamma_{\mu_0}$. We therefore seek to define a family of functions on $\Hsch$ that agree with the polymer eigenstates at the lattice points at which those states are defined, but fills in the points in between in some suitable fashion.  
There is obviously no unique prescription for this ``reverse discretization", so following \cite{physhilbspace,continuumlimit} we choose the blandest one in which the states defined on the continuous ring $\CircT$ assume the value of the corresponding graph state everywhere in an interval of width $\mu_0$ surrounding each lattice point.%
\footnote{For a more sophisticated treatment of the continuum limit from the point of view of coarse-graining of the underlying lattice, see \cite{continuumlimit}. The continuum limit for the polymer quantum harmonic oscillator has been investigated in, e.g., \cite{physhilbspace,gravitationalwave}. 
} %

In doing so, we recognize that the dimensions of wave functions in $\Hsch$ differ from those defined on the graph Hilbert space $\Hgraphmuo$. The counting measure doesn’t carry units, and therefore the discrete graph position basis states $\ket{x_i}$ don’t carry units either, while the base kets $\ket{x}$ defined on $\Hsch$ do. We may use the normalization of the states in their respective Hilbert spaces to find the correct mapping between them. We will therefore define wave functions on $\Hsch$ that, up to a dimensionful factor, agree with the wave functions defined on the lattice $\gamma_{\mu_0}$, and then investigate how those states behave as the lattice spacing, or more properly, $\mu_0/R$, goes to zero \emph{i.e.~}as the number of lattice points $N$ goes to infinity.  

Consider a normalized graph state $\ket{\psi}\in\Hgraphmuo$,
\begin{equation}
\ket{\psi} = \sum_{k=1}^N \psi_k \ket{x_k} ~,    
\label{eq:cylring}
\end{equation}
where $\sum_{k=1}^N |\psi_k|^2=1$. The position-basis ``wave function" is
\begin{subequations}
\begin{align}
\psi(x_j) &= \bracket{x_j}{\psi} \\
   &= \sum_{k=1}^N \psi_k\,\delta_{jk} ~.
\end{align}    
\label{eq:cylringQWF}%
\end{subequations}
In order to define a corresponding state in $\Hsch$, we first specify a family of intervals $\Delta_k$ that cover the ring, $\cup_{k=1}^N\Delta_k=\CircT$, and contain each lattice point $x_k$,
\begin{equation}
    \Delta_k \equiv (x_k-\mu_0,x_k] ~.
\end{equation}
(Any set of intervals that similarly cover the lattice points and the ring would do just as well for what follows.) 
Define the attendant characteristic functions
\begin{equation}
\delta^c_{x,x_k} = 
  \begin{cases}
      1 & x\in\Delta_k \\
      0 & \text{else} 
  \end{cases}    ~.
\end{equation}
These functions satisfy
\begin{subequations}
\begin{align}
  \delta^c_{x,x_j}\delta^c_{x,x_k} &= \delta_{jk}\, \delta^c_{x,x_k}  \\
  \sum_{k=1}^N \delta^c_{x,x_k} &= 1  ~.
\end{align}    
\end{subequations}
Finally, given a basis of position eigenkets $\{ \ket{x} \}$ for $\Hsch$, define the smeared position basis kets
\begin{subequations}
\begin{align}
\ket{x^c_k} & = \frac{1}{\mu_0}\int dx\, \delta^c_{x,x_k} \ket{x} \\
            &= \frac{1}{\mu_0}\int_{\Delta_k} dx\, \ket{x} ~.
\end{align}   
\label{eq:smearedx}%
\end{subequations}
It is straightforward to verify that these satisfy
\begin{equation}
 \bracket{x^c_j}{x^c_k} = \frac{1}{\mu_0} \delta_{jk} ~.   
\end{equation}
This immediately suggests the correspondence $\ket{x_k}\leftrightarrow\sqrt{\mu_0}\,\ket{x^c_k}$, 
as may have been anticipated from the need to reconcile the dimensions carried by the continuum base kets with the dimensionless discrete base kets on $\Hgraphmuo$.  Also, we have
\begin{equation}
    \bracket{x}{x^c_k} = \frac{1}{\mu_0} \delta^c_{x,x_k} ~,    
\end{equation}
so that, as expected, $\lim_{\mu_0\rightarrow0}\bracket{x}{x^c_k} = \delta(x-x_k)$.

\begin{figure}
    \centering
    \includegraphics[width=0.65\textwidth]{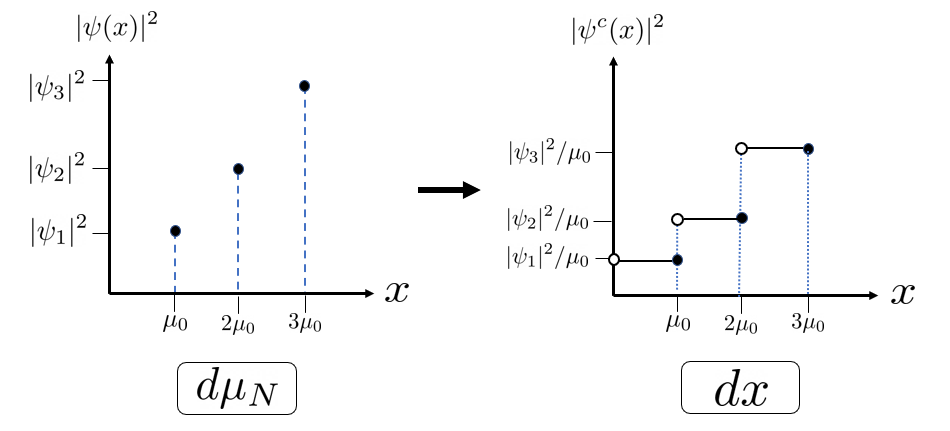}
    \caption{The mapping from the wave function for a state $\ket{\psi}$ defined on the lattice $\gamma_{\mu_0}$ endowed with the discrete counting measure $d\mu_N$ to a corresponding state $\ket{\psi^c}\in\Hsch$ defined everywhere on the ring and normalized in the Lebesgue measure. 
    }
\label{fig:kronecker}
\end{figure}

Accordingly, given a normalized graph state such as Eq.~(\ref{eq:cylring}), define the corresponding coarse-grained state in $\ket{\psi^c}\in\Hsch$ to be
\begin{equation}
    \ket{\psi^c} = \sum_{k=1}^N \sqrt{\mu_0}\, \psi_k\,\ket{x^c_k} ~,
\end{equation}
with wave function 
\begin{equation}
    \psi^c(x) = \frac{1}{\sqrt{\mu_0}} \sum_{k=1}^N \psi_k\, \delta^c_{x,x_k} ~.
\label{eq:contQWF}    
\end{equation}
The correspondence with Eq.~(\ref{eq:cylringQWF}) should be clear. It is straightforward to check that $\bracket{\psi^c}{\psi^c}=\int_0^{2\pi R}dx\, |\psi^c(x)|^2 =1$, as required, or more generally, that inner products on $\Hgraphmuo$ are preserved under this mapping to $\Hsch$, $\bracket{\phi}{\psi} = \bracket{\phi^c}{\psi^c}$. In essence, by embedding the graph $\gamma_{\mu_0}$ in the ring $\CircT$ we may ``lift" graph states normalized in the counting measure defining the graph Hilbert space to states defined on the continuous ring normalized in the Lebesgue measure. The correspondence between the wave functions $\psi(x_j)$ and $\psi^c(x)$ is illustrated in Fig.~\ref{fig:kronecker}.

\begin{figure}
    \centering
    \includegraphics[width=0.6\textwidth]{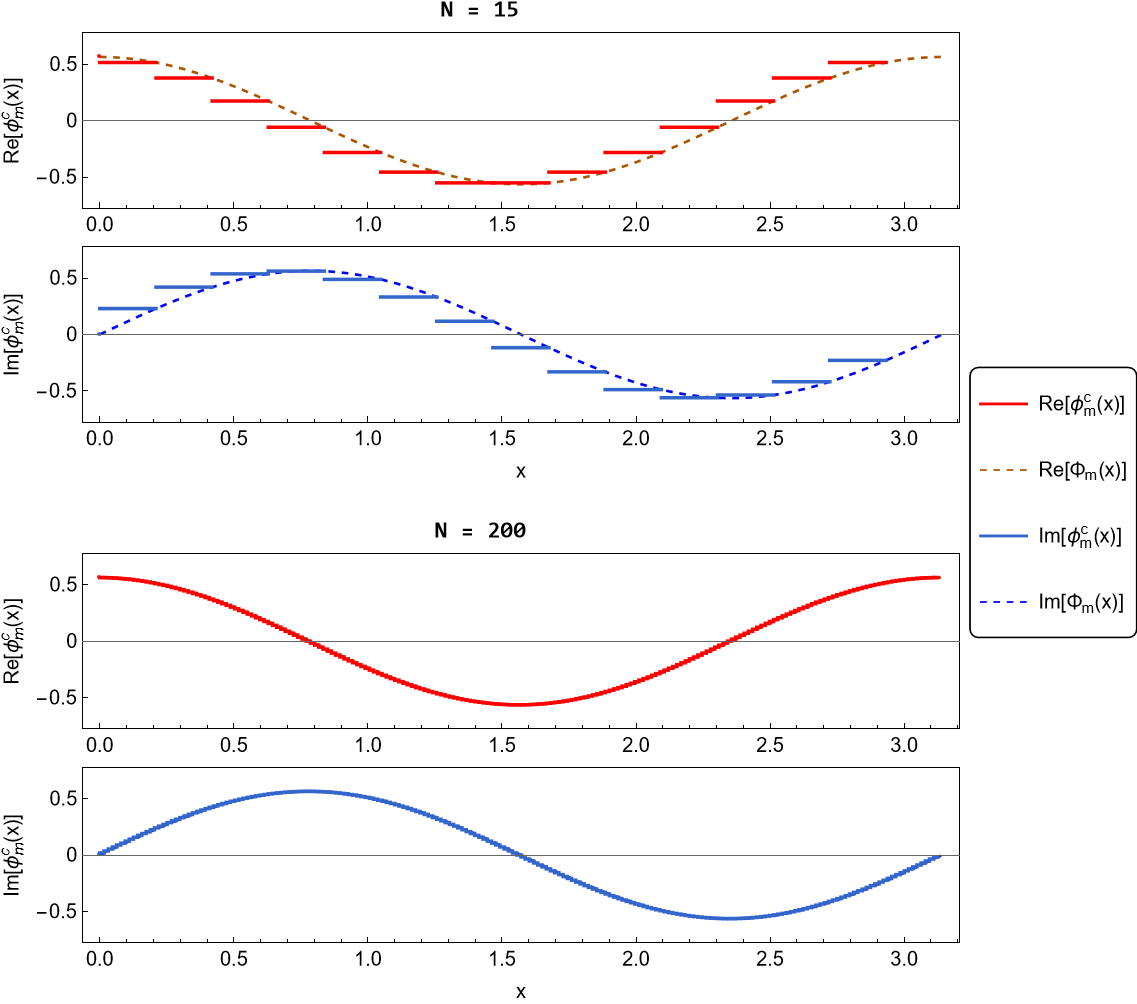}
    \caption{Convergence of polymer energy eigenfunctions to their Schr\"odinger-quantized counterparts for the polymer particle on a ring. The red and blue step functions are the real and imaginary parts of a  polymer eigenfunction $\phi^c_m(x)$. The orange and purple curves are similarly the real and imaginary components of the corresponding Schr\"odinger eigenstate $\Phi_m(x)$.  The top plot corresponds to a polymer graph with $N=15$, and the bottom plot to $N=200$. As $N\rightarrow\infty$ so that $\mu_0/R\rightarrow 0$, the polymer eigenstates become arbitrarily close to the Schr\"odinger eigenstates.
    }
\label{fig:stepfunction}
\end{figure}

Applied to the polymerized ring eigenfunctions $\ket{v_m}$ of Eq.~(\ref{eq:ringefns}), define the Schr\"odinger states
\begin{equation}
  \ket{v^c_m} = \sqrt{\frac{\mu_0}{N}}\, \sum_{k=1}^N e^{im\varphi_k}\,\ket{x^c_k}   
\end{equation}
with wave functions
\begin{subequations}
\begin{align}
 \phi^c_m(x) &\equiv \bracket{x}{v^c_m} \\
          &= \frac{1}{\sqrt{\mu_0 N}}\, \sum_{k=1}^N e^{im\varphi_k}\,\delta^c_{x,x_k}  \\ 
          &= \frac{1}{\sqrt{2\pi R}}\, \sum_{k=1}^N e^{im\varphi_k}\,\delta^c_{x,x_k}   
\end{align}    
\end{subequations}
using Eq.~(\ref{eq:NtoR}).  Thus we see that
\begin{equation}
    \phi^c_m(x\in\Delta_k) = \Phi_m(x_k) ~,
\label{eq:phicPhi}
\end{equation}
where the $\Phi_m(x)$ are the Schr\"odinger eigenstates
\begin{equation}
   \Phi_m(x) = \frac{1}{\sqrt{2\pi R}}\,e^{\frac{imx}{R}} ~. 
\end{equation}
The polymer eigenstates $\phi^c_m(x)$ and their Schr\"odinger counterparts $\Phi_m(x)$ are illustrated in Fig.~\ref{fig:stepfunction}. It is clear from the figure that the polymer energy eigenstates converge to the eigenstates of the Schr\"odinger-quantized ring as $\mu_0/R$ goes to zero.
 
In fact, this convergence is uniform. The exponentials $\Phi_m(x)$ are uniformly continuous in the standard norm $||\psi(x_0)||=\sqrt{\psi^*(x_0)\psi(x_0)}$, so that for all $\epsilon>0$, a $\delta>0$ may be found such that if $|b-a|<\delta$, then $|| \Phi_m(b)-\Phi_m(a) ||<\epsilon$ \cite{abbott2e}. Now, if we choose $\mu_0 < \delta$, any pair of points $\{a,b\}$ that are both in the same interval $\Delta_k$ are closer than $\delta$. Therefore, for any point $x\in\Delta_k$, using Eq.~(\ref{eq:phicPhi}) we have
\begin{equation}
 || \phi^c_m(x)-\Phi_m(x)|| = ||\Phi_m(x_k)-\Phi_m(x)|| < \epsilon ~,   
\end{equation}
and thus the functions $\phi^c_m(x)$ converge uniformly to $\Phi_m(x)$ as $\mu_0$ goes to zero.

To summarize, we see that in the limit as $\mu_0/R \rightarrow 0$, the polymer eigenstates approach the continuum eigenstates, the polymer eigenvalues approach their continuum values, and the UV cutoff on the theory lifts.

\subsection{Time evolution} 
\label{sec:ringtime}

To give some life to our polymer quantum states, let us consider the time evolution generated by the polymer Hamiltonian, Eq.~(\ref{eq:Hringpoly}). In contrast to the situation expected in quantum gravity, we will take time to be a continuous parameter in the sequel, so that quantum states satisfy the standard Schr\"odinger equation
\begin{equation}
    i \hbar \frac{d}{d t}\ket{\phi (t)} = \hat{H}_{\mathrm{r}}\,\ket{\phi(t)} ~, 
\end{equation}
so that in particular we have for the energy eigenstates $\phi_m(x_k,t)=\melt{x_k}{e^{{-i\hat{H}_{\mathrm{r}}\,t}/{\hbar}}}{v_m}$
\begin{equation}
    \phi_m(x_k,t) = \frac{1}{\sqrt{N}}e^{i(m\varphi_k- E_m t/\hbar)} ~.
\end{equation}




\subsubsection{Dispersion of a localized state}

To illustrate this discrete dynamics, we will investigate the dispersion of a localized state around the ring and briefly compare the results to the results of Schr\"odinger-quantization.

\begin{figure}
    \centering
    \includegraphics[width=0.35\textwidth]{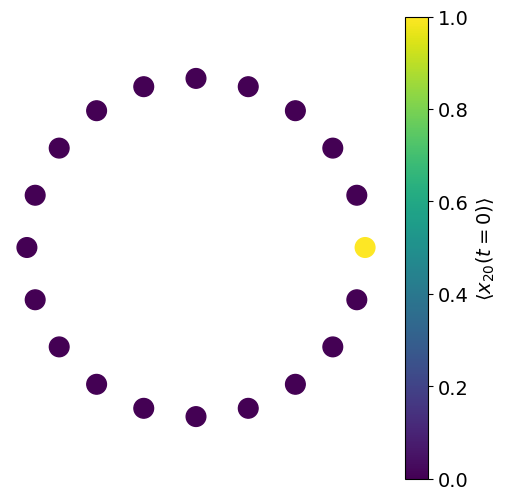}
    \caption{Real part of the polymer quantum state of a particle localized at a single lattice site on a ring graph with $N=20$.
    }
\label{fig:dispersion0}
\end{figure}

\begin{figure}[hbtp!]
    \centering 
    
    \begin{minipage}{0.6\linewidth}
        \centering
        \subfloat[]{
            \includegraphics[width=\linewidth]{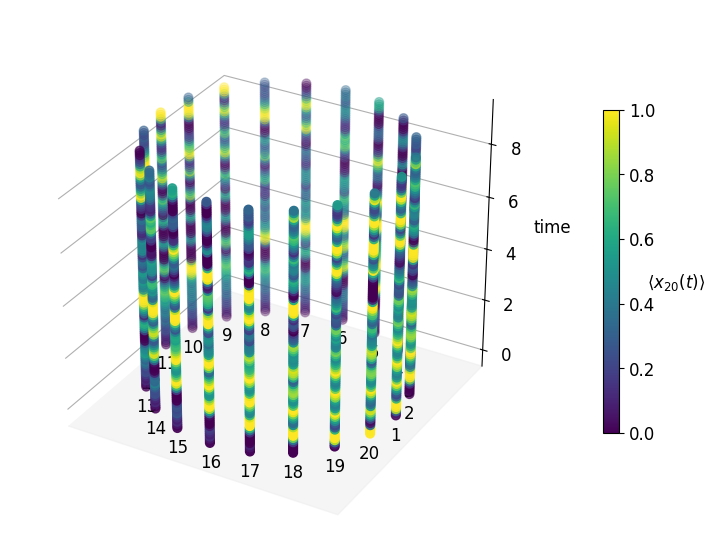}
            \label{fig:dispersion9circ}
        }
    \end{minipage}%
    \hspace{1em} 
    \begin{minipage}{0.30\linewidth}
        \centering
        \subfloat[]{
            \includegraphics[width=\linewidth]{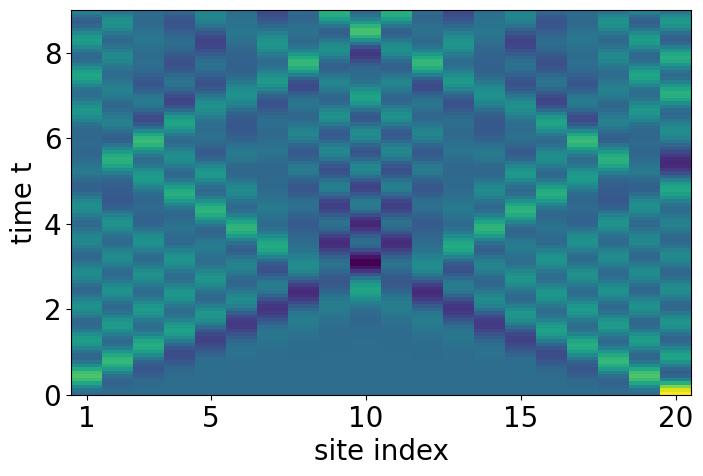}
            \label{fig:dispersion9flat}
        }
    \end{minipage}

    \caption{Two representations of the time evolution of the wave function of a state initially entirely concentrated on a single lattice site (in this case $k=20$) for a graph with $N=20$ lattice sites. One can see the state disperse symmetrically about its initial location at site $N=20$.    
    }
    \label{fig:dispersion9}
\end{figure}

We will take our initial state to to be localized at a single lattice point $x_c$. An example is illustrated in Fig.~\ref{fig:dispersion0}. We find $\ket{x_c(t)}$ in the usual fashion, 
\begin{subequations}
\begin{align}
\ket{x_c} 
    &= \sum_{m=1}^N \ket{v_m}\bracket{v_m}{x_c} \\
    &= \frac{1}{\sqrt{N}}\sum_{m=1}^N e^{-im\varphi_c}\ket{v_m} ~,
\end{align}    
\end{subequations}
so that
\begin{subequations}
\begin{align}
\ket{x_c(t)} 
    &= \frac{1}{\sqrt{N}}\sum_{m=1}^N e^{-im\varphi_c} e^{-iE_mt/\hbar}\ket{v_m} \\
    &= \frac{1}{\sqrt{N}}\sum_{m=1}^N\sum_{k=1}^N 
        e^{im(\varphi_k-\varphi_c)}e^{-iE_mt/\hbar}\ket{x_k} ~.
\end{align}    
\end{subequations}
An example of the evolution of such an initially localized state is given in Fig.~\ref{fig:dispersion9}.

\begin{figure}
    \centering
    \includegraphics[width=0.85\textwidth]{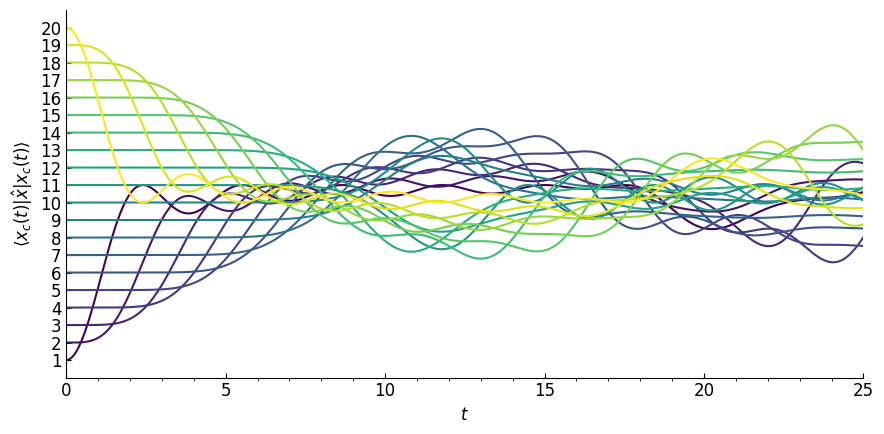}
    \caption{Expectation value of the position (in units of $\mu_0$) as a function of time for states initially localized at different points $x_c$ on a graph with $N=20$. Note that these expectation values all eventually oscillate around $N/2$, the average of all the graph labels. 
    }
\label{fig:differentstarts}
\end{figure}

Using $\xop\ket{x_k}=k\mu_o\ket{x_k}$, the expectation value of the position is then
\begin{subequations}
\begin{align}
\expct{\xop}_c(t) &= \melt{x_c(t)}{\xop}{x_c(t)} \\
  &= \frac{\mu_0}{N^2} \sum_{n,m=1}^N\sum_{k=1}^N k 
      e^{i(m-n)(\varphi_k-\varphi_c)}e^{-i(E_m-E_n)t/\hbar} ~.
\end{align}    
\end{subequations}
The behavior of this expectation value is plotted for various values of $x_c$ in Fig.~\ref{fig:differentstarts}. 
It may at first seem surprising that this expectation value oscillates around $N/2$, the ``average'' graph position, regardless of the initial location of the localized state. However, this behavior is simply reflective of the fact that as each of these localized states disperses around the ring, the probability distribution becomes (on average) uniform around the ring, so the expectation value naturally returns the average graph label.  This is consistent with what finds for the continuous ring if one calculates the expectation value of a putative%
\footnote{For discussion of phase operators in quantum theory, see, for example, \cite{barnvac07}.  We have already seen that it is usually best to consider the corresponding translation operators instead, such as in Eqs.~(\ref{eq:WeyldefPQMring}).
} %
angle operator $\hat{\varphi}$ in an energy eigenstate $\ket{m}$, for which the probability density is uniform around the ring:
\begin{subequations}
\begin{align}
    \melt{m}{\hat{\varphi}}{m} 
      &= \int_{0}^{2 \pi}d{\varphi}\, \frac{1}{\sqrt{2 \pi}}e^{-i m \varphi} \varphi 
          \frac{1}{\sqrt{2 \pi}}e^{i m \varphi} \\
    &= \pi  ~,
\end{align}
\end{subequations}
irrespective of the origin of coordinates. 
(The analogous polymer calculation gives $\expct{\hat\varphi}_m \equiv\melt{v_m}{\xop}{v_m}/R=\pi\frac{N+1}{N}$, which agrees with the result for the continuous ring for large $N$.)

\section{Discussion}
\label{sec:disc}

Polymer quantization has by now been widely applied to develop models of a quantum universe in which spacetime is fundamentally discrete. In much of the literature, however, the details of this procedure are presented only very briefly. ``Polymer quantum mechanics" was itself developed to explore these methods in simplified settings such as applications to familiar non-relativistic quantum systems. The present work has two primary goals. First, we have aimed to provide a reasonably complete pedagogical presentation of polymer quantum mechanics in order to make the subject accessible to a wider audience. Thus we include some of the details of the physical motivation and mathematical background for an audience of physicists who may be less familiar with this material. Along the way, we offer some new perspectives on polymer quantization, including an explicit account of the quantum momentum space as the group of translations dual to the configuration space; see Appendix \ref{app:appdiscretedual}-\ref{app:qmomspace}. Second, by way of example, we apply these methods to the study of the polymer quantum mechanics of systems with compact classical configuration spaces. In particular, we develop the polymer quantum theory of a particle confined to move on a ring.  This is novel because most applications of polymer quantization have so far concentrated on systems with a classical phase space in which both the position and momentum take values on the full real line.  The compactness of the classical configuration space introduces new elements to the corresponding polymer quantum theory, leading in particular to a quantum theory defined on a \emph{finite} graph Hilbert space.

Polymer quantum mechanics, in its essence, is a new choice of representation of the canonical commutation relations of the theory in the form of its Weyl algebra. Canonical quantization involves mapping the algebra of phase space variables defined by a classical theory's Poisson brackets to an algebra of quantum operators defined by commutators. It is necessary to choose a Hilbert space to carry a representation of this algebra. The Stone-von Neumann theorem places strong restrictions on how any such choice of Hilbert space may differ from the standard Schr\"odinger representation by square-integrable functions. Polymer quantum mechanics bypasses the conditions of the Stone-von Neumann theorem by identifying a representation of the Weyl algebra in which the Weyl operators are not weakly continuous due to the discrete topology of the underlying configuration (equivalently, momentum) space. For the case of a classical phase space where both position and momentum take values on the real line, the resulting Hilbert space is the space of square-integrable functions on the discrete real line, $L^2(\mathbb{R}_d,d\mu_c)$. Meanwhile, the Hilbert space representing functions of the conjugate variable is the space of square-integrable functions on the Bohr compactification of the real line, $L^2(\mathbb{R}_B,d\mu_H )$. In order to have well-defined dynamics, further restrictions must be imposed on the configuration space resulting in a theory defined on a discrete graph. The physics of polymer quantum mechanics unfolds on this lattice subspace, with dynamical equations that are discrete difference equations taking the place of the differential equations governing the dynamics of Schr\"odinger-quantized systems. The theory thus naturally breaks into ``super-selection" sectors analogous to those that arise in loop quantum cosmology.

(As an aside to this discussion of super-selection, it is worth commenting that in our analysis we were not overly concerned with the specific choice of graph, since in the systems discussed, there is no physical preference for one choice of lattice over another. In quantum cosmology, however, symmetry requirements may dictate the choice of super-selection sector. The particular example in mind is the homogeneous and isotropic Friedmann-Lema\^itre-Robertson-Walker universe with a massless scalar field and no fermionic matter. In the loop-quantization of this system the spatial volume $V$ takes on discrete values. Due to a large gauge symmetry under reversal of the triad in terms of which the spatial metric is expressed, it is necessary for the physical Hilbert space to be symmetric about $V\rightarrow -V$, necessitating the inclusion of the point $V=0$, thereby fixing the super-selection sector \cite{Ashtekar:2007em} .)

In this paper we extend the framework of polymer quantization to other topologies for the classical phase space, namely systems in which the classical configuration space is compact. The theory of Fourier analysis on topological groups tells us that the polymerized momentum space is the Bohr compactification of the Schr\"odinger momentum space. We show how this requires that the polymer quantum theory must be defined on a \emph{finite} graph with a characteristic scale $\mu_0$, meaning that polymer quantization of a system with a compact configuration space effectively reduces to the quantum mechanics of a finite dimensional system.

Applied to the polymer particle on a ring, the polymer position space and momentum space are respectively the circle group with the discrete topology, $\mathbb{T}_d$, and the Bohr compactification of the integers, $\mathbb{Z}_B$. The corresponding graph Hilbert space is finite dimensional, with dimension $N$ equal to the number of lattice sites defining the graph. We explicitly compute the energy eigenvalues and eigenfunctions of a free particle on the polymer quantized ring using two different methods, one exploiting the expression of the Hamiltonian in terms of the Weyl operators, the other by deriving and solving a recurrence relation satisfied by the coefficients of the eigenstates in the position basis.  The presence of the lattice scale $\mu_0$ leads to a UV-cutoff on the energy spectrum. We then demonstrate that in the continuum limit as the lattice scale $\mu_0$ shrinks to zero relative to the overall size of the ring, the UV cutoff lifts and both the eigenvalues and eigenstates approach their Schr\"odinger-quantized counterparts.  Finally, adopting a model in which time remains a continuous parameter, we briefly explore the dynamics of localized polymer quantum states on the ring.

In an appendix we apply the same methods to the study of a polymer particle in an infinite well. This is a system that has been previously studied, but the presentation and solution we offer here does have some new features.

Polymer quantum mechanics was originally introduced to study in a simplified setting the physics of some of the mathematical structures underlying loop quantum gravity and loop quantum cosmology, not as a proposal for a model of reality. That said, it is interesting that study of polymerized quantum theory makes it clear that it is possible to model space as fundamentally discrete, but yet the quantum theory may have ordinary Schr\"{o}dinger quantum mechanics as its continuum limit, which we \emph{know} does an excellent job of modeling the behavior of the physical world. However, this convergence to Schr\"{o}dinger quantum theory is not a generic feature of polymer-quantized systems. In particular, loop quantum cosmology is rooted in the framework of polymer quantization, in contrast to conventionally-quantized ``Wheeler-de Witt" quantum cosmological models. The standout prediction from loop quantum cosmology is the replacement of the initial big bang singularity with a ``big bounce", a global result which fundamentally differs from the Wheeler-de Witt theory, in which the classical singularity remains in the quantum theory \cite{aps,Ashtekar:2007em}. It has been demonstrated that in certain simple, exactly solvable symmetry-reduced models, the presence of the big bounce in loop quantum cosmology and the presence of the big bang singularity in the Wheeler-de Witt theory are guaranteed \cite{CS10c,CS13a,dac13a}. 
Thus, while the simple systems investigated in this manuscript are observationally indistinguishable from Schr\"{o}dinger quantum theory, this is not the case in general.

The distinctive predictions of polymer quantization are rooted in the emergence of a \emph{non-separable} Hilbert space, as opposed to the separable Hilbert spaces that typically arise in quantum theory. This in turn may be traced to the choice of a discrete topology on the classical configuration space. While non-separable Hilbert spaces are uncommon in physics, 
it is perhaps unsurprising that less familiar mathematical structures emerge when modeling physical space in unfamiliar ways. Because all infinite-dimensional separable Hilbert spaces are equivalent up to isometric isomorphism \cite{conway90}, the Stone-von Neumann theorem imposes strong restrictions on the representation of physical operators in quantum mechanics \cite{vanenk25} that are evaded in polymer-quantized systems.  Interestingly, once this is done, the restriction of the theory's dynamics to ``super-selection sectors" defined on a graph Hilbert space reduces the physical Hilbert space once again to a separable Hilbert space. Table \ref{table:hilbertspaces} summarizes the various Hilbert spaces arising in our study.

\begin{table}[h]
\label{table:hilbertspaces}
\centering
\setlength{\extrarowheight}{3pt}
\begin{tabular}{|c|c||c|c|c|c|c|c|}
\hline
\multicolumn{2}{|c||}{} & \multicolumn{3}{c|}{ Particle on a Line} & \multicolumn{3}{c|}{ Particle on a Ring} \\
\hline
Quantization & Representation & Hilbert Space & Dimension & Separable? & 
  Hilbert Space & Dimension & Separable? \\
\hline\hline
Schr\"odinger & Position  &  
$\Hsch^x=L^2(\Re_{|\cdot|},dx)$  & $\aleph_0$ & yes   &  
$\Hsch^{\varphi}=L^2(\mathbb{T}_{|\cdot|},d\varphi)$  &  $\aleph_0$ & yes  \\
\hline
  & Momentum  & $\Hsch^p=L^2(\Re_{|\cdot|},dp)$   &  $\aleph_0$ & yes
  &  $\Hsch^{L_z}=L^2(\mathbb{Z},d\mu_c)$   &   $\aleph_0$ & yes \\
\hline
Polymer & Position  & 
$\Hpoly^x = L^2(\Re_d,d\mu_c)$ & $\mathfrak{c}$ & no   & 
$\Hpoly^{\varphi} = L^2(\mathbb{T}_d,d\mu_c)$ & $\mathfrak{c}$ & no   \\
\hline
  & Momentum  & $\Hpoly^p = L^2(\Re_B,d\mu_{H})$  & $\mathfrak{c}$ & no 
  &  $\Hpoly^{L_z} = L^2(\Int_B,d\mu_{H})$  &  $\mathfrak{c}$ & no  \\
\hline
  & Position Graph &  $\mathcal{H}_{\gamma_{\mu_0}} = L^2(\Int,d\mu_c)$  &  $\aleph_0$ & yes &  $\mathcal{H}_{\gamma_{\mu_0}} = L^2(\Int_N , d\mu_c ) $  &  $N\in\mathbb{N}$ & yes (finite)  \\
\hline
\end{tabular}
\caption{Comparison of Hilbert spaces arising in Schr\"odinger and polymer quantizations of one-dimensional quantum systems in both position and momentum representations (``polarizations"). For reference, the dimension $\aleph_0$ (the cardinality of the integers) means the space is ``countably infinite'', while $\mathfrak{c}$  (the cardinality of the continuum, e.g. $\Re$) means the space is ``continuously infinite''. }
\end{table}

Future directions to extend this work include the exploration of polymer quantization methods to higher-dimensional systems such as a particle bound to a sphere or cylinder, in which ambiguities in the construction of lattice graphs and their corresponding graph Hilbert spaces arise, as well as other technical difficulties. Additionally, though in this work we took time to be a continuous parameter, polymer quantum mechanics is motivated by quantum gravity, in which we may expect both space and time to take on discrete characteristics. While this does not arise naturally in ordinary quantum theory because there is no self-adjoint operator associated with time, we might take the cue from canonical quantum cosmology, in which time is typically defined relationally in terms of matter fields via the Page-Wootters mechanism \cite{ashsingh11}, and explore relational time in polymer quantum mechanics.

\appendix 

\section{Bohr compactification and the quantum momentum space}
\label{app:bohr}

This appendix is dedicated to a brief account of the main ideas behind the Bohr compactification%
\footnote{The ``Bohr'' to whom this notion is originally due is Harald Bohr, younger brother to the renowned Niels.  Harald Bohr pioneered the study of almost-periodic functions we discuss briefly below. 
} %
and its relationship to the quantum momentum space.
A basic familiarity is assumed with elementary definitions and ideas from point-set topology such as open, closed, and compact sets and continuity of functions, as well as basics of group theory such as the definition of an Abelian (commutative) group (see, for example, \cite{szekeres,Pal19} for the relevant background). We summarize the main ideas and results only, saving the details 
for the references; as will be clear, a relatively significant (relative to the background of most physicists) degree of mathematical sophistication is required to fully unpack all of the ideas in play in the definition and analysis of the Bohr compactification.

A nice summary of the definition and main properties of the Bohr compactification may be found in \cite{whatisbohr}.  See also \cite{Thiemann,bohrpaper} for many details from the vantage point of loop quantum gravity.

\subsection{Bohr compactification}
\label{app:bohrcompact}

There are two alternative characterizations of the Bohr compactification that shed light on different aspects of the underlying idea, one from the perspective of harmonic analysis, and the other from that of $C*$-algebras. We begin with harmonic analysis.

A \emph{topological group} is a group endowed with a topology on its elements in which the group operation is a continuous map.  
A familiar example is the real line with addition as the group operation and its usual topology. Harmonic analysis \cite{RudinFourier,Harmonic3e} is a generalization of Fourier analysis to any locally compact Abelian group, that is to say, an Abelian topological group for which the topology is such that
every point has a neighborhood contained in a compact set.  We will refer to such groups as LCA groups for brevity. The additive group of real numbers is the template for an LCA group, so harmonic analysis can be thought of as the generalization of Fourier analysis on the reals to a general locally compact Abelian group.

As we have noted in the body of the paper, the Bohr compactification $\Reb$ of the real line $\Re$ is the ``dual'' group to the additive group of real numbers with the discrete topology $\Red$. Let us explain what this means. The idea of a group dual to a topological group is parallel to the related idea of the dual space of a vector space. The topological dual $\mathcal{V}^*$ of a vector space $\mathcal{V}$ is defined to be the vector space of continuous linear functionals on $\mathcal{V}$, that is, the space of continuous linear maps from vectors in $\mathcal{V}$ to (usually complex) numbers. 
In quantum mechanics on a Hilbert space $\mathcal{H}$, these are simply the ``bras'' $\bra{\varphi}\in\mathcal{H}^*$ that map ``kets'' $\ket{\psi}\in\mathcal{H}$ to complex numbers $\bracket{\varphi}{\psi}$. The Riesz representation theorem guarantees that the dual space, the space of bras, is isomorphic to the original Hilbert space of kets \cite{szekeres,RSI}, and so physicists learn to treat them on an equal footing.%

The dual group to an LCA group is defined in a similar way, as a group of maps from the original group to complex numbers, but instead of preserving the linear structure of a vector space, the maps preserve the group structure of the group. Let us make this precise. 

A \emph{character} $\xi$ of an LCA group $G$ is a continuous homomorphism from $G$ to the multiplicative group of norm-1 complex numbers (the ``circle group'' $\Circ$, also often denoted $\CircT$ or $\CircS$).  Thus
\begin{equation}
\xi(x+y) = \xi(x)\xi(y) \quad\text{and}\quad |\xi(x)| = 1 
\label{eq:character}
\end{equation}
for $x,y\in G$. The set $\hat{G}$ of all characters on $G$ is an Abelian group under pointwise multiplication (so $\chi(x)=\xi(x)\zeta(x)$ is a character if $\xi$ and $\zeta$ are) called the \emph{dual group} of $G$. 
If $\hat{G}$ is endowed with what is called the ``compact-open'' topology, in which convergence of sequences of characters in $\hat{G}$ is defined by the requirement that, considered as functions on $G$, the sequence converges uniformly on every compact subset of $G$, then $\hat{G}$ is an LCA group in its own right. 

The group of characters $\hat{G}$ is called the group ``dual'' to $G$ because of the \emph{Pontryagin duality theorem}, which states that if $\hat{G}$ is the dual group of $G$, then $G$ is the dual group of $\hat{G}$ \cite{RudinFourier,Harmonic3e}.  (Put another way, $\dhatG = G$
, similarly to how for duals of Hilbert spaces we have $(\mathcal{H}^*)^* = \mathcal{H}$.)
In view of this duality, it is customary to write $\xi(x)$ as $(x,\xi)$, in a manner similar to the way we are accustomed in quantum mechanics to writing $\psi(x)=\bracket{x}{\psi}$. %
This notation puts elements of $G$ and $\hat{G}$ on a more equal footing, in the same way as Dirac's notation does for bras and kets.  (Note, however, there is no equivalent of the Riesz representation theorem: $G$ and $\hat{G}$ are not in general isomorphic, though they may be in some cases.)

While all of that sounds very fancy, for the (additive group of) real numbers with the standard topology, the characters are simply the exponentials 
\begin{equation}
(x,\xi) = e^{i\xi x}
\label{eq:Rchar}
\end{equation}
for $x\in\Re$, where $\xi$ is also some real number,%
\footnote{Note the slight (but common) abuse of notation in which the real number $\xi$ is \emph{also} used to label the character itself.
} %
and $\hat{\Re}$ has the same topology as $\Re$.  The dual group $\hat{\Re}$ is thus in this case isomorphic to $\Re$ via the mapping $e^{i\xi x} \mapsto \xi$. (Take note of the move here: the group $\hat{\Re}$ of \emph{characters} $\{e^{i\xi x}\}$ is naturally isomorphic to the collection $\Re$ of \emph{labels} $\{\xi\}$ on the characters. The perspective on which of these sets the dual group ``is" then becomes a matter of convenience.)

With that result in mind, we mention in passing that the dual group to an LCA group $G$ is the starting point for Fourier analysis on LCA groups, because the natural domain for Fourier transforms defined on $G$ is the dual group $\hat{G}$. In fact, the Fourier transform on an LCA group $G$ is defined to be
\begin{equation}
\tilde{f}(\xi) = \int_{G} (x,\xi)^* f(x)\,dx  ~,
\label{eq:LCAFourier}
\end{equation}
where $\xi\in\hat{G}$ and $dx$ is the Haar measure on $G$.  The Fourier transform so-defined satisfies most of the nice properties that we associate with the Fourier transform on the real numbers \cite{RudinFourier,Harmonic3e}. The case $G=\Re$ is an  instance where it is more natural to think of $\hat{G}=\hat{\Re}=\Re$ as the group of \emph{labels} on the characters, so that the domain of the Fourier transform $\tilde{f}$ is most naturally thought of as the real numbers. (This applies also to the other examples considered in this paper.)

The parallel between Eq.~(\ref{eq:qp}) and Eq.~(\ref{eq:Rchar}) is thus not a coincidence. Indeed, the Fourier transform can even be \emph{defined} as the representation of an LCA group that diagonalizes translations \cite{RudinFourier,Harmonic3e}.  Note that group characters are (by construction) eigenfunctions of translations, $\xi(x+a)=\xi(a)\xi(x)$. When the momentum operator exists, they are thus also momentum eigenstates. This is one perspective on the reason why the group dual to $G$ is naturally identified with the set of momentum eigenfunctions over $G$. We will have more to say about this identification below.

A perhaps more telling example of the Pontryagin dual is $G=\CircT$, the circle group.  The characters on $\CircT$ are the functions $(\theta,n)=e^{in\theta}$ from $\CircT$ into itself. Here $e^{i\theta}\in \CircT$ with $\theta\in[0,2\pi)$, $n$ is an integer, and the topology on $\CircThat$ in which these functions converge uniformly on compact subsets  of $\CircT$ is the discrete topology.  $\CircThat$ is thus isomorphic to the integers $\Int$. Conversely, it can be checked that $\hat{\Int}=\CircT$, as required by Pontryagin duality. This is an illustration of an important theorem about dual groups: \emph{the dual group of a compact group is discrete, and the dual group of a discrete group is compact.}

Finally, this leads us to the Bohr compactification \cite{RudinFourier,Harmonic3e,whatisbohr,Thiemann}. If $G$ is an LCA group, consider its dual group $\hat{G}$, but equipped instead with the discrete topology, $\Gdhat$. The \emph{Bohr compactification} $\Gb$ of $G$ is by definition the dual group to $\Gdhat$, $\Gb\equiv\dhatGd$.%
\footnote{The notation ``$bG$'' for the Bohr compactification is commonly encountered in the mathematics literature.
} %
If $G$ is a compact group, $\Gb=G$. For non-compact groups, $\Gb$ is called a \emph{compactification} of $G$ because $G$ is dense in the compact group $\Gb$ -- $\Gb$ is the group $G$ with (as it turns out) infinitely many 
``points at infinity'' added, in a sense to be described further below.%
\footnote{Recall that if the topology on $\hat{G}$ not been replaced by the discrete topology, we have $\dhatG = G$ instead. 
} %
In the case of the real numbers, since $\hat{\Re} = \Re$, \emph{the Bohr compactification of the reals is just the dual group of the discrete real line, $\Reb=\Redhat$}. 

There is an alternative characterization of the Bohr compactification 
that is more commonly found in the physics literature, one framed in the language of $C^*$ algebras. We use the example of the real numbers, but the same constructions apply for a general LCA group.

A $*$-algebra is an algebra $\mathcal{A}$ equipped with an involution operation $*$ (most commonly thought of as complex- or Hermitian-conjugation) that is an antilinear map on $\mathcal{A}$ satisfying $A^{**}=A$ and $(A_1A_2)^* = A_2^*A_1^*$. A $C^*$-algebra is a $*$-algebra that is complete in a norm $\Vert\cdot\Vert$ on $\mathcal{A}$ that satisfies $\Vert A^*\Vert = \Vert A\Vert$ and $\Vert A^*A\Vert = \Vert A\Vert^2$ \cite{moretti2e,blankexner2e,Wald94}.

The space of finite linear combinations of characters of $\Re$,
\begin{equation}
    f(x) = \sum_{n=1}^N c_n e^{ik_nx}  ~,
\label{eq:AP}
\end{equation}
where $c_n \in \mathbb{C}$, $k_n \in \mathbb{R}$, and $N$ is finite, constitute the space $AP(\Re)$ of \emph{almost periodic} functions on $\Re$ when closed under the sup (or uniform) norm, 
\begin{equation}
   \Vert f(x)\Vert_{\mathrm{sup}} = \sup_{x \in \Re}|f(x)| ~.  
\label{eq:supnorm}
\end{equation}
Such functions are called ``almost periodic'' because one can show that there is always an interval over which they come as close as desired  
to repeating themselves. As it happens, these are precisely the continuous functions on the Bohr compactification of the real numbers $\Re$ \cite{RudinFourier,Harmonic3e,whatisbohr}
as a consequence of the Peter-Weyl theorem \cite{bojo13,moretti2e}.

The space $AP(\Re)$ is a $C^*$ algebra \cite{whatisbohr,bohrpaper}. For an algebra $\mathcal{A}$, similarly to how characters are defined on a group, the (Gelfand) \emph{spectrum} of $\mathcal{A}$, denoted $\Delta(\mathcal{A})$, is defined to be the set of all homomorphisms $\phi: \mathcal{A} \to \Co$ that also respect the $*$-operation (``$*$-homomorphisms''). These maps are also called characters. Notably, however, the definition of the characters on an algebra does not include the requirement of continuity, only that the algebra's multiplicative and involutive structure is preserved. 

An alternative definition of the Bohr compactification of the real numbers is that it is the spectrum of the $C^*$ algebra of almost periodic functions, $\Reb \equiv \Delta(AP(\Re))$ \cite{bohrpaper,Thiemann}.  The connection of this definition with the previous one can be seen as follows. In the first definition, the Bohr compactification of a group $G$ is the dual of the dual $\hat{G}$ of $G$ equipped with the discrete topology, $\Gb=\dhatGd$, or in other words, the set of continuous homomorphisms from the space of characters of $G$ (with the discrete topology) into $\Circ$ . In the present case, the $C^*$ algebra $AP(\Re)$ includes linear combinations of the characters, and therefore it is natural that the definition of the spectrum of $AP(\Re)$ allows these linear combinations to be mapped to arbitrary complex numbers instead of just those with unit norm. The lack of a requirement that these homomorphisms on $AP(\Re)$ are continuous is appropriate because the topology on $\hat{\Re} = \Re$ in the first definition is the discrete topology, $\Reb=\Redhat$. 
Since every element of $\Red$ is by definition an open set, every function defined on $\mathbb{R}_d$ is continuous,
and therefore \emph{all} homomorphisms on $\hat{\Re}$ (the space of characters of $\Re$) are encompassed by the first definition of the Bohr compactification, and therefore all homomorphisms on $AP(\Re)$ should be included in the alternative definition. One can see that the definitions align.

With this outline of how the Bohr compactification $\Reb$ is constructed technically, we can say a few words about how to think about the resulting space. A ``compactification'' of a topological space is a new, compact topological space that contains the original space as a dense subset. The intuitive idea behind compactification is that we have made the space compact by adding ``points at infinity'' to give every infinite sequence in the space a point \emph{in the space} to converge to. 
Compactification of a non-compact space is not unique.  For example, there are (among others) two commonly encountered compactifications of the real line.  The one-point compactification adds a single point at infinity, and can be thought of as wrapping the real line into a circle, with the new ``point at infinity'' the point at which the two ends join.  The \emph{two}-point compactification adds a point at infinity on either end (``positive'' and ``negative'' infinity). This space is homeomorphic%
\footnote{A \emph{homeomorphism} is just a map 
that preserves all the topological properties of a space. The required example is the map from a coffee cup to a donut.
} %
to a closed interval in $\Re$.

The Bohr compactification of $\Re$ is much more difficult to picture, because it is constructed by adding \emph{infinitely} many ``points at infinity'' to $\Re$. These ``points at infinity'' represent all possible limits of sequences of almost-periodic functions, and there are infinitely many of them because there are infinitely many ways for functions to be almost-periodic. The structure of these points at infinity is complex, and indeed, $\Reb$ is not even path-connected, let alone simply-connected: there are points in $\Reb$ that cannot be joined by a continuous path \cite{Thiemann}.
%

In spite of this, because $\Re$ is dense in $\Reb$ and all of the complex structure is ``at infinity'', this complexity plays little direct role in the polymer representation once constructed as described in Sec.\ref{sec:pqm} using the Haar measure given by Eq.(\ref{eq:BohrInt}) on $\Reb$.

\subsection{Dual Group of a Discrete Space}
\label{app:appdiscretedual}

Given this understanding of the Bohr compactification of the real numbers, let us investigate the general situation concerning the quantum momentum space (that is, the set of translation eigenfunctions) dual to a discretized configuration space, and what this implies for compact configuration spaces such as the particle on a ring.

Consider first the case where the configuration space is the ring $\mathcal{C}=\CircT$ 
analyzed in the main text. We know from earlier in this appendix that the dual group is $\CircThat=\Int$, with characters (translation eigenfunctions) $e^{im\theta}$.
Due to the compactness of the configuration space, the momentum is quantized: we're really considering the \emph{angular} momentum, the momentum conjugate to the position coordinate on the circle.  We also know (by Pontryagin duality) that $\hat{\Int}=\CircT$. 

When we polymerize the ring, the configuration space is $\mathcal{C}_d=\CircT_d$,
 which makes the polymerized momentum space $\widehat{\CircT_d}$.  What is this space?  Notice that
\begin{equation}
  \widehat{\CircT_d} = \widehat{(\hat{\Int})_d} \equiv \Int_{\mathrm{B}} ~.
\end{equation}
\emph{The polymer momentum space dual to the discretized ring is the Bohr compactification of the integers}, consistent with the general argument below. Just as with the Bohr compactification of the real numbers, $\IntB$ has a complex structure at infinity \cite{RudinFourier}, but fortunately that complexity is again mostly hidden from the corresponding polymer quantum mechanics.

The general situation is as follows. Let $C$ be an LCA group and and $D$ its Pontryagin dual: $\Cop=D$ and $\Dop=C$. In that case, notice
\begin{alignat*}{2}
C_B &\equiv \widehat{\Cop_d} &&= \widehat{D_d} \\
D_B &\equiv \widehat{\Dop_d} &&= \widehat{C_d} ~.
\end{alignat*}
Since when we polymerize the set of eigenfunctions of translations is the group dual to the configuration space with the discrete topology, if our configuration space is $C$ this tells us that in general 
\emph{the set of polymer momentum eigenfunctions ($\widehat{C_d}$) is the Bohr compactification ($D_B$) of the set of Schr\"{o}dinger momentum eigenfunctions ($D$).}  This is consistent both with ``standard" polymer quantum mechanics (where the Schr\"{o}dinger momentum space is $\Re$ and the polymer momentum space is $\Reb$), and with the case $\mathcal{C}=\CircT$ discussed just above (where the Schr\"{o}dinger momentum space is $\Int$ and the polymer momentum space is $\Int_B$). 

Suppose now that the configuration space $C$ is compact. Then $\Cop=D=D_d$ is discrete, so that
\begin{equation}
   C_B = \widehat{D_d} = \Dop = C ~, 
\end{equation}
perhaps not a surprising result.  Additionally, as shown in Sec. \ref{sec:posgraphc}, graphs satisfying the AFW conditions on compact configuration spaces are \emph{finite}, a fact heavily exploited in the analysis of the particle on a ring.

\subsection{Quantum Momentum Space}
\label{app:qmomspace}

To close this appendix, let us review the perspective we have taken on the role of the dual group to a configuration space as the space of momentum eigenfunctions over that configuration space. The physics embodied in the identification of the Pontryagin dual of a discrete configuration space with the translation (momentum) eigenfunctions on it may be summarized as follows:   
\begin{enumerate}[wide, leftmargin=*, labelindent=0pt, nosep, label=(\roman*)]
\item Group characters are (by construction) eigenfunctions of translations, $\xi(x+a)=\xi(a)\xi(x)$, and are thus eigenstates of the Weyl operator $\Vop(\beta)$. When the momentum operator exists, they are thus also momentum eigenstates. As previously noted, this is why the Fourier transform on an LCA group is naturally defined in terms of the group characters, Eq.~(\ref{eq:LCAFourier}).  Thus, the group dual to $G$ is the set of translation (momentum) eigenfunctions.
\item The domain of the Fourier transform Eq.(\ref{eq:LCAFourier}) is the space of characters, but since these are naturally isomorphic to their labels, the labels effectively become the domain of the Fourier transform.  Thus, for example, we normally consider the domain of the Fourier transform on $\Re$ to be $p\in\Re$, not the characters $e^{ipx}$ per se.  More generally, functions on the characters can be naturally regarded as functions on the labels of the characters, whether $\Reb$ or $\IntB$. Therefore we somewhat loosely use this isomorphism to switch back and forth between thinking of the dual group as the set of characters (when we are interested in the translation eigenfunctions) or as the set of labels on those characters (when considering functions of those characters). 
(This is really just the familiar distinction between the eigenstates $\ket{p}$ and their position-space representatives $\varphi_p(x)\equiv\bracket{x}{p}$.)
\item The space of almost-periodic functions $AP(G)$ on $G$ is the space of \emph{linear combinations} of $G$'s characters, the eigenfunctions of translations on $G$, and thus is naturally identified with the set of momentum-space states on the configuration space $G$.
\item We have learned that $AP(G)$ is the space of continuous functions on the Bohr compactification $\Gb$. In the case that $G$ is discrete, we have also learned that the polymer momentum (really, translation) eigenfunctions are identified with the Bohr compactification $\Gb$ of $G$.  We have just argued that the space of almost-periodic functions $AP(G)$ on $G$ is the space of quantum momentum-space states on the configuration space $G$. 
More properly, since $AP(G)$ comprises the continuous functions on $\Gb$, it is dense in the square-integrable functions $L^2(\Gb,d\mu_H)$ on $\Gb$, the Hilbert space we normally associate with momentum-space states.  
\end{enumerate}
For a general discussion of canonically conjugate pairs in Hilbert spaces of various types see \cite{vanenk25}.

\section{Polymer particle in a box}
\label{app:pqmbox}

The solution for the polymer particle on a ring presented in the body of this paper is new. A related system with a compact configuration space which \textit{has} been previously analyzed in the context of polymer quantum mechanics is the particle in a box \cite{polymerbox,polymerStatThermo}.%
\footnote{The discrete particle in a box has also been explored in other contexts. For example, in \cite{dQMBox} a discrete version of the infinite well is investigated with a mathematical setup similar to that of polymer quantum mechanics, and \cite{solidStateBox} solves the same system in the context of solid-state physics. See also \cite{vanenk25,vourdas04}.
} %
Instead of a particle which is bound to the circle $\mathbb{T}$, the system consists of a particle confined to a closed interval $[0,L]\subset \Re$.

This system does not fit tidily into the framework we have developed in this paper because the underlying configuration space (a closed interval) is not a group under translations, unlike the ring. There are various strategies one might try to work around this, but for present purposes we choose to take the naive configuration space at face value. We will not, however, try to define a quantum momentum space or a corresponding Weyl algebra. (These issues are not peculiar to polymer quantization, and are related to the well-known lack of an essentially self-adjoint momentum operator for a particle in a box \cite{vanenk25,Bonneau:1999zq}). One may nonetheless formulate a well-defined quantum theory of position-space wave functions. In the non-polymer case, this is of course simply $L^2([0,L],dx)$. As with the particle on a ring, an AFW graph on a closed interval will once again be finite, and methods similar that that employed for the particle on a ring may be employed to solve the polymer Schr\"odinger equation on it.  

Previous treatments \cite{polymerbox, polymerStatThermo} have found the energy eigenvalues 
\begin{equation}\label{eq:litboxeigenvalues}
    E_n = \frac{\hbar^2}{m\mu_0^2}\left( 1- \cos\left(\frac{n\pi}{N}\right)  \right),
\end{equation}
where the quantum number $n$ can take on values $1,2,\dots,N-1$, while the corresponding eigenstates are
\begin{equation}\label{eq:litboxeigenstates}
    \ket{\phi_n} = \sqrt{\frac{2}{N}}\,\sum_{k=0}^N \sin\left(\frac{n\pi k}{N}\right)\ket{x_k} ~.
\end{equation}
In \cite{polymerbox}, these solutions were found using the solution for a free polymer particle and restricting to a finite box by imposing the boundary condition for an infinite potential well,  $\psi(x_0)=\psi(x_N)=0$, while the solution in \cite{polymerStatThermo}, reviewed below, employs a recurrence relation approach very similar to what we used for the particle on the ring. As in the ring case, the finite-dimensionality of the graph Hilbert space automatically introduces a UV cutoff on the polymerized spectrum.

In this appendix, our goal is to apply the same methods used to analyze the polymer particle on a ring to analyze the particle in an infinite well, and show that they lead to the same results as previous work. Additionally, we will demonstrate how in the continuum limit the polymer eigenstates converge to the corresponding eigenstates of the Schr\"odinger-quantized infinite well.

\subsection{Solution by recurrence}
\label{sec:boxrecur}

Here we analyze the particle in a box using a recurrence method equivalent to the solution presented in \cite{polymerStatThermo}; we include it here to emphasize the similarity to the recurrence solution for the particle on the ring given in Sec.~\ref{sec:ringrecur}. We begin with a polymer particle on a line described by states in the graph Hilbert space $\Hgraphmuo$ defined in Sec.~\ref{sec:posgraph}.
The graph lattice spacing is chosen to divide the width $L$ of the well so that the number of lattice points $N$ inside the well is given by $L = (N+1) \mu_0$. The infinite well is then defined in the usual way by the potential
\begin{equation}
    \mathcal{V}(x_k) = 
    \begin{cases}
        0 & \text{if } 1 \leq k \leq N\\
        \infty & \text{otherwise}
    \end{cases}  ~,
\end{equation}
so that the walls of the box are located at the lattice sites with $k=0$ and $k=N+1$. 
(Note that with this choice, we are allowing one additional lattice graph site inside the box compared to \cite{polymerbox, polymerStatThermo}.)

The Hamiltonian and energy eigenvalue equation may be written
\begin{subequations}
\begin{align}
    \hat{H} &= \frac{\hbar^2}{2M\mu_0^2} [2\Id - \Vop(\mu_0) - \Vop(-\mu_0)] + \mathcal{V} \label{eq:boxham} \\
    \lambda\ket{\phi} &= [2\Id - \Vop(\mu_0) - \Vop(-\mu_0)]\ket{\phi} + \mathcal{V}\ket{\phi}  ~, \label{eq:BoxTISE}
\end{align}
\end{subequations}
where, as in section \ref{sec:ringrecur}, we have defined $\alpha = \frac{\hbar^2}{2M\mu_0^2}$, $\lambda = \frac{E}{\alpha}$, and set $\mathcal{V}/\alpha \rightarrow \mathcal{V}$ in the time-independent Schr\"odinger equation. With $\ket{\phi} = \sum_k c_k \ket{x_k}$, this gives 
\begin{equation}
    \lambda c_k = 2 c_k - c_{k-1} - c_{k+1} + \mathcal{V}(x_k) c_k ~.
    \label{eq:boxrecurr}
\end{equation}

Since the potential is piecewise-defined, we treat the cases where $\mathcal{V} = 0$ and $\mathcal{V} = \infty$ separately. For $k<1$ and $k>N$, $\mathcal{V}(x_k) = \infty$, and the only solution to Eq.~(\ref{eq:boxrecurr}) is $c_k = 0$. For $1 \leq k \leq N$, $\mathcal{V}(x_k) = 0$, giving
\begin{equation}
    c_{k+1} = (2-\lambda)c_{k} - c_{k-1}  ~.
\end{equation}
Since the particle is free inside the well, this is the same recurrence relation we found for the ring in Sec.~\ref{sec:ringrecur}, but now the boundary conditions are $c_0 = c_{N+1} = 0$. Just like on the ring, there are 3 cases to consider. 

In the first case, if $\lambda < 0$ or $\lambda > 4$, the solutions will have the form 
\begin{equation}
    c_k = At_+{}^k + Bt_-{}^k ~,
\end{equation}
where $t_\pm$ are the roots of the characteristic polynomial $t^2 + (\lambda - 2)t + 1 = 0$. This can only satisfy $c_0 = 0$ if $A = -B$. Then to satisfy $c_{N+1} = 0$, we must have $|t_+| = |t_-|$, which is not possible in this range of $\lambda$ values. Therefore, the only solution for $\lambda < 0$ or $\lambda > 4$ is the trivial solution with $A = B = 0$. 

Next, if $\lambda = 0$ or $\lambda = 4$, the solutions will look like 
\begin{equation}
    c_k = At_*{}^k + Bkt_*{}^k ~.
\end{equation}
Requiring $c_0 = 0$ gives $A = 0$. Then, regardless of the value of $t_*$, we can only have $c_{N+1} = 0$ if $B = 0$, which is again the trivial solution. 

Finally, if $0 < \lambda < 4$, we have solutions of the form
\begin{equation}
    c_k = r^k\left(A\cos(k\tau) + B\sin(k\tau)\right) ~.
    \label{eq:boxunsimplified}
\end{equation}
As we found in Sec.~\ref{sec:ringrecur}, we have $r = 1$ and $\cos(\tau) = \frac{2 - \lambda}{2}$. Setting $c_0 = 0$ then requires that $A = 0$, and $c_{N+1} = 0$ yields the condition
\begin{equation}
    (N+1)\tau = \pi n ~, \ \ n \in \Int ~.
\end{equation}
Inserting this information into Eq.~(\ref{eq:boxunsimplified}) results in the energy eigenstates
\begin{align}
    \ket{\phi_n} &= A \sum_{k = 1}^N \sin\left(\frac{\pi n}{N+1}k\right) \ket{x_k}
\label{eq:boxefns}    
\end{align}
and energy eigenvalues $E = \alpha \lambda$
\begin{align}
    E_n &= \alpha(2 - 2\cos(\tau)) = \frac{\hbar^2}{m\mu_0^2}\left(1 - \cos\left(\frac{\pi  n}{N+1}\right)\right) ~.
\label{eq:boxevals}    
\end{align}
Normalizing $\ket{\phi_n}$,  we find $A=\sqrt{\frac{2}{N+1}}$. Since we have chosen a graph with one additional lattice site compared to \cite{polymerbox, polymerStatThermo}, these are equivalent to Eqs.~(\ref{eq:litboxeigenvalues}) and (\ref{eq:litboxeigenstates}).

\subsection{Operator solution}
\label{sec:boxop}

We now consider an alternative analysis of the polymerized infinite well, a direct solution of the time-independent Schr\"odinger equation considered as an operator (matrix) equation. In contrast to the polymer particle on a ring, however, this system is not translation invariant, and we must take $\Vop(\mu_0) \ket{x_N} = 0$ and $\Vop(-\mu_0)\ket{x_1} = 0$ in place of $\Vop(\mu_0)^N\ket{x_k}= \ket{x_k}$, in which case $\Vop(\mu_0)$ and $\Vop(-\mu_0)$ no longer commute with each other, or with the Hamiltonian. Instead, we will find the eigenvalues and eigenvectors of the full Hamiltonian directly.
(To our knowledge, this approach to solving the discrete particle in a box is new.)

Representing the position eigenstates $\{\ket{x_1},\ldots,\ket{x_N}$ corresponding to lattice sites within the well by the standard basis, 
$\ket{x_1} = \begin{bmatrix}
        1 & 0 & \dots & 0
\end{bmatrix}^T$ (\emph{etc}), the Hamiltonian, Eq.~(\ref{eq:boxham}), may be represented by the $N \times N$ matrix
\begin{equation}
\label{eq:matrixhamiltonian}
    H \doteq \frac{\hbar^2}{2m\mu_0^2} 
    \begin{bmatrix}
      2 & -1 & 0 & \dots & \\
      -1 & 2 & -1 & 0 & \dots \\
     \ddots & \ddots & \ddots & \ddots & \ddots \\
     \dots & 0 & -1 & 2 & -1 \\
      & \dots & 0& -1 & 2 
  \end{bmatrix}  ~.
\end{equation}
This takes the form of a tridiagonal Toeplitz matrix (TTM), a well-studied class of matrices with known eigenvalues and eigenvectors. TTM matrices have the general form
\begin{equation}
  T = \begin{bmatrix}
      a & b & 0 & \dots & \\
      c & a & b & 0 & \dots \\
     \ddots & \ddots & \ddots & \ddots & \ddots \\
     \dots & 0 & c & a & b \\
      & \dots & 0& c & a 
  \end{bmatrix} ~.  
\end{equation}
The eigenvalues of a TTMs are 
\begin{equation}
    \lambda_n = a - 2 \sqrt{bc}\cos\left(\frac{n\pi}{N+1}  \right) ~,
\end{equation}
a result which stems from the relationship between the eigenvalues of these matrices and the zeroes of Chebyshev polynomials \cite{ttoeplitz1}. Meanwhile, the components of the corresponding eigenvectors $\mathbf{v}_n = \begin{bmatrix}
    x_{n,1} & \dots & x_{n,k} & \dots x_{n,N}
\end{bmatrix}^T$ of a TTM are given by \cite{ttoeplitz2}
\begin{equation}
    x_{n,k} = A\sin\left(\frac{nk\pi}{N+1} \right) ~.
\end{equation}
In our notation, the eigenfunctions of the box Hamiltonian can be therefore be expressed as
\begin{equation}
    \ket{\phi_n} = A \sum_{k=1}^N \sin\left( \frac{nk\pi}{N+1} \right)\ket{x_k} ~,
\end{equation}
where $A = \sqrt{\frac{2}{N+1}}$, with energy eigenvalues
\begin{equation}
    E_n = \frac{\hbar^2}{m\mu_0^2}\left(1-\cos\left({\frac{\pi  n}{N+1}}\right)\right) ~,
\end{equation}
in complete agreement with Eqs.~(\ref{eq:boxefns}) and (\ref{eq:boxevals}).

This approach, while in agreement with the results of \cite{polymerbox, polymerStatThermo}, enabled us to find the spectrum of the polymer particle in a box dealing exclusively with the graph Hilbert space $\Hgraphmuo$ defined on the compact interval $\left[ 0, L \right]$.

\subsection{Continuum limit}
\label{sec:boxcont}

Finally, we may show that in a manner entirely parallel to the polymer particle on a ring that in the limit that $\mu_0/L \rightarrow 0$, the polymer eigenvalues and eigenstates approach that of the Schr\"odinger theory.  We have chosen a graph $\gamma_{\mu_0}$ with $N$ lattice sites inside the well, so that $L=\mu_0(N+1)$, and the walls of the well are at $x=0$ and $x=(N+1)\mu_0$.

In the limit that $\mu_0/L \rightarrow 0$, $N\rightarrow\infty$, giving
\begin{subequations}
\begin{align}
\lim_{N \to \infty}E_n 
  &= \lim_{N \to \infty} \frac{\hbar^2}{M\mu_0^2}\left(1-\cos\left({\frac{n\pi}{N+1}}\right)\right) \\
  & = \lim_{N \to \infty} \frac{\hbar^2(N+1)^2}{mL^2}\left(1-\cos\left({\frac{n\pi}{N+1}}\right)\right)   \\
  &= \lim_{N \to \infty} \frac{\hbar^2 (N+1)^2}{mL^2} \left( \frac{\pi^2 n^2}{2!(N+1)^2} - \frac{\pi^4n^4}{4!(N+1)^4}+\dots    \right) \\
  &= \frac{(n\pi\hbar)^2}{2ML^2} ~,
\end{align}    
\end{subequations}
the expected result. As for the polymer particle on a ring, the UV cutoff lifts as the lattice graph approaches a continuum.

Similarly, defining the continuum counterparts to the polymerized wave functions for Eq.~(\ref{eq:boxefns}) in the same way as Eq.~(\ref{eq:contQWF}), 
\begin{subequations}
\begin{align}
\phi^c_n(x) 
  &= \sqrt{\frac{2}{\mu_0(N+1)}}\sin\left( \frac{\pi n x}{(N+1)\mu_0} \right)\, \delta^c_{x,x_k} \\
  &=  \sqrt{\frac{2}{L}}\sin\left( \frac{n\pi x}{L} \right) \delta^c_{x,x_k} ~.
\end{align}    
\end{subequations}
Once again we see that polymerized wave functions converge to their continuum Schr\"odinger counterparts in the limit as the lattice spacing $\mu_0/L\rightarrow 0$. (The proof that this convergence is uniform is identical to that for the ring.)

\begin{acknowledgments}
The authors wish to thank Parampreet Singh and Meysam Motaharfar for enlightening discussions and Elaine Cozzi for a helpful conversation. 
M.S.~and B.S.~would like to offer special thanks to Ethan Minot and Dublin Nichols for much thoughtful feedback on the Oregon State University Honors theses which serve as the foundation of this paper. M.S.~is supported by NSF grant PHY-2409543.

\textbf{Author contributions:} 
Siebersma and Seibert formulated the mathematical model for polymer quantum mechanics on a ring, carried out the associated calculations, and wrote the first draft of the paper.  Shuman contributed the solution by recurrence.  Craig developed the theory of the polymer-quantized momentum space, contributed to the mathematical formulation of polymer quantum mechanics on compact configuration spaces, wrote the technical part of the paper introducing PQM and its physical and mathematical background based on the earlier draft by Seibert and Siebersma, and supervised the undergraduate thesis projects that served as the foundation for the work.  All authors contributed to final editing and improvements.

\textbf{Conflicts of interest:} All authors declare no conflicts of interest. 

\end{acknowledgments}


\ifthenelse{\arxiv=1}{%
\bibliography{PQMCompact}
}{%
\bibliography{PQMCompact}

@book{abbott2e,
	author = {Abbott, Stephen},
	edition = {2nd},
	place = {New York},
	publisher = {Springer},
	title = {Understanding Analysis},
	year = {2015}}

@article{abl03,
	archiveprefix = {arXiv},
	author = {Ashtekar, Abhay and Bojowald, Martin and Lewandowski, Jerzy},
	eprint = {gr-qc/0304074},
	journal = {Adv. Theor. Math. Phys.},
	pages = {233-268},
	primaryclass = {gr-qc},
	slaccitation = {%%CITATION = GR-QC/0304074;%%},
	title = {Mathematical structure of loop quantum cosmology},
	volume = {7},
	year = {2003}}

@article{afwPQM,
	author = {Ashtekar, Abhay and Fairhurst, Stephen and Willis, Joshua L},
	doi = {10.1088/0264-9381/20/6/302},
	journal = {Class. Quantum Grav.},
	number = {6},
	pages = {1031--1061},
	title = {Quantum gravity, shadow states and quantum mechanics},
	volume = {20},
	year = {2003}}

@article{aps,
  title = {Quantum nature of the big bang: An analytical and numerical investigation},
  author = {Ashtekar, Abhay and Pawlowski, Tomasz and Singh, Parampreet},
  journal = {Phys. Rev. D},
  volume = {73},
  issue = {12},
  pages = {124038},
  numpages = {33},
  year = {2006},
  month = {Jun},
  publisher = {American Physical Society},
  doi = {10.1103/PhysRevD.73.124038},
  url = {https://link.aps.org/doi/10.1103/PhysRevD.73.124038}
}

@article{ashsingh11,
	archiveprefix = {arXiv},
	author = {Ashtekar, Abhay and Singh, Parampreet},
	doi = {10.1088/0264-9381/28/21/213001},
	eprint = {1108.0893},
	journal = {Class. Quantum Grav.},
	pages = {213001},
	primaryclass = {gr-qc},
	slaccitation = {%%CITATION = ARXIV:1108.0893;%%},
	title = {Loop quantum cosmology: a status report},
	volume = {28},
	year = {2011}}

@book{basictopology,
	author = {Armstrong, Mark A.},
	place = {New York},
	publisher = {Springer},
	title = {Basic Topology},
	year = {1983}}

@book{blankexner2e,
	author = {Blank, Ji\v{r}i and Exner, Pavel and Havli\v{c}ek, Miloslav},
	edition = {2nd},
	place = {New York},
	publisher = {Springer},
	title = {{H}ilbert Space Operators in Quantum Physics},
	year = {2008}}

@article{bohrpaper,
	author = "Velhinho, J. M.",
    title = "{The Quantum configuration space of loop quantum cosmology}",
    eprint = "0704.2397",
    archivePrefix = "arXiv",
    primaryClass = "gr-qc",
    doi = "10.1088/0264-9381/24/14/013",
    journal = "Class. Quant. Grav.",
    volume = "24",
    pages = "3745--3758",
    year = "2007"
}

@book{bojo10,
	author = {Bojowald, Martin},
	booktitle = {Canonical gravity and applications: cosmology, black holes, and quantum gravity},
	publisher = {Cambridge University Press},
	title = {Canonical Gravity and Applications: Cosmology, Black Holes, and Quantum Gravity},
	year = {2010}}

@article{bojo13,
	author = {Bojowald, Martin},
	journal = {SIGMA},
	pages = {082},
	title = {Mathematical structure of loop quantum cosmology: homogeneous models},
	volume = {9},
	year = {2013}}

@article{chung2023quantum,
	author = "Chung, Won Sang and Haouam, Ilyas and Hassanabadi, Hassan",
    title = "{Quantum mechanics on a circle with a finite number of {\ensuremath{\alpha}}-uniformly distributed points}",
    eprint = "2304.03176",
    archivePrefix = "arXiv",
    primaryClass = "quant-ph",
    doi = "10.1016/j.physleta.2023.129098",
    journal = "Phys. Lett. A",
    volume = "485",
    pages = "129098",
    year = "2023"
}

@article{continuumlimit,
  author = "Corichi, Alejandro and Vukasinac, Tatjana and Zapata, Jose A.",
    title = "{Polymer Quantum Mechanics and its Continuum Limit}",
    eprint = "0704.0007",
    archivePrefix = "arXiv",
    primaryClass = "gr-qc",
    reportNumber = "IGPG-07-03-2",
    doi = "10.1103/PhysRevD.76.044016",
    journal = "Phys. Rev. D",
    volume = "76",
    pages = "044016",
    year = "2007"
}

@article{CS10c,
  author = "Craig, David A. and Singh, Parampreet",
    title = "{Consistent Probabilities in Wheeler-DeWitt Quantum Cosmology}",
    eprint = "1006.3837",
    archivePrefix = "arXiv",
    primaryClass = "gr-qc",
    reportNumber = "PI-QG-191, PI-QG-191",
    doi = "10.1103/PhysRevD.82.123526",
    journal = "Phys. Rev. D",
    volume = "82",
    pages = "123526",
    year = "2010"
}

@article{dac13a,
	archiveprefix = {arXiv},
	author = {Craig, David A.},
	doi = {http://dx.doi.org/10.1088/0264-9381/30/3/035010},
	eprint = {1207.5601},
	journal = {Class. Quantum Grav.},
	pages = {035010},
	preprint = {pi-qg-250},
	primaryclass = {gr-qc},
	slaccitation = {%%CITATION = ARXIV:1207.5601;%%},
	title = {Dynamical eigenfunctions and critical density in loop quantum cosmology},
	volume = {30},
	year = {2013}}

@book{differenceEqns,
	address = {New York},
	author = {Elaydi, Saber N.},
	publisher = {Springer},
	title = {{An Introduction to Difference Equations}},
	year = {1996}}

@book{DiracQM4e,
	author = {Dirac, {P.A.M.}},
	edition = {4th revised},
	publisher = {Oxford University Press},
	title = {Principles of Quantum Mechanics},
	year = {1958}}

@misc{dQMBox,
      title={Free particle trapped in an infinite quantum well examined through the discrete calculus model}, 
      author={Dušan Popov},
      year={2023},
      eprint={2303.08212},
      archivePrefix={arXiv},
      primaryClass={quant-ph},
      url={https://arxiv.org/abs/2303.08212}, 
}

@article{gravitationalwave,
	author = {Stargen, D. Jaffino and Shankaranarayanan, S. and Das, Saurya},
	doi = {10.1103/physrevd.100.086007},
	journal = {Phys. Rev. D},
	number = {8},
	title = {Polymer quantization and advanced gravitational wave detector},
	volume = {100},
	year = {2019}}

@book{haarjoy,
	author = {Diestel, Joe and Spalsbury, Angela},
	publisher = {American Mathematical Society},
	series = {Graduate Studies in Mathematics},
	title = {The Joys of {H}aar Measure},
	volume = {150},
	year = {2014}}

@book{hallLie2e,
	author = {Hall, Brian C.},
	edition = {2nd},
	publisher = {Springer},
	title = {{L}ie Groups, {L}ie Algebras, and Representations},
	year = {2015}}

@book{hallQM,
	author = {Hall, Brian C.},
	place = {New York},
	publisher = {Springer},
	title = {Quantum Theory for Mathematicians},
	year = {2013}}

@book{Harmonic3e,
	author = {Katznelson, Yitzhak},
	edition = {3rd},
	publisher = {Cambridge University Press},
	title = {An Introduction to Harmonic Analysis},
	year = {2004}}

@book{HellSah21,
	author = {Helliwell, T. M. and Sahakian, V. V.},
	publisher = {Cambridge University Press},
	title = {Modern Classical Mechanics},
	year = {2021}}

@book{HewittRossI,
	author = {Hewitt, Edwin and Ross, Kenneth A.},
	edition = {2nd},
	place = {New York},
	publisher = {Springer-Verlag},
	title = {Abstract Harmonic Analysis},
	volume = {I},
	year = {1979}}

@book{kiefer25,
	author = {Kiefer, Claus},
	edition = {4th},
	publisher = {Oxford University Press},
	title = {Quantum Gravity},
	year = {2025}}

@article{liv-rev,
	author = {Bojowald, Martin},
	journal = {Living Rev. Rel.},
	pages = {4},
	slaccitation = {%%CITATION = 00222,11,4;%%},
	title = {{Loop quantum cosmology}},
	volume = {11},
	year = {2008}}

@book{McIntyre,
	author = {McIntyre, David H.},
	place = {Cambridge},
	publisher = {Cambridge University Press},
	title = {Quantum Mechanics: A Paradigms Approach},
	year = {2023}}

@book{moretti2e,
	author = {Moretti, Valter},
	edition = {2nd},
	place = {Milano},
	publisher = {Springer},
	title = {Spectral Theory and Quantum Mechanics},
	year = {2017}}

@article{physhilbspace,
	author = {Corichi, Alejandro and Vuka{\v s}inac, Tatjana and Zapata, Jos{\'e} A},
	doi = {10.1088/0264-9381/24/6/008},
	journal = {Class. Quantum Grav.},
	number = {6},
	pages = {1495--1511},
	title = {Hamiltonian and physical {Hilbert} space in polymer Quantum Mechanics},
	volume = {24},
	year = {2007}}

@article{polymerbox,
	author = {Flores-Gonz{\'{a}}lez, Ernesto and Morales-T{\'{e}}cotl, Hugo A. and Reyes, Juan D.},
	doi = {10.1016/j.aop.2013.05.005},
	journal = {Ann. Phys.},
	month = {sep},
	pages = {394--412},
	publisher = {Elsevier {BV}},
	title = {Propagators in polymer quantum mechanics},
	volume = {336},
	year = 2013}

@article{polymerStatThermo,
	archiveprefix = {arXiv},
	author = {Chac{\'{o}}n-Acosta, Guillermo and Manrique, Elisa and Dagdug, Leonardo and Morales-T{\'{e}}cotl, Hugo A.},
	doi = {10.3842/SIGMA.2011.110},
	eprint = {1109.0803},
	journal = {Symmetry, Integrability, and Geometry: Methods and Applications},
	pages = {110},
	primaryclass = {gr-qc},
	title = {Statistical Thermodynamics of Polymer Quantum Systems},
	volume = 7,
	year = 2011
}

@article{propagator,
    author = "Hossain, Golam Mortuza and Husain, Viqar and Seahra, Sanjeev S.",
    title = "{The Propagator in polymer quantum field theory}",
    eprint = "1007.5500",
    archivePrefix = "arXiv",
    primaryClass = "gr-qc",
    doi = "10.1103/PhysRevD.82.124032",
    journal = "Phys. Rev. D",
    volume = "82",
    pages = "124032",
    year = "2010"
}

@book{prugovecki2e,
	author = {Prugove\v{c}ki, Eduard},
	edition = {2nd},
	place = {New York},
	publisher = {Academic Press},
	title = {Quantum Mechanics in {H}ilbert Space},
	year = {1981}}

@book{RSI,
	author = {Reed, Michael and Simon, Barry},
	edition = {Revised and enlarged},
	place = {Orlando},
	publisher = {Academic Press},
	title = {Methods of Modern Mathematical Physics {I}: Functional Analysis},
	volume = {I},
	year = {1980}}

@book{RudinFourier,
	author = {Rudin, Walter},
	place = {New York u.a.},
	publisher = {Interscience},
	title = {Fourier Analysis on Groups},
	year = {1970}}

@book{Sakurai3e,
    author = "Sakurai, Jun John and Napolitano, Jim",
    title = "{Modern Quantum Mechanics}",
    edition = "3",
    doi = "10.1017/9781108587280",
    isbn = "978-0-8053-8291-4, 978-1-108-52742-2, 978-1-108-58728-0",
    publisher = "Cambridge University Press",
    series = "Quantum physics, quantum information and quantum computation",
    month = "10",
    year = "2020"
}

@book{schutzgeo,
	author = {Schutz, Bernard F.},
	publisher = {Cambridge University Press},
	title = {Geometrical Methods of Mathematical Physics},
	year = {1980}}

@article{solidStateBox,
	author = {Timothy B Boykin and Gerhard Klimack},
	doi = {10.1088/0143-0807/25/4/006},
	journal = {Euro. J. Phys.},
	number = 4,
	title = {The discretized Schr{\"{o}}dinger equation and simple models for semiconductor quantum wells},
	volume = 25,
	year = 2004}

@book{StoneGoldbart,
	author = {Stone, Michael and Goldbart, Paul},
	publisher = {Cambridge University Press},
	title = {Mathematics for Physics, A Guided Tour for Graduate Students},
	year = {2009}}

@book{szekeres,
	author = {Szekeres, Peter},
	publisher = {Cambridge University Press, Cambridge},
	title = {A Course in Modern Mathematical Physics: Groups, {H}ilbert Space and Differential Geometry},
	year = {2004}}

@book{Thiemann,
	author = {Thiemann, Thomas},
	place = {Cambridge},
	publisher = {Cambridge University Press},
	title = {Modern Canonical Quantum General Relativity},
	year = {2007}}

@article{ttoeplitz1,
	author = {Gover, M.J.C.},
	doi = {10.1016/0024-3795(94)90481-2},
	journal = {Linear Algebra and its Applications},
	pages = {63--78},
	title = {The eigenproblem of a tridiagonal {2-Toeplitz} matrix},
	volume = {197--198},
	year = {1994}}

@article{ttoeplitz2,
	author = {Noschese, Silvia and Pasquini, Lionello and Reichel, Lothar},
	doi = {10.1002/nla.1811},
	journal = {Numerical Linear Algebra with Applications},
	number = {2},
	pages = {302--326},
	title = {Tridiagonal {Toeplitz} matrices: Properties and novel applications},
	volume = {20},
	year = {2012}}

@article{velhinho04,
	author = "Velhinho, J. M.",
    title = "{Comments on the kinematical structure of loop quantum cosmology}",
    eprint = "gr-qc/0406008",
    archivePrefix = "arXiv",
    doi = "10.1088/0264-9381/21/15/L01",
    journal = "Class. Quant. Grav.",
    volume = "21",
    pages = "L109",
    year = "2004"
}

@book{Wald94,
	author = {Wald, Robert M.},
	publisher = {University of Chicago Press},
	title = {Quantum Field Theory in Curved Spacetime and Black Hole Thermodynamics},
	year = {1994}}

@misc{whatisbohr,
	author = {Staten, Corey},
	howpublished = {\url{https://math.osu.edu/sites/math.osu.edu/files/whatis_bohr.pdf}},
	note = {Accessed: 2025-08-16},
	title = {{Bohr} Compactifications},
	year = {2009}}

@techreport{ZachosCBH,
	author = {Zachos, Cosmas},
	institution = {Argonne National Laboratory},
	title = {Crib notes on {Campbell-Baker-Hausdorff} expansions},
	url = {https://www.researchgate.net/publication/236173438_Crib_Notes_on_Campbell-Baker-Hausdorff_expansions},
	year = {1999}}

@article{Bonneau:1999zq,
    author = "Bonneau, Guy and Faraut, Jacques and Valent, Galliano",
    title = "{Selfadjoint extensions of operators and the teaching of quantum mechanics}",
    eprint = "quant-ph/0103153",
    archivePrefix = "arXiv",
    reportNumber = "PAR-LPTHE-99-33, PAR/LPTHE/99-33",
    doi = "10.1119/1.1328351",
    journal = "Am. J. Phys.",
    volume = "69",
    pages = "322",
    year = "2001"
}

@article{Ashtekar:2007em,
    author = "Ashtekar, Abhay and Corichi, Alejandro and Singh, Parampreet",
    title = "{Robustness of key features of loop quantum cosmology}",
    eprint = "0710.3565",
    archivePrefix = "arXiv",
    primaryClass = "gr-qc",
    reportNumber = "IGC-07-10-01, PI-QG-61",
    doi = "10.1103/PhysRevD.77.024046",
    journal = "Phys. Rev. D",
    volume = "77",
    pages = "024046",
    year = "2008"
}

@article{vanenk25,
	archiveprefix = {arXiv},
	author = {{van Enk}, S.J and Steck, Daniel A.},
	doi = {https://doi.org/10.1103/fzdt-p75h},
	eprint = {2502.08494},
	journal = {Phys. Rev. Research},
	pages = {033117},
	primaryclass = {quant-ph},
	title = {All {Hilbert} spaces are the same: {Consequences} for generalized coordinates and momenta},
	url = {https://arxiv.org/abs/2502.08494},
	volume = {8},
	year = {2026}
}

@article{Wilson:1974sk,
    author = "Wilson, Kenneth G.",
    editor = "Taylor, J. C.",
    title = "{Confinement of Quarks}",
    reportNumber = "CLNS-262",
    doi = "10.1103/PhysRevD.10.2445",
    journal = "Phys. Rev. D",
    volume = "10",
    pages = "2445--2459",
    year = "1974"
}

@article{STOVICEK1984157,
title = {Quantum mechanics in a discrete space-time},
journal = {Reports on Mathematical Physics},
volume = {20},
number = {2},
pages = {157-170},
year = {1984},
issn = {0034-4877},
doi = {https://doi.org/10.1016/0034-4877(84)90030-2},
url = {https://www.sciencedirect.com/science/article/pii/0034487784900302},
author = {P. Šťovíček and J. Tolar},
}

@article{de2002quantum,
    author = "de la Torre, A. C. and Goyeneche, D.",
    title = "{Quantum mechanics in finite dimensional Hilbert space}",
    eprint = "quant-ph/0205159",
    archivePrefix = "arXiv",
    doi = "10.1119/1.1514208",
    journal = "Am. J. Phys.",
    volume = "71",
    pages = "49--54",
    year = "2003"
}

@incollection{lorente1997quantum,
  title={Quantum mechanics on discrete space and time},
  author={Lorente, Miguel},
  booktitle={New Developments on Fundamental Problems in Quantum Physics},
  pages={213--224},
  year={1997},
  publisher={Springer}
}

@article{carroll2023completely,
  title={Completely discretized, finite quantum mechanics},
  author={Carroll, Sean M},
  journal={Foundations of Physics},
  volume={53},
  number={6},
  pages={90},
  year={2023},
  publisher={Springer}
}

@article{Ashtekar:2006wn,
    author = "Ashtekar, Abhay and Pawlowski, Tomasz and Singh, Parampreet",
    title = "{Quantum Nature of the Big Bang: Improved dynamics}",
    eprint = "gr-qc/0607039",
    archivePrefix = "arXiv",
    reportNumber = "IGPG-06-7-2",
    doi = "10.1103/PhysRevD.74.084003",
    journal = "Phys. Rev. D",
    volume = "74",
    pages = "084003",
    year = "2006"
}

@article{LandscapePQC,
  title = {Landscape of polymer quantum cosmology},
  author = {Amadei, Lautaro and Perez, Alejandro and Ribisi, Salvatore},
  journal = {Phys. Rev. D},
  volume = {107},
  issue = {8},
  pages = {086007},
  numpages = {23},
  year = {2023},
  month = {Apr},
  publisher = {American Physical Society},
  doi = {10.1103/PhysRevD.107.086007},
  url = {https://link.aps.org/doi/10.1103/PhysRevD.107.086007}
}

@article{Zulfiqar:2025aef,
    author = "Zulfiqar, Aleena and Hassan, Syed Moeez",
    title = "{Polymer cosmology with polymer matter: effective dynamics}",
    eprint = "2502.04875",
    archivePrefix = "arXiv",
    primaryClass = "gr-qc",
    doi = "10.1088/1475-7516/2025/09/018",
    journal = "JCAP",
    volume = "09",
    pages = "018",
    year = "2025"
}

@article{Kreienbuehl:2013toa,
    author = "Kreienbuehl, Andreas and Paw{\l}owski, Tomasz",
    title = "{Singularity resolution from polymer quantum matter}",
    eprint = "1302.6566",
    archivePrefix = "arXiv",
    primaryClass = "gr-qc",
    doi = "10.1103/PhysRevD.88.043504",
    journal = "Phys. Rev. D",
    volume = "88",
    number = "4",
    pages = "043504",
    year = "2013"
}

@article{Garcia-Chung:2022pdy,
    author = "Garcia-Chung, Angel and Carney, Matthew F. and Mertens, James B. and Parvizi, Aliasghar and Rastgoo, Saeed and Tavakoli, Yaser",
    title = "{What do gravitational wave detectors say about polymer quantum effects?}",
    eprint = "2208.09739",
    archivePrefix = "arXiv",
    primaryClass = "gr-qc",
    doi = "10.1088/1475-7516/2022/11/054",
    journal = "JCAP",
    volume = "11",
    pages = "054",
    year = "2022"
}

@article{crowe2025,
    author = {Crowe, Sean and Evans, Stefan and Smolyaninov, Alexei},
    year = {2025},
    month = {12},
    pages = {},
    title = {Analysis of polymerized superconducting circuits},
    volume = {38},
    journal = {Superconductor Science and Technology},
    doi = {10.1088/1361-6668/ae28e6},
    eprint={2509.18016},
    archivePrefix={arXiv},
    primaryClass={quant-ph},
    url={https://arxiv.org/abs/2509.18016} 
}

@misc{StackHaarBohr,
    TITLE = {What is the {Haar} measure on the {Bohr} compactification $b\mathbb{Z}$ of the integers?},
    AUTHOR = {Mike F (https://math.stackexchange.com/users/6608/mike-f)},
    HOWPUBLISHED = {Mathematics Stack Exchange},
    NOTE = {(version: 2025-10-14)},
    EPRINT = {https://math.stackexchange.com/q/5101741},
    URL = {https://math.stackexchange.com/q/5101741},
    lastchecked = {20 December, 2025},
    urldate = {2025-12-20},
    year = {2025}

}

@book{conway90,
	author = {Conway, John B.},
	edition = {Second},
	publisher = {Springer-Verlag},
	title = {A Course in Functional Analysis},
	year = {1990}}

@article{vourdas04,
	author = {Vourdas, A.},
	doi = {10.1088/0034-4885/67/3/R03},
	journal = {Rep. Prog. Phys.},
	pages = {267-320},
	title = {Quantum systems with finite {Hilbert} space},
	volume = {67},
	year = {2004}
}

@book{barnvac07,
	editor = {Barnett, Stephen M. and Vaccaro, John A.},
	publisher = {Taylor and Francis},
	title = {The quantum phase operator: {A} review},
	year = {2007}
}

@article{CS13a,
	archiveprefix = {arXiv},
	author = {Craig, David A. and Singh, Parampreet},
	doi = {10.1088/0264-9381/30/20/205008},
	eprint = {1306.6142},
	journal = {Class. Quantum Grav.},
	pages = {205008},
	preprint = {pi-qg-249},
	primaryclass = {gr-qc},
	slaccitation = {%%CITATION = ARXIV:1306.6142;%%},
	title = {Consistent probabilities in loop quantum cosmology},
	volume = {30},
	year = {2013}
    }

@article{rovvid15,
	archiveprefix = {arXiv},
	author = {Rovelli, Carlo and Vidotto, Francesca},
	doi = {10.48550/arXiv.1502.00278},
	eprint = {1502.00278},
	journal = {Phys. Rev.},
	pages = {084037},
	primaryclass = {gr-qc},
	slaccitation = {%%CITATION = ARXIV:.1502.00278;%%},
	title = {Compact phase space, cosmological constant, and discrete time},
	volume = {D91},
	year = {2015}
}

@book{pal19,
	author = {Pal, Palash B.},
	publisher = {Cambridge University Press},
	title = {A Physicist's Introduction to Algebraic Structures},
	year = {2019}
}
}%

\end{document}